\documentclass[journal]{IEEEtran}
\IEEEoverridecommandlockouts
\usepackage{cite}
\usepackage{amsmath,amssymb,amsfonts}
\usepackage{graphicx}
\usepackage[tight,footnotesize]{subfigure}
\usepackage{textcomp}
\usepackage{diagbox}
\usepackage{colortbl,hhline}
\usepackage{booktabs}
\usepackage{pifont}
\usepackage{xcolor}
\usepackage{setspace}
\usepackage{multirow}
\usepackage{epstopdf}
\usepackage{caption}
\usepackage{threeparttable}
\usepackage{extarrows}
\usepackage{setspace}
\usepackage{footmisc}
\usepackage{tabularx}
\usepackage{makecell}
\usepackage{bbding}
\usepackage{pifont}
\usepackage{wasysym}
\usepackage[citecolor=blue]{hyperref}

\def\BibTeX{{\rm B\kern-.05em{\sc i\kern-.025em b}\kern-.08em
    T\kern-.1667em\lower.7ex\hbox{E}\kern-.125emX}}

\definecolor{red}{rgb}{1,0,0} %urgent to be fixed
\definecolor{green}{rgb}{0,0.7,0}  %parallel changes
\definecolor{blue}{rgb}{0,0,0.7} %from others
\MakeRobust{\overrightarrow}
\begin{document}

% \title{A Survey on Mobile Edge-Cloud AI-Generated Content (AIGC) Networks: Generative Models, Creative Applications, Collaborative Infrastructure}
\title{Unleashing the Power of Edge-Cloud Generative AI in Mobile Networks: A Survey of AIGC Services}
\author{Minrui Xu, Hongyang Du*, Dusit Niyato, \emph{Fellow, IEEE}, Jiawen Kang, Zehui Xiong, Shiwen Mao, \emph{Fellow, IEEE},\\  Zhu Han, \emph{Fellow, IEEE}, Abbas Jamalipour, \emph{Fellow, IEEE}, Dong In Kim, \emph{Fellow, IEEE}, Xuemin~(Sherman)~Shen, \emph{Fellow, IEEE}, Victor C. M. Leung, \emph{Life Fellow, IEEE}, and H.~Vincent~Poor, \emph{Life Fellow, IEEE}
\thanks{M.~Xu, H.~Du, and D.~Niyato are with the School of Computer Science and Engineering, Nanyang Technological University, Singapore 639798, Singapore (e-mail: minrui001@e.ntu.edu.sg; hongyang001@e.ntu.edu.sg; dniyato@ntu.edu.sg).}
\thanks{J.~Kang is with the School of Automation, Guangdong University of Technology, and Key Laboratory of Intelligent Information Processing and System Integration of IoT, Ministry of Education, Guangzhou 510006, China, and also with Guangdong-HongKong-Macao Joint Laboratory for Smart Discrete Manufacturing, Guangzhou 510006, China (e-mail: kavinkang@gdut.edu.cn).}
\thanks{Z.~Xiong is with the Pillar of Information Systems Technology and Design, Singapore University of Technology and Design, Singapore 487372, Singapore (e-mail: zehui\_xiong@sutd.edu.sg).}
\thanks{S.~Mao is with the Department of Electrical and Computer Engineering, Auburn University, Auburn, AL 36849-5201 USA (email: smao@ieee.org).}
\thanks{Z.~Han is with the Department of Electrical and Computer Engineering, University of Houston, Houston, TX 77004 USA, and also with the Department of Computer Science and Engineering, Kyung Hee University, Seoul 446-701, South Korea (e-mail: zhan2@uh.edu).}
\thanks{A.~Jamalipour is with the School of Electrical and Information Engineering,
University of Sydney, Sydney, NSW 2006, Australia (e-mail: a.jamalipour@ieee.org).}
\thanks{D.~I.~Kim is with the Department of Electrical and Computer Engineering, Sungkyunkwan University, Suwon 16419, South Korea (email:dikim@skku.ac.kr).}
\thanks{X.~Shen is with the Department of Electrical and Computer Engineering, University of Waterloo, Waterloo, ON N2L 3G1, Canada (e-mail: sshen@uwaterloo.ca).}
\thanks{V.~C.~M.~Leung is with the College of Computer Science and Software Engineering, Shenzhen University, Shenzhen 518061, China, and also with the Department of Electrical and Computer Engineering, The University of British Columbia, Vancouver BC V6T 1Z4, Canada (E-mail: vleung@ieee.org).}
\thanks{H.~V.~Poor is with the Department of Electrical and Computer Engineering, Princeton University, Princeton, NJ 08544, USA (e-mail: poor@princeton.edu).}
% offloading
% security
% performance analysis
% incentive
% energy
% mobile
% training
% FL
% AI compression
% hardware - physical layer

% table / compare of aigc surveys dl 

% the second survey: focus on cloud
}

\maketitle

\begin{abstract}
Artificial Intelligence-Generated Content (AIGC) is an automated method for generating, manipulating, and modifying valuable and diverse data using AI algorithms creatively. This survey paper focuses on the deployment of AIGC applications, e.g., ChatGPT and Dall-E, at mobile edge networks, namely mobile AIGC networks, that provide personalized and customized AIGC services in real time while maintaining user privacy. We begin by introducing the background and fundamentals of generative models and the lifecycle of AIGC services at mobile AIGC networks, which includes data collection, training, fine-tuning, inference, and product management. We then discuss the collaborative cloud-edge-mobile infrastructure and technologies required to support AIGC services and enable users to access AIGC at mobile edge networks. Furthermore, we explore AIGC-driven creative applications and use cases for mobile AIGC networks. Additionally, we discuss the implementation, security, and privacy challenges of deploying mobile AIGC networks. Finally, we highlight some future research directions and open issues for the full realization of mobile AIGC networks.
% Artificial Intelligence Generated Content (AIGC) is an automated method for generating, manipulating, and modifying valuable and diverse data using AI algorithms creatively. However, without a comprehensive understanding of existing models, applications, and challenges, it is impossible to fully realize the potential of AIGC in transforming people's lifestyles and workflows. This survey paper aims to address this gap by focusing on the deployment of AIGC applications, e.g., ChatGPT, at mobile edge networks, i.e., mobile AIGC networks that provide personalized and customized AIGC services in real-time while maintaining user privacy. We begin by introducing the background and fundamentals of generative models and the lifecycle of AIGC services at mobile AIGC networks, which includes data collection, training, fine-tuning, inference, and product management. We then discuss the collaborative cloud-edge-mobile infrastructure and technologies required to support AIGC services and enable users to access AIGC at mobile edge networks. Furthermore, we explore AIGC-driven creative applications and use cases for mobile AIGC networks. Additionally, we discuss the implementation, security, and privacy challenges of deploying mobile AIGC networks. Finally, we highlight some future research directions and open issues for the full realization of mobile AIGC networks.
\end{abstract}


\section{Introduction}

\subsection{Background}

\begin{figure}[t]
    \centering
    \includegraphics[width=1\linewidth]{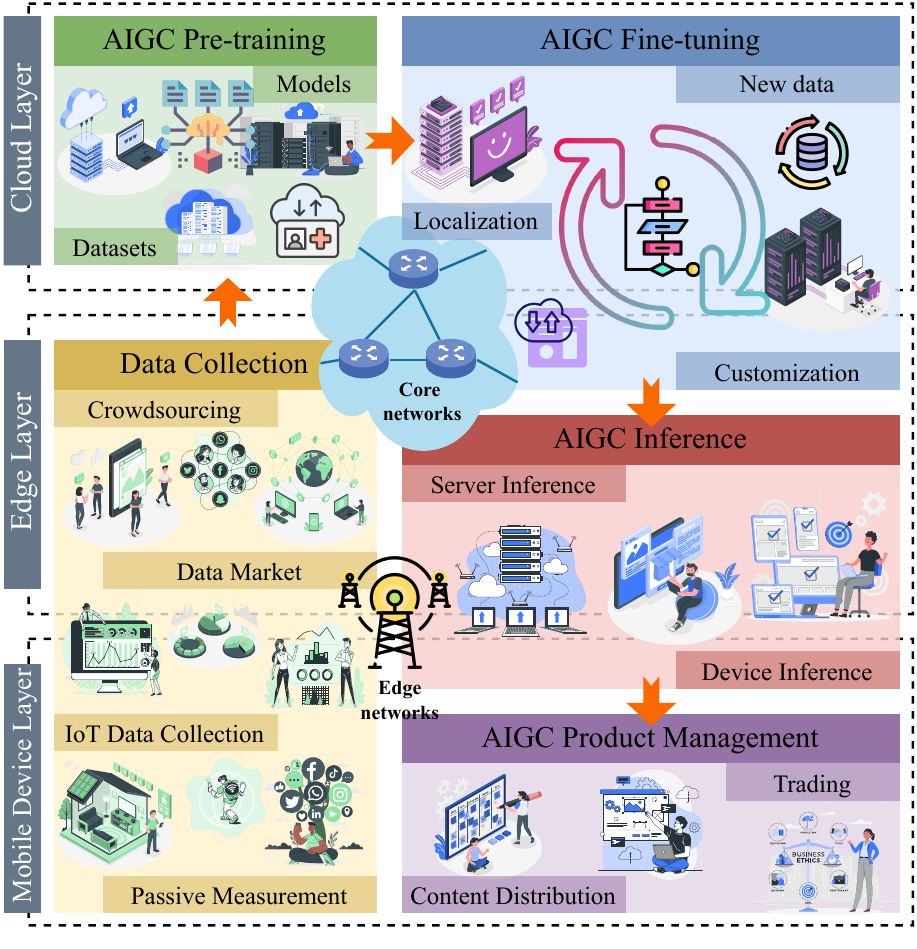}
    \caption{The overview of mobile AIGC networks, including the cloud layer, the edge layer, and the mobile device layer. The lifecycle of AIGC services, including data collection, pre-training, fine-tuning, inference, and product management, is circulated among the core networks and edge networks.}
    \label{fig:system}
\end{figure}

In recent years, artificial intelligence-generated content (AIGC) has emerged as a novel approach to the production, manipulation, and modification of data~\cite{du2023beyond}. By utilizing AI technologies, AIGC automates content generation alongside traditionally professionally-generated content (PGC) and user-generated content (UGC)~\cite{cetinic2022understanding, lee2021creators,wu2022ai}. With the marginal cost of data creation reduced to nearly zero, AIGC, e.g., ChatGPT~\cite{wang2023survey}, promises to supply a vast amount of synthetic data for AI development and the digital economy, offering significant productivity and economic value to society. The rapid growth of AIGC capabilities is driven by the continuous advancements in AI technology, particularly in the areas of large-scale and multimodal models~\cite{bond2021deep,radford2021learning}. A prime example of this progress is the development of the transformer-based DALL-E~\cite{ramesh2021zero} which is designed to generate images by predicting successive pixels. In its latest iteration, DALL-E2~\cite{ramesh2022hierarchical}, a diffusion model is employed to reduce noise generated during the training process, leading to more refined and novel image generation. In the context of text-to-image generation using {\color{black} generative AI models}, the language model serves as a guide, enhancing semantic coherence between the input prompt and the resulting image. Simultaneously, the {\color{black} generative AI model} processes existing image attributes and components, generating limitless synthesis images from existing datasets.

Based on large-scale pre-trained models with billions of parameters, AIGC services are designed to enhance knowledge and creative work fields that employ billions of people. By leveraging generative AI, these fields can achieve at least a 10\% increase in efficiency for content creation, potentially generating trillions of dollars in economic value~\cite{GenerativeAI2022}. AIGC can be applied to various forms of text generation, ranging from practical applications, such as customer service inquiries and messages, to creative tasks like activity tracking and marketing copywriting~\cite{crothers2022machine}. For example, OpenAI's ChatGPT~\cite{ChatGPT2022} can automate the generation of socially valuable content based on user-provided prompts. Through extended and coherent conversations with ChatGPT, individuals from diverse professions from all walks of life, can seek assistance in debugging code, discovering healthy recipes, writing scripts, and devising marketing campaigns. In the realm of image generation, {\color{black} generative AI models} can process existing images according to their attributes and components, enabling end-to-end image synthesis, such as generating complete images directly from existing ones~\cite{ramesh2022hierarchical}. Moreover, {\color{black} generative AI models} hold immense potential for cross-modal generation, as they can spatially process existing video attributes and simultaneously process multiple video clips automatically~\cite{ho2022imagen}.

% AIGC is designed for knowledge and creative work fields that employ billions of people, and generative AI can make them at least 10\% more efficient at creating new content and generating trillions of dollars in economic value~\cite{GenerativeAI2022}. AIGC can be used for texting, including applied texting, such as for customer service questions and messages, and creative texting, such as for activity tracking and marketing copywriting~\cite{crothers2022machine}. For instance, ChatGPT~\cite{ChatGPT2022}, developed by OpenAI, automates the generation of socially valuable content based on prompts entered by the user. In long-term and fluent conversations with ChatGPT, people from all walks of life, such as software engineers, chefs, screenwriters, and marketers, can ask ChatGPT to debug their code, search for healthy recipes, write scripts, and plan marketing campaigns. For image generation, the {\color{black} generative AI model} can process existing images based on their attributes and parts and enable end-to-end image generation, such as generating complete images directly from images~\cite{ramesh2022hierarchical}. The {\color{black} generative AI model} has great potential for cross-modal generation and can spatially process existing video attributes and process multiple video clips automatically and simultaneously~\cite{ho2022imagen}.

The benefits of AIGC in content creation, when compared to PGC and UGC, are already apparent to the public. Specifically, generative AI models can produce high-quality content within seconds and deliver personalized content tailored to users' needs~\cite{lee2021creators, kim2023bargaining}. Over time, the performance of AIGC has significantly improved, driven by enhanced models, increased data availability, and greater computational power~\cite{wang2020convergence}. On one hand, superior models~\cite{bond2021deep}, such as diffusion models, have been developed to provide more robust tools for cross-modal AIGC generation. These advancements are attributed to the foundational research in generative AI models and the continuous refinement of learning paradigms and network structures within generative deep neural networks (DNNs). On the other hand, data and computing power for generative AI training and inference have become more accessible as networks grow increasingly interconnected~\cite{westerlund2019emergence, crothers2022machine, yuan2023computing}. For instance, {\color{black} generative AI models} that require thousands of GPUs can be trained and executed in cloud data centers, enabling users to submit frequent data generation requests over core networks.
\begin{table*}[t]
\small\centering
\caption{Summary of related works versus our survey.}
\label{tab:related}
\begin{tabular}{|m{0.04\textwidth}<{\centering}|m{.07\textwidth}<{\centering}|m{.47\textwidth}<{\raggedright}|m{0.09\textwidth}<{\centering}|m{0.09\textwidth}<{\centering}|m{0.09\textwidth}<{\centering}|}
\hline
Year &
Ref. &
Contributions &
AIGC Algorithms &
AIGC Applications &
Edge Intelligence\\ \hline
2019 &
  \cite{zhang2019mobile} &
  Introduce mobile edge   intelligence, and discuss the infrastructure, implementation methodologies,   and use cases &
  \color{red}\ding{55} &
  \color{red}\ding{55} &
  \color{green}\ding{51} \\ \hline
\multirow{2}{*}{2020} & \cite{lim2020federated}           & Present the   implementation challenges of FL at mobile edge networks               & \color{green}\ding{51} & \color{red}\ding{55} & \color{green}\ding{51} \\ \cline{2-6} 
 &
  \cite{wang2020convergence} &
  Discuss the visions,   implementation details, and applications of the convergence of edge computing   and  DL &
  \color{green}\ding{51} &
  \color{red}\ding{55} &
  \color{green}\ding{51} \\ \hline
\multirow{6}{*}{2021} & \cite{makhmutov2020survey}  & Investigate the copyright laws   regarding AI-generated music                                      & \color{green}\ding{51} & \color{green}\ding{51} & \color{red}\ding{55} \\ \cline{2-6} 
 &
  \cite{cetinic2022understanding} &
  Illustrate the   interaction of art and AI from two perspectives, i.e., AI for art analysis   and AI for art creation &\color{red}\ding{55}
   &\color{green}\ding{51}
   &\color{red}\ding{55}
   \\ \cline{2-6} 
    & \cite{lee2021creators}            & Discuss the   application of computational arts in Metaverse to create surrealistic   cyberspace    & \color{green}\ding{51} & \color{green}\ding{51} & \color{red}\ding{55} \\ \cline{2-6} 
    & \cite{chen2021distributed}            & Investigate the deployment of distributed learning in wireless networks    & \color{red}\ding{55} & \color{red}\ding{55} & \color{green}\ding{51} \\ \cline{2-6} 
    & \cite{zhan2021multimodal}              & Provide a comprehensive overview of the major approaches, datasets, and metrics used to synthesize and process multimodal images          & \color{green}\ding{51} & \color{green}\ding{51} & \color{red}\ding{55} \\ \cline{2-6} 
    & \cite{shen2021holistic}            & Propose a novel conceptual architecture for 6G networks, which consists of holistic network virtualization and pervasive network intelligence    & \color{red}\ding{55} & \color{red}\ding{55} & \color{green}\ding{51} \\ \hline
\multirow{6}{*}{2022} 
% & Zhan \textit{et al.} \cite{zhan2021multimodal}              & Provide a comprehensive overview of the major approaches, datasets, and evaluation metrics for multimodal image synthesis and processing             & \color{green}\ding{51} & \color{green}\ding{51} & \color{red}\ding{55} \\ \cline{2-6}
& \cite{letaief2021edge}              & Discusses the visions and potentials of low-power, low-latency, reliable, and trustworthy edge intelligence for 6G wireless networks             & \color{red}\ding{55} & \color{red}\ding{55} & \color{green}\ding{51} \\ \cline{2-6}
& \cite{bond2021deep}              & Provide comprehensive guidance and comparison among advanced generative models, including GAN, energy-based models, VAE, autoregressive models, flow-based models, and diffusion models              & \color{green}\ding{51} & \color{red}\ding{55} & \color{red}\ding{55} \\ \cline{2-6}

& \cite{cao2022survey}              & Present fundamental   algorithms, classification and applications of diffusion models               & \color{green}\ding{51} & \color{red}\ding{55} & \color{red}\ding{55} \\ \cline{2-6} 

                      &  \cite{crothers2022machine} & Provide a   comprehensive overview of generation and detection methods for machine-generated text & \color{green}\ding{51} & \color{green}\ding{51} & \color{red}\ding{55} \\ \cline{2-6}
 &
  \cite{wang2022integrating} &
  Provide   a comprehensive examination of what, why, and how edge intelligence and   blockchain can be integrated & \color{red}\ding{55}
   & \color{red}\ding{55}
   &\color{green}\ding{51} \\ \cline{2-6} 
 &
  \cite{xu2022full}&
  Introduce the  architecture of edge-enabled Metaverse and discuss enabling technologies in communication, computing, and blockchain &\color{red}\ding{55}
   &\color{green}\ding{51}
   &\color{green}\ding{51} \\ \hline
\multirow{2}{*}{2023}          
 & \cite{nyatsanga2023comprehensive}                          &          Summarize existing works on the generation of gestures with simultaneous speeches based on deep generative models                                                                                           &\color{green}\ding{51}  &\color{green}\ding{51}  &\color{red}\ding{55}  \\ \hhline{~|-----|}
 \multirow{2}{*}{}          
 & \cite{du2023beyond}                          &          A comprehensive tutorial on applying generative diffusion model in various network optimization tasks. Case studies explore integrating the diffusion model with DRL, incentive mechanism design, semantic communications, and Internet of Vehicles (IoV) networks.                                                        &\color{green}\ding{51}  & \color{red}\ding{55}  &\color{green}\ding{51}  \\ \hhline{~|-----|}
& \cellcolor{blue!10}Ours                            &                                \cellcolor{blue!10}Investigate the deployment of mobile AIGC networks via collaborative cloud-edge-mobile infrastructure, discuss creative mobile applications and exemplary use cases, and identify existing implementation challenges                                                                 &\cellcolor{blue!10}\color{green}\ding{51}  &\cellcolor{blue!10}\color{green}\ding{51}  &\cellcolor{blue!10}\color{green}\ding{51}  \\ \hline
\end{tabular}
\end{table*}

\subsection{Motivation}

Although AIGC is acknowledged for its potential to revolutionize existing production processes, users accessing AIGC services on mobile devices currently lack support for interactive and resource-intensive data generation services~\cite{du2023beyond,zhou2019edge, zhang2019mobile}. Initially, the robust computing capabilities of cloud data centers can be utilized to pre-train generative AI models, such as GPT-3 for ChatGPT and GPT-4 for ChatGPT Plus. Subsequently, users can access cloud-based AIGC services via the core network by executing {\color{black} generative AI models} on cloud servers. However, due to their remote nature, cloud services exhibit high latency. Consequently, deploying interaction-intensive AIGC services on mobile edge networks, i.e., mobile AIGC networks, as shown in Fig.~\ref{fig:system}, should be considered a more practical option~\cite{mao2017survey, chen2020computation,kang2017neurosurgeon}. {\color{black} In mobile AIGC networks, the cloud layer handles the pre-training and fine-tuning of AIGC models, which require a significant amount of computing and storage resources. In addition, the edge layer is responsible for data collection, inference, and product management, requiring specialized hardware and software, as well as efficient communication protocols. Finally, the mobile device layer is crucial for data collection, inference, and product management with low latency, presenting unique challenges that can be addressed with specialized techniques such as federated learning and differential privacy.}
In detail, the motivations for developing mobile AIGC networks include
\begin{itemize}
    \item \emph{Low-latency:} Instead of directing requests for AIGC services to cloud servers within the core network, users can access low-latency services in mobile AIGC networks~\cite{zhang2022intelligent}. For example, users can obtain AIGC services directly in radio access networks (RANs) by downloading pre-trained models to edge servers and mobile devices for fine-tuning and inference, thereby supporting real-time, interactive AIGC. 
    \item \emph{Localization and Mobility:} In mobile AIGC networks, base stations with computing servers at the network's edge can fine-tune pre-trained models by localizing service requests~\cite{huang2020location, zhang2022toward}. Furthermore, users' locations can serve as input for AIGC fine-tuning and inference, addressing specific geographical demands. Additionally, user mobility can be integrated into the AIGC service provisioning process, enabling dynamic and reliable AIGC service provisioning.
    \item \emph{Customization and Personalization:} Local edge servers can adapt to local user requirements and allow users to request personalized services based on their preferences while providing customized services according to local service environments. On one hand, edge servers can tailor AIGC services to the needs of the local user community by fine-tuning them accordingly~\cite{lee2021creators}. On the other hand, users can request personalized services from edge servers by specifying their preferences.
    \item \emph{Privacy and Security:} AIGC users only need to submit service requests to edge servers, rather than sending preferences to cloud servers within the core network. Therefore, the privacy and security of AIGC users can be preserved during the provisioning, including fine-tuning and inference, of AIGC services.
\end{itemize}

As illustrated in Fig.~\ref{fig:system}, when users access AIGC services on mobile edge networks through edge servers and mobile devices, limited computing, communication, and storage resources pose challenges for delivering interactive and resource-intensive AIGC services. First, resource allocation on edge servers must balance the tradeoff among accuracy, latency, and energy consumption of AIGC services at edge servers. In addition, computationally intensive AIGC tasks can be offloaded from mobile devices to edge servers, improving inference latency and service reliability. Moreover, AI models that generate content can be cached in edge networks, similar to content delivery networks (CDNs)~\cite{wang2014cache,huang2022fine}, to minimize delays in accessing the model. Finally, mobility management and incentive mechanisms should be explored to encourage user participation in both space and time~\cite{liew2023economics}. Compared to traditional AI, AIGC technology requires overall technical maturity, transparency, robustness, impartiality, and insightfulness of the algorithm for effective application implementation. From a sustainability perspective, AIGC can use both existing and synthetic datasets as raw materials for generating new data. However, when biased data are used as raw data, these biases persist in the knowledge of the model, which inevitably leads to unfair algorithm results. Finally, static {\color{black} generative AI models} rely primarily on templates to generate machine-generated content that may have similar text and output structures.

\begin{figure*}[t]
    \centering
    \includegraphics[width=0.95\linewidth]{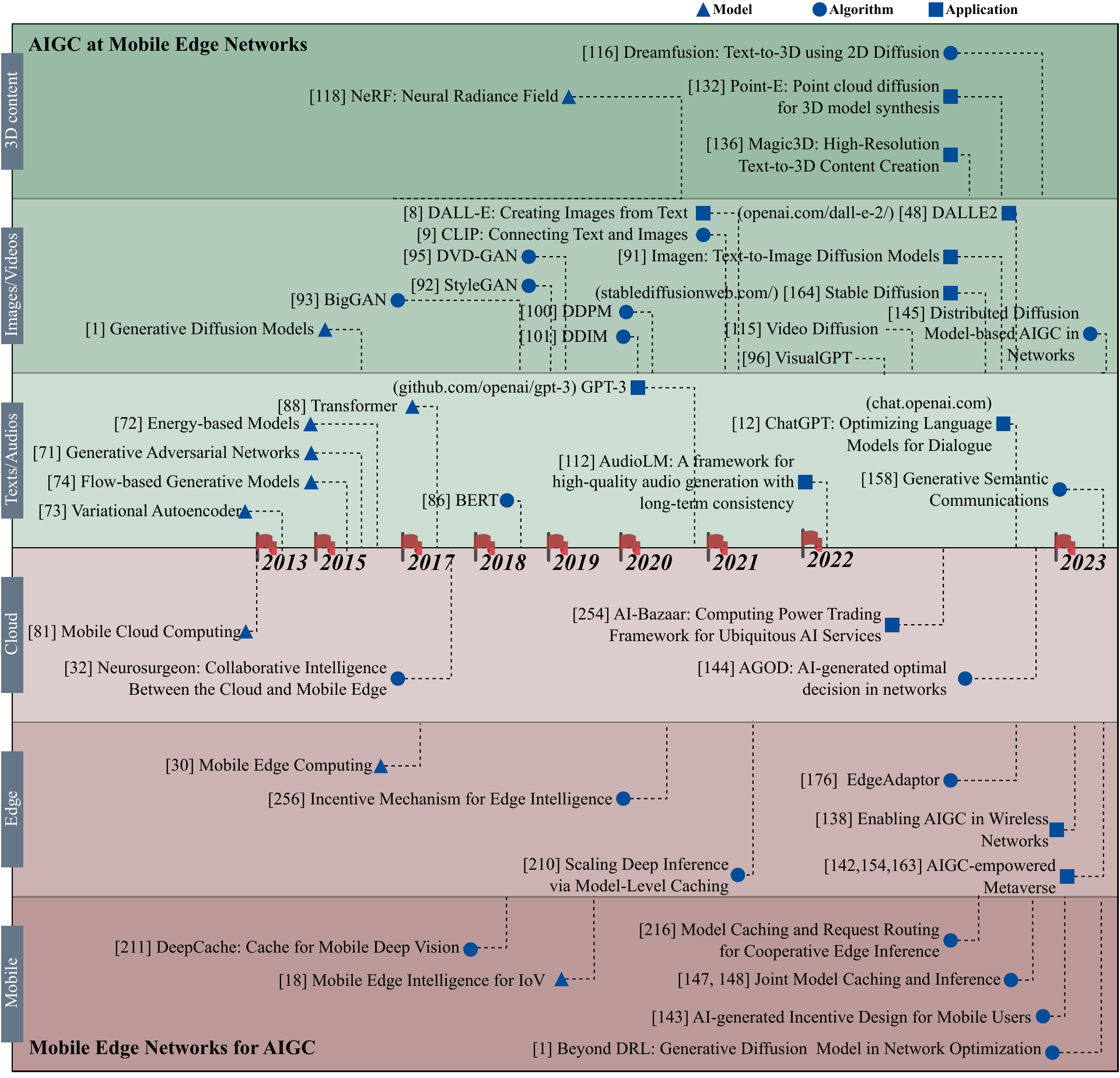}
    \caption{The development roadmap of AIGC and mobile edge networks from 2013 to Oct 2023. From the perspective of AIGC technology development, AIGC has evolved from generating text and audio to generating 3D content. From the perspective of mobile edge computing, computing has gradually shifted from cloud data centers to mobile device computing.}
    \label{fig:roadmap}
\end{figure*}

\subsection{Related Works and Contributions}
\begin{figure*}[t]
    \centering
    \includegraphics[width=1\linewidth]{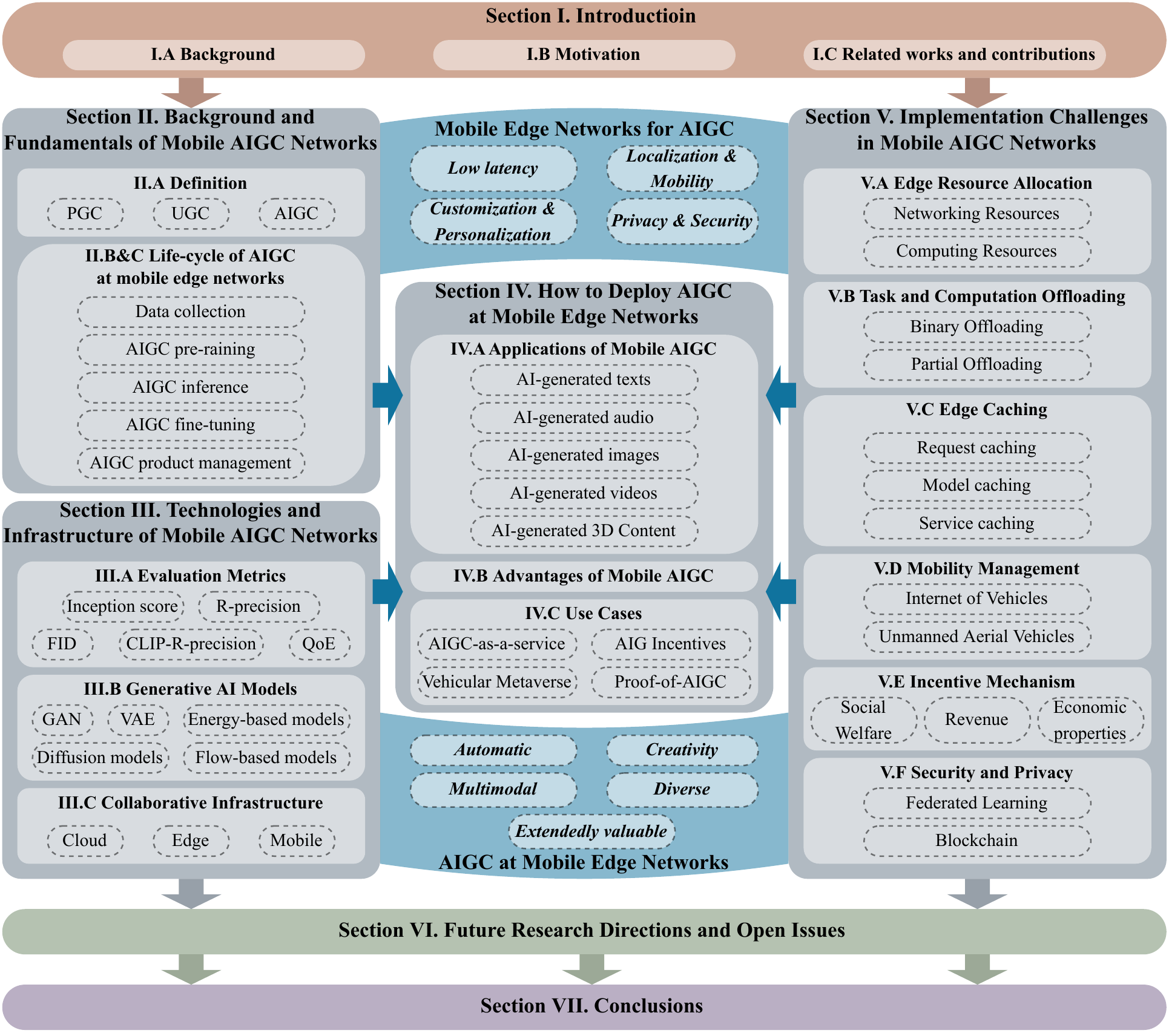}
    \caption{The outline of this survey, where we introduce the provisioning of AIGC services at mobile edge networks and highlight some essential implementation challenges about mobile edge networks for provisioning AIGC services.}
    \label{fig:outline}
\end{figure*}
In this survey, we provide an overview of research activities related to AIGC and mobile edge intelligence, as illustrated in Fig.~\ref{fig:roadmap}. Given the increasing interest in AIGC, several surveys on related topics have recently been published. Table~\ref{tab:related} presents a comparison of these surveys with this paper.

{\color{black}The study by \cite{du2023beyond} offers a focused exploration of Generative Diffusion Models (GDMs) in network optimization tasks\footnote{The code is available at \url{https://github.com/HongyangDu/GDMOPT}.}. Commencing with an essential background on GDMs, it outlines their ability to model complex data distributions effectively. This enables them to excel in diverse tasks, ranging from image generation to reinforcement learning. The paper advances by presenting case studies that integrate GDMs with Deep Reinforcement Learning, incentive mechanism design, Semantic Communications, and Internet of Vehicles networks. These case studies substantiate the model's practical utility in solving complex network optimization problems.} The study in~\cite{gozalo2023chatgpt} provides a comprehensive overview of the current {\color{black} generative AI models} published by researchers and the industry. The authors identify nine categories summarizing the evolution of generative AI models, including text-to-text, text-to-image, text-to-audio, text-to-video, text-to-3D, text-to-code, text-to-science, image-to-text, and other models. In addition, they reveal that only six organizations with enormous computing power and highly skilled and experienced teams can deploy these state-of-the-art models, which is even fewer than the number of categories. Following the taxonomy of generative AI models developed in~\cite{gozalo2023chatgpt}, other surveys discuss generative AI models in detail subsequently. The study in~\cite{crothers2022machine} examines existing methods for generating text and detecting models. The study in \cite{zhan2021multimodal} provides a comprehensive overview of the major approaches, datasets, and evaluation metrics for multimodal image synthesis and processing. Based on techniques of speech and image synthesis, the study in~\cite{nyatsanga2023comprehensive} summarizes existing works on the generation of gestures with simultaneous speeches based on deep generative models. The study in~\cite{makhmutov2020survey} investigates the copyright laws regarding AI-generated music, which includes the complicated interactions among AI tools, developers, users, and the public domain.
The study in~\cite{bond2021deep} provides comprehensive guidance and comparison among advanced generative models, including GANs, energy-based models, variational autoencoder (VAE), autoregressive models, flow-based models, and diffusion models.
As diffusion models draw tremendous attention in generating creative data, the study in~\cite{cao2022survey} presents fundamental algorithms and comprehensive classification for diffusion models. Based on these algorithms, the authors~\cite{cetinic2022understanding} illustrate the interaction of art and AI from two perspectives, i.e., AI for art analysis and AI for art creation. In addition, the authors in~\cite{lee2021creators} discuss the application of computational arts in the Metaverse to create surrealistic cyberspace.

In 6G~\cite{shen2021holistic}, mobile edge intelligence based on edge computing systems, including edge caching, edge computing, and edge intelligence, for intelligent mobile networks, is introduced in \cite{zhang2019mobile, lin2023split}. The study in~\cite{chen2021distributed} investigates the deployment of distributed learning in wireless networks. The study~\cite{lim2020federated} provides a guide to federated learning (FL) and a comprehensive overview of implementing FL at mobile edge networks. The authors offer a detailed analysis of the challenges of implementing FL, including communication costs, resource allocation, privacy, and security. In~\cite{wang2020convergence}, various application scenarios and technologies for edge intelligence and intelligent edges are presented and discussed in detail. In addition, the study~\cite{letaief2021edge} discusses the visions and potentials of low-power, low-latency, reliable, and trustworthy edge intelligence for 6G wireless networks. The study~\cite{wang2022integrating} explores how blockchain technologies can be used to enable edge intelligence and how edge intelligence can support the deployment of blockchain at mobile edge networks. The authors provide a comprehensive review of blockchain-driven edge intelligence, edge intelligence-amicable blockchain, and their implementation at mobile edge networks. 

% Mobile edge intelligence based on edge computing systems, including edge caching, edge computing, and edge intelligence, for intelligent mobile networks, is introduced in \cite{zhang2019mobile}. The study in~\cite{chen2021distributed} investigates the deployment of distributed learning in wireless networks. The study~\cite{lim2020federated} provides a guide to federal learning and a comprehensive overview of implementing FL at mobile edge networks. The authors provide a detailed analysis of the challenges of implementing FL, including communication costs, resource allocation, privacy, and security. In~\cite{wang2020convergence}, various application scenarios and technologies for edge intelligence and intelligent edges are presented and discussed in detail. In addition, the study~\cite{letaief2021edge} discusses the visions and potentials of low-power, low-latency, reliable, and trustworthy edge intelligence for 6G wireless networks. The study~\cite{wang2022integrating} explores how blockchain technologies can be used to enable edge intelligence and how edge intelligence can support the deployment of blockchain at mobile edge networks. The authors provide a comprehensive review of blockchain-driven edge intelligence, edge intelligence-amicable blockchain, and their implementation at mobile edge networks. The authors~\cite{xu2022full} provide a vision of realizing the Metaverse at mobile edge networks. In detail, enabling technologies and challenges are discussed, including communication and networking, computing, and blockchain.

Distinct from existing surveys and tutorials, our survey concentrates on the deployment of mobile AIGC networks for real-time and privacy-preserving AIGC service provisioning. We introduce the current development of AIGC and collaborative infrastructure in mobile edge networks. Subsequently, we present the technologies of deep generative models and the workflow of provisioning AIGC services within mobile AIGC networks. Additionally, we showcase creative applications and several exemplary use cases. Furthermore, we identify implementation challenges, ranging from resource allocation to security and privacy, for the deployment of mobile AIGC networks. The \emph{contributions of our survey} are as follows.

% \cite{xu2022edge, bond2021deep}
% 做一个表格, 我们全部都做了 分成两部分, 第一个是AI的, 另一个是网络的
% Creators meet the metaverse~\cite{lee2021creators}
% Understanding and creating art with AI: Review and outlook~\cite{cetinic2022understanding}
% Machine Generated Text: A Comprehensive Survey of Threat Models and Detection Methods~\cite{crothers2022machine}
% The emergence of deepfake technology: A review~\cite{westerlund2019emergence}
% Survey on Copyright Laws about Music Generated by Artificial Intelligence~\cite{makhmutov2020survey}

% Generative Diffusion Models~\cite{croitoru2022diffusion, cao2022survey}
% 3D-aware Image Synthesis~\cite{xia2022survey}

\begin{itemize}
    \item We initially offer a tutorial that establishes the definition, lifecycle, models, and metrics of AIGC services. Then, we propose the mobile AIGC networks, i.e., provisioning AIGC services at mobile edge networks with collaborative mobile-edge-cloud communication, computing, and storage infrastructure.
    \item We present several use cases in mobile AIGC networks, encompassing creative AIGC applications for text, images, video, and 3D content generation. We summarize the advantages of constructing mobile AIGC networks based on these use cases.
    \item We identify crucial implementation challenges in the path to realizing mobile AIGC networks. The implementation challenges of mobile AIGC networks stem not only from dynamic channel conditions but also from the presence of meaningless content, insecure content precepts, and privacy leaks in AIGC services.
    \item Lastly, we discuss future research directions and open issues from the perspectives of networking and computing, machine learning (ML), and practical implementation considerations, respectively.
\end{itemize}

As the outline illustrated in Fig.~\ref{fig:outline}, the survey is organized as follows. Section~\ref{sec:background} examines the background and fundamentals of AIGC. Section~\ref{sec:technologies} presents the technologies and collaborative infrastructure of mobile AIGC networks. The applications and advantages of mobile AIGC networks are discussed in Section~\ref{sec:applications}, and potential use cases are shown in Section~\ref{sec:case}. Section~\ref{sec:challenges} addresses the implementation challenges. Section~\ref{sec:future} explores future research directions. Section~\ref{sec:conclusion} provides the conclusions.

\section{Background and Fundamentals of AIGC}\label{sec:background}
In this section, the background and fundamentals of AIGC technology are presented. Specifically, we examine the definition of AIGC, its classification, and the technological lifecycle of AIGC in mobile networks. Finally, we introduce ChatGPT as a use case, which is the most famous and revolutionary application of AIGC.
% \begin{table*}[t]
% \centering
% \caption{Comparison of PGC, UGC, and AIGC}
% \label{tab:comparison}
% \begin{tabular}{|c|c|c|c|c|c|}
% \hline
%    \diagbox{Types}{Degrees}{Attributes}  & Automatic & Creativity & Multimodal & Diverse & Extendedly valuable \\ \hline
% PGC  & Medium    & Low        & Medium     & Low     & High              \\ \hline
% UGC  & Low       & Medium     & Low        & Medium  & Medium            \\ \hline
% AIGC & High      & High       & High       & High    & Low               \\ \hline
% \end{tabular}
% \end{table*}

\subsection{Definitions of PGC, UGC, and AIGC}
In the next generation of the Internet, i.e. Web 3.0 and Metaverse~\cite{du2022attention, yang2022fusing, ren2023building}, there are three primary forms of content~\cite{cetinic2022understanding}, including PGC, UGC, and AIGC.
\subsubsection{Professionally-generated Content}
PGC refers to professional-generated digital content~\cite{tiago2019youtube}. Here, the generators are individuals or organizations with professional skills, knowledge, and experience in a particular field, e.g., journalists, editors, and designers. As these experts who create PGC are typically efficient and use specialized tools, PGC has the advantages in terms of {\textit{automation}} and {\textit{multimodality}}. However, because PGC is purposeful, the {\textit{diversity}} and {\textit{creativity}} of PGC can be limited.

% One of the most critical applications of PGC is content marketing. Businesses employ PGC to develop and distribute practical, pertinent, and engaging content to their target audience to attract and keep consumers. Additionally, PGC can educate and enlighten people about a brand's goods, services, and values, fostering customer trust and loyalty~\cite{cetinic2022understanding}. For example, when Meta released their Metaverse application {\textit{Meta Horizon}}, a series of blog posts named ``what-is-horizon'' were simultaneously released to introduce the {\textit{Meta Horizon}}\footnote{Introduction of {\textit{Meta Horizon}}: https://www.meta.com/help/accounts/what-is-horizon/}. The blog posts are written by the company's in-house experts, who have extensive knowledge and experience in the industry and business.

% As the experts who create PGC are typically efficient and use specialized tools, PGC has the advantages in terms of {\textit{automation}} and {\textit{multimodality}}. However, because PGC is purposeful, the {\textit{diversity}} and {\textit{creativity}} of PGC can be limited.

\subsubsection{User-generated Content}
UGC refers to digital material generated by users, rather than by experts or organizations~\cite{krumm2008user}. The users include website visitors and social media users. UGC can be presented in any format, including text, photos, video, and audio. 
% UGC can be widely shared on social media platforms, online forums, and other online communities, and is seen by other users as a useful source of information. An example of UGC in the context of Non-Fungible Tokens (NFTs)~\cite{chohan2021non} would be the creation and selling of digital art or other creative works by individual artists or creators.
% With the emergence of NFTs, many artists and producers have been able to monetize their digital content through the sale of unique, one-of-a-kind tokens representing their creations.
% In this instance, the digital art or creative work itself is UGC since it is made by individual users and shared on blockchain platforms for sale as NFTs.
% The ownership of the NFTs is recorded on the blockchain, offering a safe and transparent way of authenticating the legitimacy and ownership of the works. 
{\color{black}The barrier for users to create UGC} is being lowered. For example, some websites\footnote{Example of a website that allows users to create their own UGC: https://ugc-nft.io/Home} allow users to create images with a high degree of freedom on a pixel-by-pixel basis. As a result, UGC is more {\textit{creative}} and {\textit{diverse}}, thanks to a wide user base. However, UGC is less {\textit{automated}} and less {\textit{multimodal} than the PGC that is generated by experts.

\subsubsection{AIGC}
AIGC is generated by using generative AI models according to input from users. Because AI models can learn the features and patterns of input data from the human artistic mind, they can develop a wide range of content. The recent success of text-to-image applications based on the diffusion model~\cite{croitoru2022diffusion} and the ChatGPT based on transformer~\cite{ChatGPT2022} has led to AIGC gaining a lot of attention. We have defined the AIGC according to its characteristics as follows
\begin{itemize}
    \item Automatic: AIGC is generated by AI models automatically. After the AI model has been trained, users only need to provide input, such as the task description, to efficiently obtain the generated content. The process, from input to output, does not require user involvement and is done automatically by the AI models.
    \item Creativity: AIGC refers to an idea or item that is innovative. For example, AIGC is believed to be leading to the development of a new profession, called Prompt Engineer~\cite{oppenlaender2022prompt}, which aims to improve human interaction with AI. In this context, the prompt serves as the starting point for the AI model, and it significantly impacts the originality and quality of the generated content. A well-crafted prompt that is specific results in more relevant and creative content than a vague or general prompt.
    \item Multimodal: The AI models to generate AIGC can handle multimodal input and output. For example, ChatGPT~\cite{ChatGPT2022} allows conversational services that employ text as input and output, DALL-E~2~\cite{marcus2022very} can create original, realistic images from a text description, and AIGC services with voice and 3D models as input or output are progressing~\cite{chi2021audio}.
    \item Diverse: AIGC is diverse in service personalization and customization. On the one hand, users can adjust the input to the AI model to suit their preferences and needs, resulting in a personalized output. On the other hand, AI models are trained to provide diverse outputs. For example, consider the DALL-E~2 as an example, the model can generate images of individuals that more correctly represent the diversity of the global population, even with the same text input.
    \item Extendedly valuable: AIGC should be extendedly valuable to society, economics, and humanity~\cite{chui2018notes}. For example, AI models can be trained to write medical reports and interpret medical images, enabling healthcare personnel to make accurate diagnoses.
\end{itemize}

AIGC provides various advantages over PGC and UGC, including better efficiency, originality, diversity, and flexibility. The reason is that AI models can produce vast amounts of material quickly and develop original content based on established patterns and principles. These advantages have led to the growing creative applications of the {\color{black} generative AI models}, which are discussed in Section~\ref{application}.
% \subsection{A Taxonomy of AIGC}

% \subsubsection{Text-based Generation} Text-based generated content mainly uses language models and generative adversarial networks to generate new texts, such as poems, novels, and essays. For text generation content, language models and generative adversarial networks learn the grammatical rules, vocabulary, and semantics of the language during training, and this knowledge can be used to generate reasonable text.

% % 段落结构: 功能总结, 实现/操作流程, 举例, 优势总结

% Photorealistic Text-to-Image Diffusion Models with Deep Language Understanding~\cite{saharia2022photorealistic}

% Hierarchical Text-Conditional Image Generation with CLIP Latents~\cite{ramesh2022hierarchical}

% \subsubsection{Image-based Generation}

% Image-based generative content mainly uses adversarial networks to generate new images, such as videos, pictures, and animations. These algorithms can generate new images and modify or synthesize new ones based on existing ones. For image generation content, generative adversarial networks are trained to learn features such as texture, structure, objects, and scenes of images, which can be used to generate reasonable images.

% Palette: Image-to-Image Diffusion Models~\cite{saharia2022palette}
\subsection{Serving ChatGPT at Mobile Edge Networks}
\begin{figure}[t]
    \centering
    \includegraphics[width=1\linewidth]{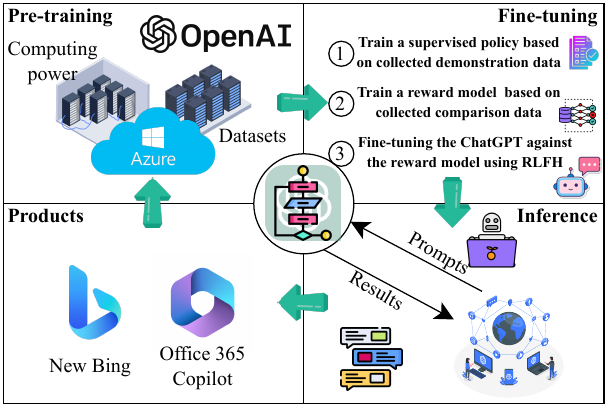}
    \caption{The four development stages of ChatGPT, including pre-training, fine-tuning, inference, and product management.}
    \label{fig:chatgpt}
\end{figure}

ChatGPT, developed by OpenAI, excels at generating human-like text and engaging in conversations~\cite{ChatGPT2022}. Based on the GPT-3~\cite{brown2020language}, this transformer-based neural network model can produce remarkably coherent and contextually appropriate text. Among its primary advantages, ChatGPT is capable of answering questions, providing explanations, and assisting with various tasks in a manner nearly indistinguishable from human responses. As illustrated in Fig.~\ref{fig:chatgpt}, the development of ChatGPT involves four main stages, including pre-training, fine-tuning, inference, and product management.

\subsubsection{Pre-training}
In the initial stage, known as pre-training, the foundation model of ChatGPT, GPT-3, is trained on a large corpus of text, which includes books, articles, and other information sources. This process enables the model to acquire knowledge of language patterns and structures, as well as the relationships between words and phrases. The base model, GPT-3, is an autoregressive language model with a Transformer architecture that has 175 billion parameters, making it one of the largest language models available. During pre-training, GPT-3 is fed with a large corpus of text from diverse sources, such as books, articles, and websites for self-supervised learning, where the model learns to predict the next word in a sentence given the context. To train the foundation model, the technique used is called maximum likelihood estimation, where the model aims to maximize the probability of predicting the next word correctly. Training GPT-3 demands significant computational resources and time, typically involving specialized hardware like graphics processing units (GPUs) or tensor processing units (TPUs). {\color{black}The exact resources and time required depend on} factors such as model size, dataset size, and optimization techniques.

\subsubsection{Fine-tuning}
The fine-tuning stage of ChatGPT involves adapting the model to a specific task or domain, such as customer service or technical support, to enhance its accuracy and relevance for that task. To transform ChatGPT into a conversational AI, a supervised learning process is employed using a dataset containing dialogues between humans and AI models~\cite{ouyang2022training}. To optimize ChatGPT's parameters, a reward model for reinforcement learning is built by ranking multiple model responses by quality. Alternative completions are ranked by AI trainers, and the model uses these rankings to improve its performance through several iterations of Proximal Policy Optimization~\cite{schulman2017proximal}. This technique allows ChatGPT to learn from its mistakes and improve its responses over time.

\subsubsection{Inference}
In the inference stage, ChatGPT generates text based on a given input or prompt, testing the model's ability to produce coherent and contextually appropriate responses relevant to the input. ChatGPT generates responses by leveraging the knowledge it acquired during pre-training and fine-tuning, analyzing the context of the input to generate relevant and coherent responses. In-context learning involves analyzing the entire context of the input~\cite{dong2022survey}, including the dialogue history and user profile, to generate responses that are personalized and tailored to the user's needs. ChatGPT employs chain-of-thought to generate responses that are coherent and logical, ensuring that the generated text is not only contextually appropriate but also follows a logical flow. The resources consumed during inference are typically much lower than those required for training, making real-time applications and services based on ChatGPT computationally feasible.

\subsubsection{Product Management}
The final product management phase involves deploying the model in a production environment and ensuring its smooth and efficient operation. In the context of mobile edge networks, the applications of AI-powered tools such as the new Bing~\cite{Newbing} and Office 365 Copilot~\cite{OfficeCopilot} could be particularly useful due to their ability to provide personalized and contextually appropriate responses while conserving resources. The new Bing offers a new type of search experience with AI-powered features such as detailed replies to complex questions, summarized answers, and personalized responses to follow-up questions, while Office 365 Copilot, powered by GPT-4 from OpenAI, assists with generating documents, emails, presentations, and other tasks in Microsoft 365 apps and services. These tools can be integrated into mobile edge networks with specialized techniques that balance performance and accuracy while preserving data integrity.

\begin{itemize}
    \item New bing: The new Bing offers a set of AI-powered features that provide a new type of search experience, including detailed replies to complex questions, summarized answers, and personalized responses to follow-up questions. Bing also offers creative tools such as assistance with writing poems and stories. In the context of mobile edge networks, Bing's ability to consolidate reliable sources across the web and provide a single, summarized answer could be particularly useful for users with limited resources. Additionally, Bing's ability to generate personalized responses based on user behavior and preferences could improve the experience of users in mobile edge networks.
    \item Office 365 copilot: Microsoft has recently launched an AI-powered assistant named Office 365 Copilot, which can be summoned from the sidebar of Microsoft 365 apps and services. Copilot can help users generate documents, emails, and presentations, as well as provide assistance with features such as PivotTables in Excel. It can also transcribe meetings, remind users of missed items, and provide summaries of action items. However, when deploying Copilot in mobile edge networks, it is important to keep in mind the limited resources of these devices and to develop specialized techniques that can balance performance and accuracy while preserving data integrity.
\end{itemize}

In addition to the previously mentioned commercial applications, ChatGPT holds substantial commercial potential owing to its capacity for producing human-like text, which is characteristically coherent, pertinent, and contextually fitting. This language model can be fine-tuned to accommodate a diverse array of tasks and domains, rendering it highly adaptable for numerous applications. ChatGPT exhibits remarkable proficiency in comprehending and generating text across multiple languages. Consequently, it can facilitate various undertakings, such as composing emails, developing code, generating content, and offering explanations, ultimately leading to enhanced productivity. By automating an assortment of tasks and augmenting human capabilities, ChatGPT contributes to a paradigm shift like human work, fostering new opportunities and revolutionizing industries. In addition to ChatGPT, more use cases developed by various generative AI models are discussed in Section~\ref{sec:case}.

\subsection{Life-cycle of AIGC at Mobile Edge Networks}
AIGC has gained tremendous attention as a technology superior to PGC and UGC. However, the lifecycle of the AIGC is also more elaborate. In the following, we discuss the AIGC lifecycle with mobile edge network enablement:
\subsubsection{Data Collection}
Data collection is an integral component of AIGC and plays a significant role in defining the quality and diversity of the material created by AI systems~\cite{yang20226g}. The data used to train AI models influences the patterns and relationships that the AI models  learn and, consequently, the output. There are several data collection techniques for AIGC:
\begin{itemize}
    \item Crowdsourcing: Crowdsourcing is the process of acquiring information from a large number of individuals, generally via the use of online platforms~\cite{daniel2018quality}. Crowdsourced data may be used to train ML models for text and image generation, among other applications. One common example is the use of Amazon Mechanical Turk\footnote{The website of Amazon Mechanical Turk as a crowdsourcing marketplace: https://www.mturk.com/}, where individuals are paid to perform tasks such as annotating text or images, which can then be used to train {\color{black} generative AI models}.
    \item Data Market: Another way to obtain data is to buy it from a data provider. For example, Datatang\footnote{The website of Datatang: https://www.datatang.ai/} is a firm that offers high-quality datasets and customized data services to assist businesses in enhancing the performance of their AI models. By giving access to varied, high-quality data, Datatang enables organizations to train AI models that are more accurate and effective, resulting in enhanced business performance and results.
    \item Internet-of-Things (IoT) data collection: In IoT, edge devices can help to collect the data, e.g., Global Positioning System (GPS) records and wireless sensing data~\cite{zhang2022holographic}. \textcolor{black}{For example, mobile phone sensors can track the device's movement and location or users~\cite{deng2019data}. The sensors can be used to collect data on the location, speed, and direction of movement of the device. These data are important for the implementation of personalized {\color{black} generative AI models}. In addition to these traditional data collection methods, large-scale datasets are specifically designed for training generative AI models. For instance, the LAION-400M dataset~\cite{schuhmann2021laion}, a large-scale, non-curated dataset consisting of 400 million English (image, text) pairs, is used in training models like CLIP.}
    \item Passive data collection can be achieved with the help of edge networks~\cite{du2022semantic}. In the smart city, sensors can be placed at strategic locations, such as on lamp posts, buildings, or other structures, to collect data on various aspects of the city environment. The data obtained by the sensors might be used to train AI models, which could subsequently be utilized to produce insights on air quality, traffic flow, and pedestrian density. \textcolor{black}{Using data obtained from air quality sensors, an AI model can be trained to forecast air quality. The model can then be used to create a real-time map of the city's air quality. This real-time map could be used to guide policy choices about the management of air quality, leading to the development of generative AI models that are capable of generating decision solutions for managing air quality.} 
\end{itemize}
After the data has been collected, the data is then used to train the {\color{black} generative AI model}.

\subsubsection{Pre-training}
The collected data is used to train the {\color{black} generative AI model}. In mobile networks, training is typically done by central servers with powerful computing power. During the training process, the generative model automatically learns the patterns and features in the data and predicts the target outcome.
We introduce several generative AI technologies in Section~\ref{GAI}, including Generative Adversarial Networks (GANs), VAE, Flow-based models, and diffusion models. These different training techniques have different strengths and weaknesses. The choice of technique depends on the specific requirements of the AIGC task, the available data, the desired output, and the computational resources available. After training is complete, cloud data centers can accept requests uploaded by network users to perform subsequent fine-tuning and inference tasks. Alternatively, cloud data centers can deliver the trained {\color{black} generative AI models} down to network edge servers, which can process user requests locally. 
It is important to note the substantial computational resources required for the pre-training of {\color{black} generative AI models}. For instance, the pre-training process of the Stable Diffusion model, a large-scale AI model developed by Stability AI, was conducted on a cloud cluster with 256 Nvidia A100 GPUs for about 150,000 hours, which equates to a cost of approximately \$600,000 (https://huggingface.co/CompVis/stable-diffusion-v1-4). This highlights the intensive computational demands of training such models.

\subsubsection{Fine-tuning}
Fine-tuning in AIGC is the process of adjusting a pre-trained {\color{black} generative AI model} to new tasks or domains by including a modest quantity of extra data. This approach can be used to enhance the model's performance on a given task or in a specific area by adjusting the AI model's parameters to suit the new data better. In mobile networks, tasks of fine-tuning can be performed by the edge network, using the small-size dataset uploaded by mobile users.

\subsubsection{Inference}
Using the trained {\color{black} generative AI model}, inference can be done, which involves generating the desired content based on the input. {\color{black} generative AI models} are traditionally managed via centralized servers, such as the Hugging Face platform~\cite{jain2022hugging}. In this setting, a large number of users make requests to the central server, wait in line, and obtain the requested services. Researchers aim to install AIGC services on edge networks to prevent request congestion and optimize service latency. Edge devices have sufficient computational capacity for AIGC inference and are closer to consumers than central servers. Therefore, users can interact with devices with a reduced transmission delay. In addition, as AIGC services are dispersed to several edge devices, the latency can be significantly reduced.

\subsubsection{Product Management}
The preceding stages cover content generation. However, as an irreplaceable online property comparable to NFT, AIGC possesses unique ownership, copyright, and worth for each content. Consequently, the preservation and management of AIGC products should be incorporated into the AIGC life cycle. Specifically, we refer to the party requesting the production of the AIGC as producers, e.g., mobile users or companies, who hire AIGC generators, e.g., network servers, to perform the AIGC tasks. Then, the main process in AIGC product management includes:
\begin{itemize}
    \item {\textit{Distribution:}} After the content is generated in network edge servers, the producers acquire ownership of the AIGC products. Consequently, they have the right to distribute these products to social media or AIGC platforms through edge networks
    \item {\textit{Trading:}} Since AIGC products are regarded as a novel kind of non-fungible digital properties, they can be traded. The trading process can be modeled as a fund ownership exchange between two parties.
\end{itemize}

To implement the aforementioned AIGC lifecycle in mobile networks, we further investigate the technical implementation of AIGC in the following section.

\section{Technologies and Collaborative Infrastructure of Mobile AIGC Networks}\label{sec:technologies}
{\color{black}
In this section, we delve into the technologies and collaborative infrastructure of mobile AIGC networks. This section aims to provide a comprehensive understanding of the rationale and objectives of edge computing systems designed to support AIGC. Before we explore the design of these systems, it is crucial to establish the performance metrics that measure whether the system can maximize user satisfaction and utility. 
}
\subsection{Evaluation Metrics of Generative AI Models and Services} \label{sec:metric}
% {\color{red} formulas for calculating these evaluation metrics.}
% \cite{kynkaanniemi2019improved, naeem2020reliable, benny2021evaluation, park2021benchmark}
We first discuss several metrics for assessing the quality of {\color{black} generative AI models}, which can be used by AIGC service providers and users in mobile networks.

\subsubsection{Inception Score} The Inception Score (IS) can be used to measure the accuracy of images generated by {\color{black} generative AI models} in the mobile network~\cite{kynkaanniemi2019improved}. The IS is based on the premise that high-fidelity generated images should have high-class probabilities, which suggest a reliable classification model, and a low Kullback-Leibler (KL) divergence between the projected class probability and a reference class distribution. To compute the IS, an exponential function is applied to the KL divergence between the anticipated class probabilities and the reference class distribution. The resulting value is then averaged over all created photos to obtain the IS. A higher IS indicates better overall image quality.

\subsubsection{Frechet Inception Distance} 
The Frechet Inception Distance (FID) has emerged as a well-established metric for evaluating the effectiveness of generative models, particularly GANs, in terms of image quality and diversity~\cite{park2021benchmark}. FID leverages a pre-trained Inception network to calculate the distance between actual and synthetic image embeddings. This metric can be used by {\color{black} generative AI model} providers to evaluate the quality of their generative models in mobile networks. Additionally, users can assess the capabilities of AIGC service providers through multiple requests for services based on FID measurements. However, when evaluating conditional text-to-image synthesis, FID only measures the visual quality of the output images, ignoring the adequacy of their conditioning on the input text~\cite{wu2023visual}. Thus, while FID is an excellent evaluation metric for assessing image quality and diversity, it is limited when applied to conditional text-to-image synthesis.

\subsubsection{R-Precision} 
R-Precision is a standard metric to evaluate how AI-generated images align with text inputs~\cite{benny2021evaluation}. In mobile networks, the {\color{black} generative AI model} producers can retrieve matching text from 100 text candidates using the AI-generated image as a query. The R-Precision measures the proportion of relevant items retrieved among the top-R retrieved items, where R is typically set to 1. Specifically, the Deep Attentional Multimodal Similarity Model (DAMSM) is commonly used to compute the text-image retrieval similarity score~\cite{xu2018attngan}. DAMSM maps each subregion of an image and its corresponding word in the sentence to a joint embedding space, allowing for the measurement of fine-grained image-text similarity for retrieval. However, it should be noted that text-to-image {\color{black} generative AI models} can directly optimize the DAMSM module used to calculate R-Precision. This results in the metric being model-specific and less objective, limiting the evaluation of {\color{black} generative AI models} in mobile networks.

\subsubsection{CLIP-R-Precision} CLIP-R-Precision is an assessment metric to address the model-specific character of the R-Precision metric~\cite{naeem2020reliable}. Instead of the conventional DAMSM, the suggested measure uses the latest multimodal CLIP model~\cite{radford2021learning} to obtain R-Precision scores. Here, CLIP is trained on a massive corpus of web-based image-caption pairings and is capable, via a contrastive aim, of bringing together the two embeddings (visual and linguistic). Thus, the CLIP-R-Precision can provide a more objective evaluation of text-to-image {\color{black} generative AI model} performance in mobile networks.

% \subsubsection{Mean Reciprocal Rank}
% Mean Reciprocal Rank (MRR) is a widely-used metric for evaluating the effectiveness of recommendation systems, which can be used in the context of AI-generated content such as personalized advertisements. MRR considers the rank of the first relevant AI-generated content item in the list of recommendations provided to the user. 
% The MRR is computed as the average of the reciprocal ranks of the first relevant item across all queries or users. For example, if the first relevant AI-generated advertisement appears at rank 3 for a specific user, the reciprocal rank would be 1/3. A higher MRR value indicates a more effective recommendation system, as relevant AI-generated content is ranked higher for users.
% By employing MRR, mobile AIGC network providers can assess the performance of their content recommendation systems and fine-tune them to deliver more accurate and relevant AI-generated content, such as personalized advertisements, to their users. This, in turn, can enhance user satisfaction and engagement with the mobile AIGC network.

% In mobile networks, service providers can utilize MRR to gauge the relevance and accuracy of their AI-generated content recommendation systems.

\subsubsection{Quality of Experience}
The Quality of Experience (QoE) metric plays a critical role in evaluating the performance of AIGC in mobile network applications~\cite{du2023rethinking}. QoE measures user satisfaction with the generated content, considering factors such as visual quality, relevancy, and utility. Gathering and analyzing user surveys, interaction, and behavioral data are standard methods used to determine QoE. In addition, the definition of QoE can vary depending on the objectives of the mobile network system designer and the user group being considered. With the aid of QoE, AIGC performance can be improved, and new models can be created to meet user expectations. It is essential to account for QoE when analyzing the performance of AIGC in mobile network applications to ensure that the generated content meets user expectations and provides a great user experience.

\begin{figure}
    \centering
    \includegraphics[width=1\linewidth]{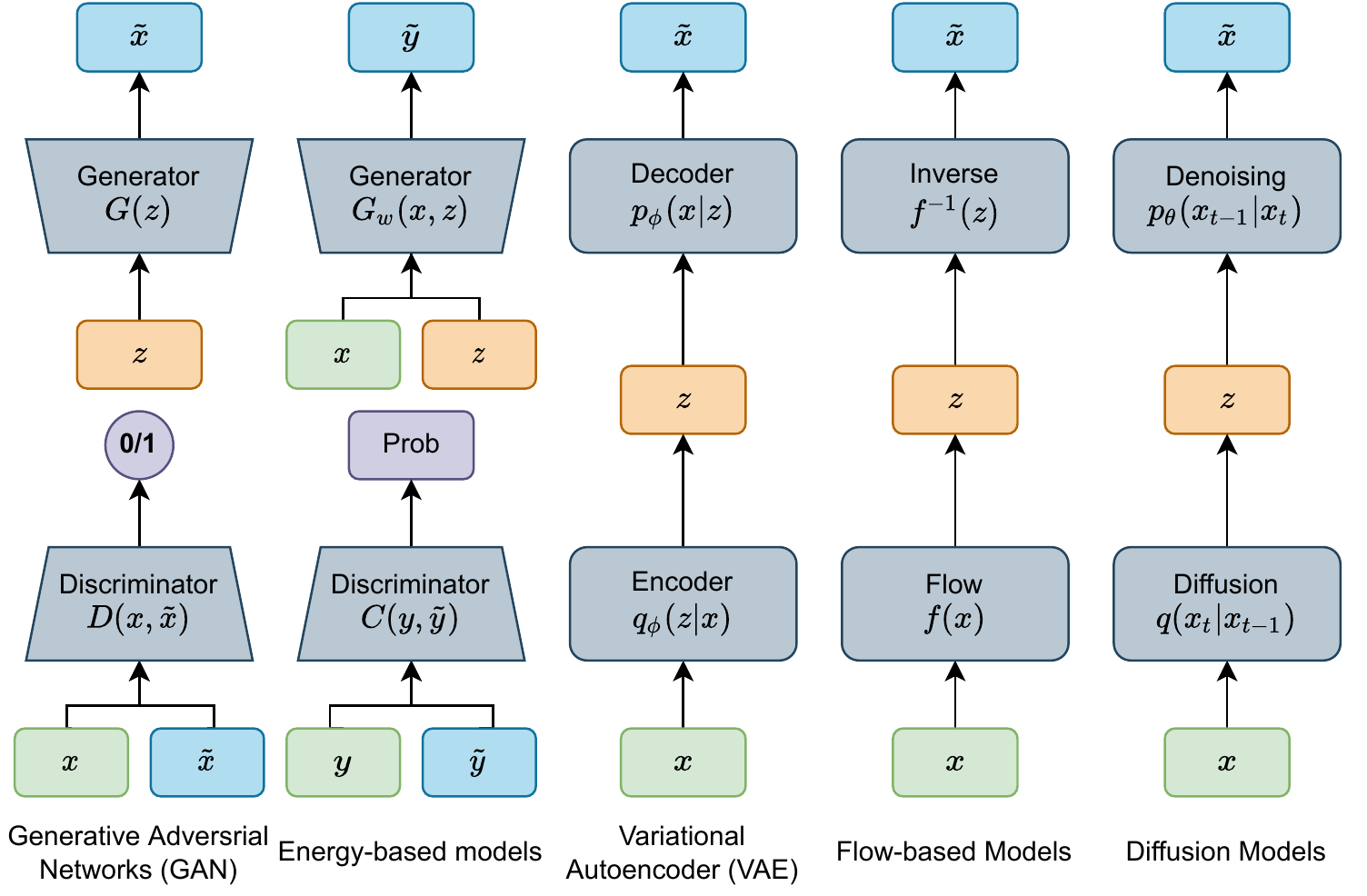}
    \caption{The model architecture of generative AI models, including generative adversarial networks, energy-based models, variational autoencoder, flow-based models, and diffusion models.}
    \label{fig:archtecture}
\end{figure}

Based on the aforementioned evaluation metrics, diverse and valuable synthetic data can be generated from deep generative models. Therefore, in the next section, we introduce several generative AI models for mobile AIGC networks.

\subsection{Generative AI Models}~\label{GAI}

Generative AI models aim to understand and replicate the true data distribution of input data through iterative training. This understanding allows the generation of novel data that closely aligns with the original distribution. As depicted in Fig.~\ref{fig:archtecture}, this section delves into five fundamental generative models: Generative Adversarial Networks (GANs), energy-based models, Variational Autoencoders (VAEs), flow-based models, and diffusion models.

\subsubsection{Generative Adversarial Networks}

{\color{black}GANs are a fundamental framework} for AIGC, comprising a generative model and a discriminative model~\cite{goodfellow2020generative}. The generative network aims to generate data that is as realistic and similar to the original data as possible to deceive the discriminative model based on the data in the original dataset. Conversely, the discriminant model's task is to differentiate between real and fake instances. During the GAN training process, the two networks continually enhance their performance by competing against each other until they reach a stable equilibrium. 
% By the end of the training process, the discriminator network is no longer able to differentiate between real and fake data. However, GANs have limited control over the output and can produce meaningless images. Moreover, they generate low-resolution images, only augment the existing dataset rather than creating new content on the original dataset, and cannot generate new content across modalities.
{\color{black} The advantages and disadvantages of GANs can be summarized as follows~\cite{goodfellow2020generative}:
\begin{itemize}
\item \textbf{Advantages:}
\begin{itemize}
\item GANs can generate new data closely resembling the original dataset, making them useful for tasks such as image synthesis and text-to-image translation.
\item The adversarial training process leads to continuous improvement in the performance of both the generative and discriminative models.
\end{itemize}

\item \textbf{Disadvantages:}
\begin{itemize}
\item GANs can be difficult to train because the two networks in a GAN, i.e., the generator and the discriminator, constantly compete against others, making training unstable and slow.
\item GANs primarily augment the existing dataset rather than creating entirely new content, limiting their ability to generate new content with other modalities.
\end{itemize}
\end{itemize}}

\subsubsection{Energy-based Generative Models}
{\color{black} Energy-based generative models are a class of generative models that represent input data using energy values~\cite{zhao2016energy}. These models define an energy function and then minimize the input data's energy value through optimization and training.} This approach is easily comprehensible, and the models exhibit excellent flexibility and generalization ability in providing AIGC services. EBMs capture dependencies by associating an unnormalized probability scalar (energy) to each configuration of the combination of observed and latent variables. Inference consists of finding latent variables that minimize the energy given a set of observed variables. The model learns a function that associates low energies with the latent variables' correct values and higher energies with incorrect values.

\subsubsection{Variational Autoencoder}

{\color{black}The VAE~\cite{kingma2019introduction} is a type of {\color{black}generative models that consist of two primary components}: an encoder and a decoder network. The encoder transforms the input data into a set of parameters (mean and variance) in a latent space. These parameters are then used to sample from the latent space, generating latent variables. The decoder takes these latent variables as input and generates new data. VAEs differ from GANs in their training methods. While GANs are trained using a supervised learning approach, VAEs employ an unsupervised learning approach.} This difference is reflected in how they generate data. VAEs generate data by sampling from the learned distribution, while GANs approximate the data distribution using the generator network.

\subsubsection{Flow-based Generative Models}
Flow-based generative models~\cite{rezende2015variational} facilitate the data generation process by employing probabilistic flow formulations. Additionally, these models compute gradients during generation using backpropagation algorithms, enhancing training and learning efficiency. Consequently, flow-based models in mobile edge networks present several benefits. One such advantage is computational efficiency. Flow-based models can directly compute the probability density function during generation, circumventing resource-intensive calculations. This promotes more efficient computation within mobile edge networks.

\subsubsection{Generative Diffusion Models}
Diffusion models are likelihood-based models trained with Maximum Likelihood Estimation (MLE)~\cite{cao2022survey}, as opposed to GANs trained with a minimax game between the generator and the discriminator. Therefore, the pattern collapses and thus the training instabilities can be avoided. Specifically, diffusion models are inspired by non-equilibrium thermodynamics theory~\cite{du2023beyond}. They learn the inverse diffusion process to construct the desired data sample from noise by defining a Markov chain of diffusion steps that gradually add random noise to the data. In addition, diffusion can mathematically transform the computational space of the model from pixel space to a low-dimensional space called latent space. This reduces the computational cost and time required and improves the training efficiency of the model. Unlike VAE or flow-based models, diffusion models are learned using a fixed procedure, and the hidden variables have high dimensions that are the same as the original data. {\color{black}This versatility and computational efficiency make diffusion models highly effective across a broad range of applications, including computer vision, natural language processing, audio synthesis, 3D modeling, and network optimization~\cite{du2023beyond}}.

{\color{black}\subsubsection{Large Language Models}
Large language models (LLM), which consist of billions of parameters, are trained on large-scale datasets~\cite{zhao2023survey}, and thus demonstrate the ability to handle various downstream tasks. LLMs can understand input prompts and generate human-like text in response. These models have greatly influenced our interaction with technology and have helped pave the way for advancements in artificial general intelligence. For instance, Google's PaLM-E~\cite{driess2023palm} is an embodied language model that can handle tasks involving reasoning, visuals, and language. It can process multimodal sentences and transfer knowledge across domains, enabling it to perform tasks such as robot planning and embodied question answering.

In wireless networks, deploying LLMs faces several important issues from the perspectives of wireless communications, computing, and storage~\cite{chen2023big}. In terms of wireless communications, efficient utilization of computing and energy resources is crucial due to the large sizes of LLMs and the need to process vast amounts of data~\cite{bariah2023large}. Compatibility with existing infrastructure is also a concern, including potential limitations in data, configuration, and transmission protocols. From a computing perspective, LLMs face challenges such as long response times, high bandwidth requirements, and data privacy concerns~\cite{lin2023pushing}. Deploying LLMs at the network edge is necessary to address these challenges. The staggering size of LLMs poses significant obstacles for mobile edge computing (MEC) systems. Balancing inference accuracy and memory usage is crucial when employing parameter sharing in LLMs. Furthermore, there are still numerous open research problems regarding the utilization of MEC systems to support LLMs. In terms of storage and caching~\cite{xu2023sparks}, managing the computation and memory-intensive nature of LLMs is essential during loading and execution on edge servers. Core network latency and congestion can be problematic when offloading services for caching and inference, particularly due to the high number of service requests. Designing effective caching algorithms that consider the frequency of use for LLMs and user preferences is important. Dynamic cache structures based on service runtime configuration, such as batch size, add complexity to cache loading and eviction. Balancing the tradeoff between latency, energy consumption, and accuracy is a key consideration when managing cached models at edge servers.}

%{\color{red}
%Model size \\
%Computation time \\
%Training + Inference \\
%Advantage \\
% Mobility
%Compression \\
%

\subsection{Collaborative Infrastructure for Mobile AIGC Networks}\label{sec:infrastructure}
\begin{figure}[t]
    \centering
    \includegraphics[width=1\linewidth]{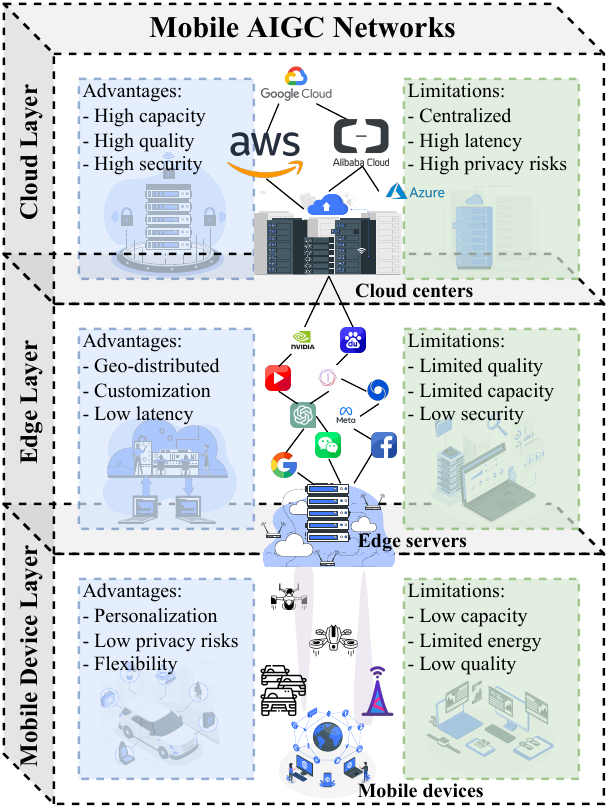}
    \caption{The collaborative cloud-edge-mobile infrastructure for mobile AIGC networks. The advantages and limitations of provisioning AIGC services in each layer are elaborated.}
    \label{fig:layers}
\end{figure}
% \cite{zhou2019edge, wang2020convergence, xu2022edge}
By asking ChatGPT the question ``Integrating AI-generated content and mobile edge networks, please define mobile AIGC networks in one sentence," we can get the answer ``\textit{Mobile AIGC networks are a fusion of AI-generated content and mobile edge networks, enabling rapid content creation, delivery, and processing at the network's edge for enhanced user experiences and reduced latency.}" (from Mar. 14 Version based on GPT-4)
To support the pre-training, fine-tuning, and inference of the aforementioned models, substantial computation, communication, and storage resources are necessary. Consequently, to provide low-latency and personalized AIGC services, a collaborative cloud-edge-mobile AIGC framework shown in Fig.~\ref{fig:layers} is essential, requiring extensive cooperation among heterogeneous resource shareholders.
\subsubsection{Cloud Computing}
In mobile AIGC networks, cloud computing~\cite{dinh2013survey} represents a centralized infrastructure supplying remote server, storage, and database resources to support AIGC service lifecycle processes, including data collection, model training, fine-tuning, and inference. Cloud computing allows users to access AIGC services through the core network where these services are deployed, rather than building and maintaining physical infrastructure. Specifically, there are three primary delivery models in cloud computing: Infrastructure as a Service (IaaS), Platform as a Service (PaaS), and Software as a Service (SaaS). In mobile AIGC networks, IaaS providers offer access to virtualized AIGC computing resources such as servers, storage, and databases~\cite{shen2021holistic}. Additionally, PaaS provides a platform for developing and deploying AIGC applications and services. Lastly, SaaS delivers applications and services over the internet, enabling users to access {\color{black}generative AI models} directly through a web browser or mobile application. In summary, cloud computing in mobile AIGC networks allows developers and users to harness the benefits of AI while reducing costs and mitigating challenges associated with constructing and maintaining physical infrastructure, playing a critical role in the development, deployment, and management of AIGC services.
\subsubsection{Edge Computing}

By providing computing and storage infrastructure at the edge of the core network~\cite{zhou2019edge}, users can access AIGC services through radio access networks (RAN). Unlike the large-scale infrastructure of cloud computing, edge servers' limited resources often cannot support {\color{black} generative AI model} training. However, edge servers can offer real-time fine-tuning and inference services that are less computationally and storage-intensive. By deploying edge computing at the network's periphery, users need not upload data through the core network to cloud servers to request AIGC services. Consequently, reduced service latency, improved data protection, increased reliability, and decreased bandwidth consumption are benefits of AIGC services delivered via edge servers. Compared to exclusively delivering AIGC services through centralized cloud computing, location-aware AIGC services at the edge can significantly enhance user experience~\cite{xu2022personalized}. Furthermore, edge servers for local AIGC service delivery can be customized and personalized to meet user needs. Overall, edge computing enables users to access high-quality AIGC services with lower latency.

\subsubsection{Mobile Computing}

Device-to-device (D2D) mobile computing involves using mobile devices for the direct execution of AIGC services by users~\cite{zhang2019mobile,du2021millimeter}. On one hand, mobile devices can directly execute {\color{black} generative AI models} and perform local AIGC inference tasks. While running {\color{black} generative AI models} on devices demands significant computational resources and consumes mobile device energy, it reduces AIGC service latency and protects user privacy. On the other hand, mobile devices can offload AIGC services to edge or cloud servers operating over wireless connections, providing a flexible scheme for delivering AIGC services. However, offloading AIGC services to edge or cloud servers for execution necessitates stable network connectivity and increases service latency. Lastly, model compression and quantization must be considered to minimize the resources required for execution on mobile devices, as {\color{black} generative AI models} are often large-scale.

Specifically, the connections among AIGC services, wireless communication, mobile edge computing, and generative AI are illustrated in Fig.~\ref{fig:scope}.

\begin{figure}
    \centering
    \includegraphics[width=0.8\linewidth]{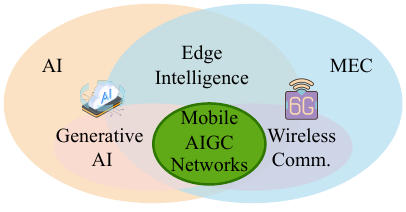}
    \caption{The connections among AIGC services, wireless communication, mobile edge computing, and generative AI.}
    \label{fig:scope}
\end{figure}

% Diffusion GAN~\cite{xiao2021tackling}

% Complex Scene Image Synthesis~\cite{fan2022frido}

% Diffusion Models Beat GANs on Image Synthesis~\cite{dhariwal2021diffusion}

% DiVAE: Photorealistic Images Synthesis with Denoising Diffusion Decoder~\cite{shi2022divae}

% GLIDE: Towards Photorealistic Image Generation and Editing with Text-Guided Diffusion Models~\cite{nichol2021glide}

% Cascaded Diffusion Models for High Fidelity Image Generation~\cite{ho2022cascaded}

% Improved denoising diffusion probabilistic models (DDPM)~\cite{nichol2021improved}

% Novel View Synthesis with Diffusion Models~\cite{watson2022novel}

\subsection{Lessons Learned}

\subsubsection{Cloud-Edge Collaborative Training and Fine-tuning for {\color{black} Generative AI Models}}

To support AIGC services with required performance evaluated based on metrics discussed in Section~\ref{sec:metric}, cloud-edge collaborative pre-training and fine-tuning are envisioned to be promising approaches. On the one hand, {\color{black} cloud data centers} can train {\color{black} generative AI models} by using powerful computing and data resources. {\color{black} Pre-training in cloud data centers enables leveraging powerful computing and data resources and pre-training on large datasets, which can help models learn general features. However, AIGC services require significant communication and bandwidth resources, and thus raise privacy concerns, and may not be as effective for fine-tuning on smaller more specific datasets.} On the other hand, utilizing a large amount of user data in the edge network, the {\color{black} generative AI model} can be fine-tuned to be more customized and personalized. {\color{black} The selection discusses the pros and cons of fine-tuning AIGC models on edge devices, including the utilization of user data available on edge devices, real-time interaction/response, and reduced privacy concerns, as well as limitations such as computing and storage resources and the need for specialized hardware and software.}
\begin{figure*}[t]
\vspace{-0.5cm}
    \centering
    \subfigure[Stable Diffusion]{\includegraphics[height=0.24\linewidth, width=0.24\linewidth]{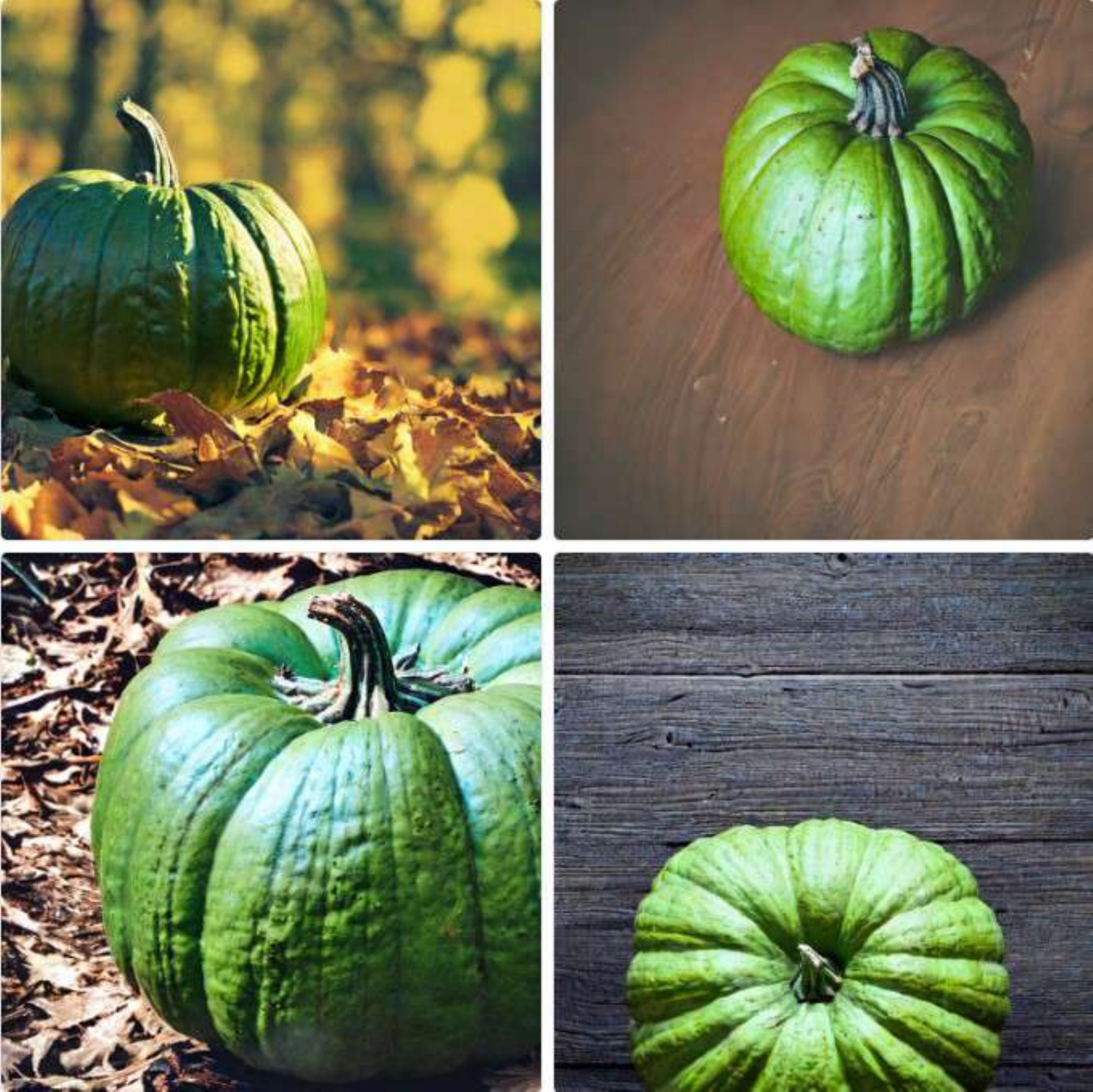}%
        \label{fig:stable}}
    % \hfil
    \subfigure[DALLE-2]{\includegraphics[height=0.24\linewidth,width=0.24\linewidth]{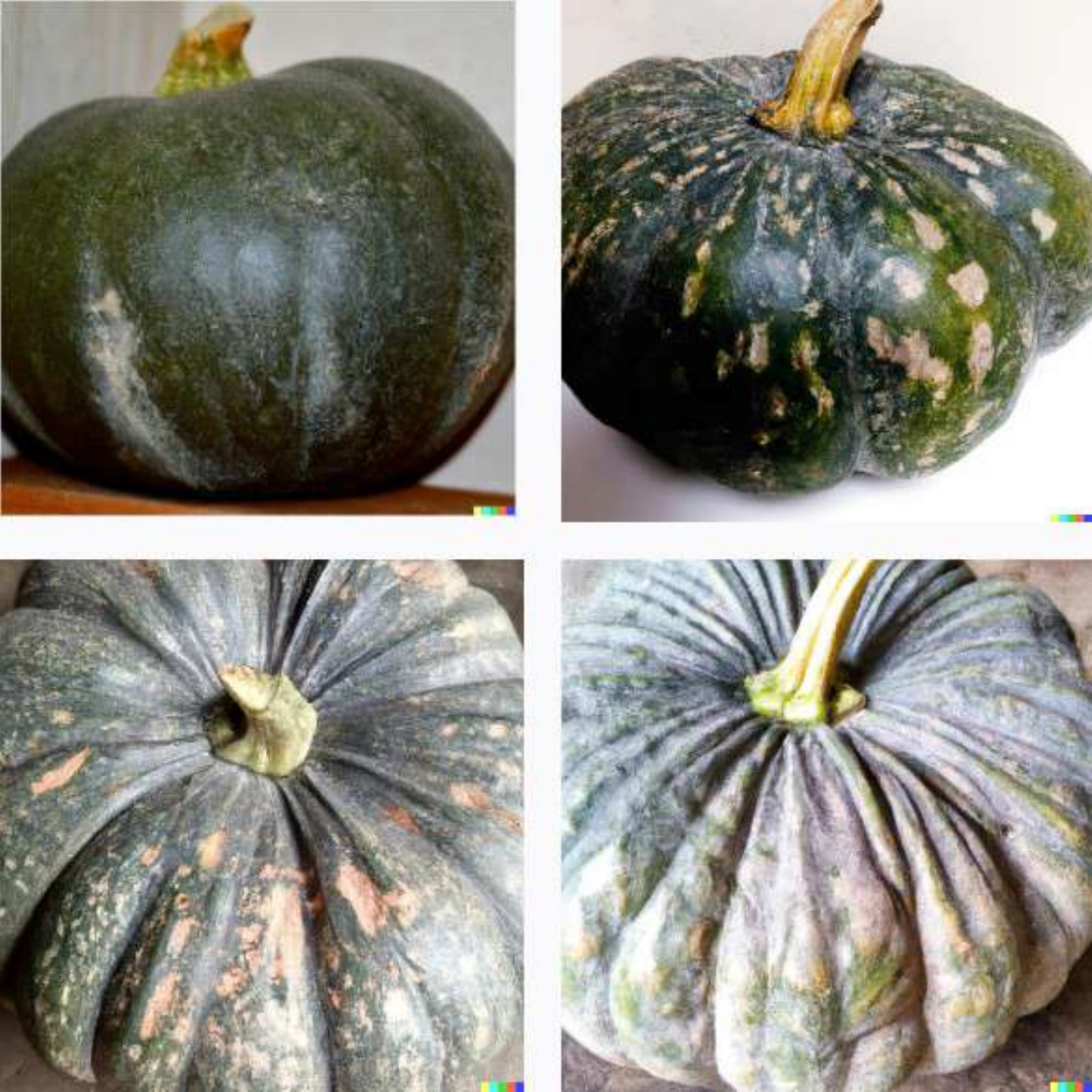}%
        \label{fig:real_revenuedt}}
    \subfigure[Visual ChatGPT]{\includegraphics[height=0.24\linewidth,width=0.24\linewidth]{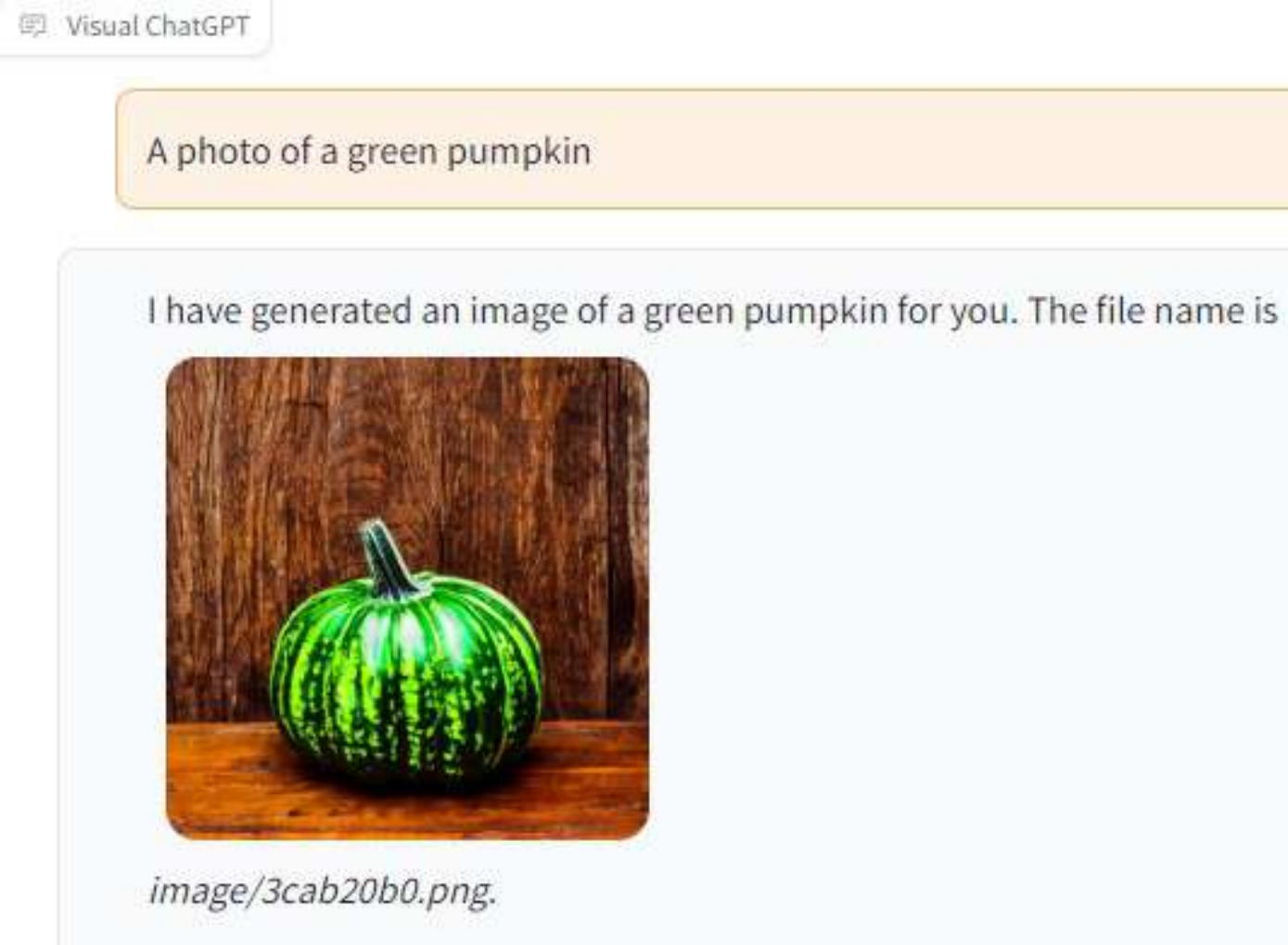}%
        \label{fig:visual}}
    \subfigure[Point-E]{\includegraphics[height=0.24\linewidth,width=0.24\linewidth]{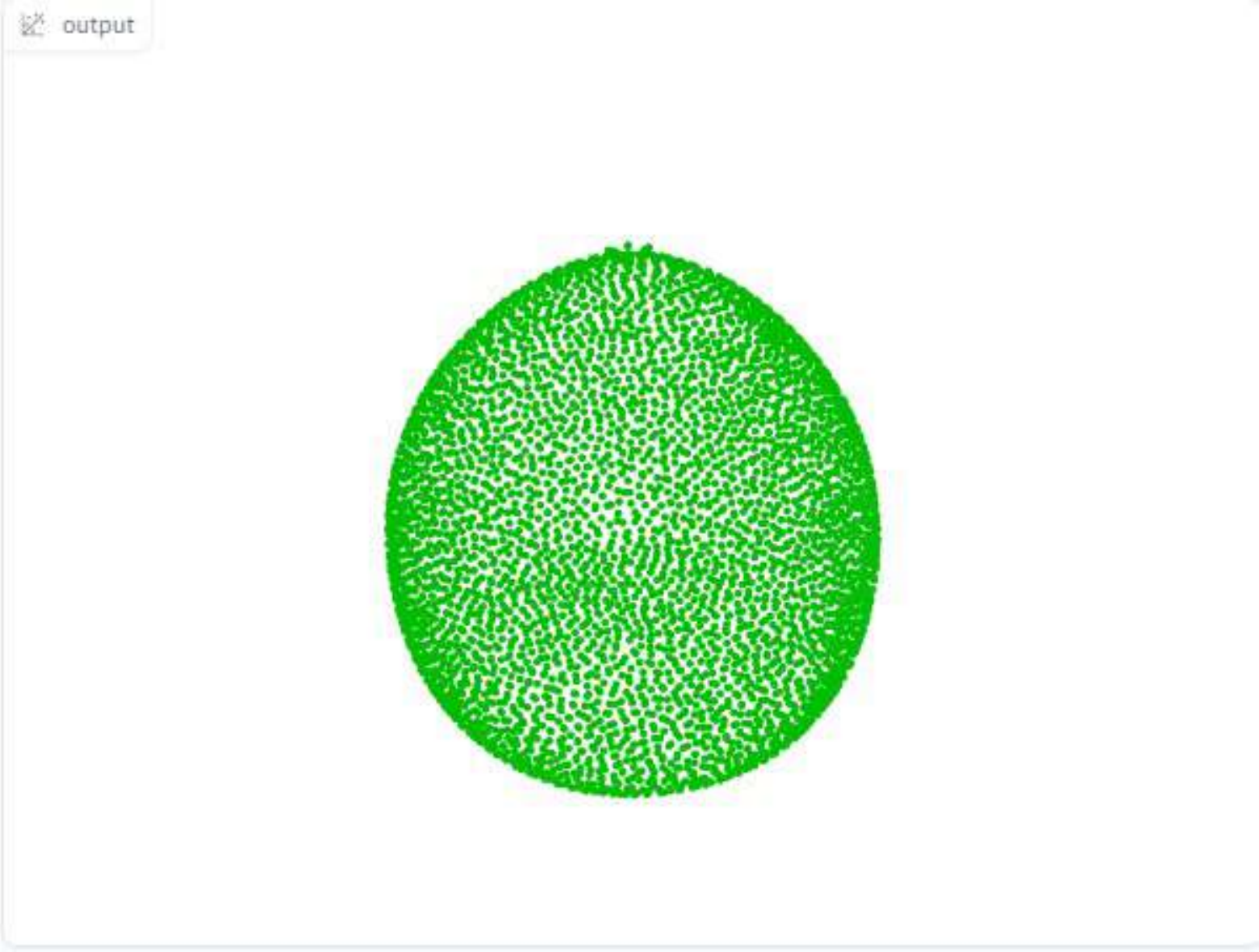}%
    \label{fig:pointe}}
    \caption{Generated images of different {\color{black} generative AI models}, including Stable Diffusion (\url{https://huggingface.co/spaces/stabilityai/stable-diffusion}), DALLE-2 (\url{https://labs.openai.com/}), Visual ChatGPT (\url{https://huggingface.co/spaces/microsoft/visual_chatgpt}), Point-E (\url{https://huggingface.co/spaces/openai/point-e}), using the prompt ``A photo of a green pumpkin".}
    \label{fig:pumpkin}
\end{figure*}
\subsubsection{Edge-Mobile Collaborative Inference for AIGC Services}

In a mobile AIGC network, the user's location and mobility change over time~\cite{wu2022hitdl}. Therefore, a large number of edge and mobile collaborations are required to complete the provision of AIGC inference services. Due to the different mobility of users, the AIGC services forwarded to the edge servers for processing are also dynamic. {\color{black} Several techniques can be leveraged to address the mobility issues in mobile AIGC networks, which include federated learning and distributed training to improve the efficiency of AIGC model updates, advanced DRL algorithms, and meta-learning techniques to optimize the AIGC provider selection strategy in response to changing network conditions, edge caching to deliver low-latency content generation and computing services, and gathering user historical requests and profiles to provide personalized services.} Therefore, dynamic resource allocation and task offloading decisions of AIGC applications are some of the challenges in deploying mobile AIGC networks, which we discuss in Section~\ref{sec:challenges}.

% \subsubsection{Cloud-Edge Collaborative Training for {\color{black} generative AI models}}

% \subsubsection{Social Aware Mobile AIGC Services}

% \subsubsection{Device-to-device Collabroative AIGC Inference}

\section{How to Deploy AIGC at Mobile Edge Networks: Applications and Advantages of AIGC}\label{sec:applications}
This section introduces creative applications and advantages of AIGC services in the mobile edge network. Then, we provide four use cases of AIGC applications of mobile AIGC networks. Some examples of {\color{black} generative AI models} are shown in Fig.~\ref{fig:pumpkin}. The applications elaborated in this section are summarized in Table \ref{table:applications}.

% {\color{red} In this section, we summarize the AIGC services in the mobile edge network. In general, this section serves as a primer for the next section. In detail, the AIGC services mentioned in this section are based on the above technologies. In addition, the challenges summarized in the next section are also based on the applications in this section. Please note that the content of this section should not be out of scope. Please do not review purely algorithmic articles in this section. To extend a little bit, we can include some cloud computing articles in this section, but a large number of articles need to include edge elements.

% Why consider AIGC at mobile edge networks, What are key challenges when using AIGC at mobile edge networks, How can AIGC be used at mobile edge networks

\subsection{Applications of Mobile AIGC Networks}

% \usepackage{multirow}
\begin{table*}[t]
\centering
\caption{Summary of State-of-the-art {\color{black} generative AI models}.}
\label{tab:summaryAIGC}
{\color{black}
\begin{tabular}{c|m{5cm}|m{2.5cm}|m{2.5cm}|m{3cm}}
\hline
\hline
\textbf{Application} & \textbf{Models} & \textbf{Network Architectures} & \textbf{Datasets} & \textbf{Evaluation Metrics} \\ \hline
Text Generation & GPT-3\cite{gpt3}, GPT-4, BERT\cite{devlin2018bert}, LaMDA\cite{thoppilan2022lamda}, ChatGPT\cite{ChatGPT2022} & Transformer\cite{vaswani2017attention}, Diffusion Model & WebText, BookCorpus\cite{zhu2015aligning}, Common Crawl & BLEU\cite{papineni2002bleu}, ROUGE\cite{lin2004rouge}, Perplexity \\ \hline
Image Generation & StyleGAN\cite{karras2019style}, BigGANs\cite{brock2018large}, StyleGANXL\cite{sauer2022stylegan}, DVD-GAN\cite{clark2019adversarial}, DALLE\cite{ramesh2021zero}, DALLE2\cite{ramesh2022hierarchical}, CLIP\cite{radford2021learning}, VisualGPT\cite{chen2022visualgpt}, VAE\cite{kingma2013auto}, Energy-based GAN\cite{zhao2016energy}, Flow-based models\cite{rezende2015variational}, Imagen\cite{saharia2022photorealistic}, diffusion probabilistic models\cite{sohl2015deep}, DDPM\cite{ho2020denoising}, DDIM\cite{song2020denoising} & Diffusion Model, GAN\cite{goodfellow2014generative}, VQ-VAE\cite{oord2017neural}, Transformer\cite{vaswani2017attention} & ImageNet\cite{deng2009imagenet}, CelebA\cite{liu2015deep}, COCO\cite{lin2014microsoft} & FID\cite{heusel2017gans}, IS\cite{salimans2016improved}, LPIPS\cite{zhang2018unreasonable} \\ \hline
Music Generation & MuseNet\cite{museNet}, Jukedeck, WaveNet\cite{oord2016wavenet}, AudioLM\cite{borsos2022audiolm} & Transformer, RNN, CNN, Diffusion Model & MIDI Dataset, MAESTRO\cite{hawthorne2018enabling} & ABC-notation, Music IS \\ \hline
Video Generation & Diffusion models beat GANs\cite{dhariwal2021diffusion}, Video Diffusion Models\cite{ho2022video}, Dreamfusion\cite{poole2022dreamfusion} & Diffusion Model\cite{du2023beyond}, GAN & Kinetics\cite{kay2017kinetics} & PSNR, SSIM  \\ \hline
3D Generation & NeRF\cite{mildenhall2021nerf} & Diffusion Model, MLP & Synthetic and real-world scenes & PSNR, SSIM, LPIPS \\ \hline
\hline
\end{tabular}
}
\label{table:applications}
\end{table*}

{\color{black}
\subsubsection{AI-generated Texts}\label{application}
Recent advancements in Natural Language Generation (NLG) technology have led to AI-generated text that is nearly indistinguishable from human-written text \cite{crothers2022machine}. The availability of powerful open-source AI-generated text models, along with their reduced computing power requirements, has facilitated widespread adoption, particularly in mobile networks. The development of lightweight NLG models that can operate on resource-constrained devices, such as smartphones and IoT devices, while maintaining high-performance levels, has made AI-generated text an essential service in mobile AIGC networks~\cite{gozalo2023chatgpt}.

One example of such a model is ALBERT (A Lite BERT), designed to enhance the efficiency of BERT (Bidirectional Encoder Representations from Transformers) while reducing its computational and memory requirements \cite{lan2019albert}. ALBERT is pre-trained on a vast corpus of text data and uses factorized embedding parameterization, cross-layer parameter sharing, and sentence-order prediction tasks to optimize BERT's performance while minimizing computational and memory demands. ALBERT has achieved performance levels comparable to BERT on various natural language processing tasks, such as question answering and sentiment analysis~\cite{ChatGPT2022}. Its lighter model design makes it more suitable for deployment on edge devices with limited resources.

MobileBERT is another model designed for deployment on mobile and edge devices with minimal resources \cite{sun2020mobilebert}. This more compact variant of the BERT model is pre-trained on the same amount of data as BERT but features a more computationally efficient design with fewer parameters. Quantization is employed to reduce the model's weight accuracy, further decreasing its processing requirements. MobileBERT is a highly efficient model compatible with various devices, including smartphones and IoT devices, and can be used in multiple mobile applications, such as personal assistants, chatbots, and text-to-speech systems~\cite{gozalo2023chatgpt}. Additionally, it can be employed in small-footprint cross-modal applications, such as image captioning, video captioning, and voice recognition. These AI-generated text models offer significant advantages to mobile edge networks, enabling new applications and personalized user experiences in real time while preserving user privacy.

\subsubsection{AI-generated Audio}
AI-generated audio has gained prominence in mobile networks due to its potential to enhance user experience, and increase efficiency, security, personalization, cost-effectiveness, and accessibility \cite{makhmutov2020survey}. For instance, AIGC-based speech synthesis and enhancement can improve call quality in mobile networks, while AIGC-based speech recognition and compression can optimize mobile networks by reducing the data required to transmit audio and automating tasks such as speech-to-text transcription. Voice biometrics powered by AI can bolster mobile network security by utilizing the user's voiceprint as a unique identifier for authentication~\cite{oord2016wavenet}. AIGC-driven audio services, such as personalized music generation, can automate tasks and reduce network load, thereby cutting costs.

{\color{black} Audio Albert~\cite{chi2021audio}, a streamlined version of the BERT model adapted for self-supervised learning of audio representations, demonstrates competitive performance levels compared to other popular AI-generated audio models in various natural language processing tasks such as speech recognition, speaker identification, and music genre classification. In terms of latency, Audio Albert shows faster inference times than previous models, with a 20\% reduction in average inference time on average, which can significantly improve response times in mobile edge networks. Additionally, Audio Albert's accuracy is comparable to BERT and achieves state-of-the-art results on several benchmarks. Furthermore, Audio Albert's model design is lighter than other models, making it suitable for deployment on edge devices with limited resources, improving computational efficiency while maintaining high-performance levels. Utilizing Audio Albert in mobile edge networks can provide several benefits, such as faster response times, reduced latency, and lower power consumption, making it a promising solution for AI-generated audio in mobile edge networks.}

\subsubsection{AI-generated Images}
AI-generated images offer numerous applications in mobile networks, such as image enhancement, image compression, image recognition, and text-to-image generation~\cite{nichol2021glide}. Image enhancement can improve picture quality in low-light or noisy environments, while image compression decreases the data required to transmit images, enhancing overall efficiency. Various image recognition applications include object detection, facial recognition, and image search. Text-to-image generation enables the creation of images from textual descriptions for visual storytelling, advertising, and virtual reality/augmented reality (VR/AR) experiences~\cite{shi2022divae, xu2022wireless, zhang2023location, du2020mec}.

Make-a-Scene, a novel text-to-image generation model proposed in \cite{gafni2022make}, leverages human priors to generate realistic images based on textual descriptions. The model consists of a text encoder, an image generator, and a prior human module trained on human-annotated data to incorporate common sense knowledge. In mobile networks, this model can be trained on a large dataset of images and textual descriptions to swiftly generate images in response to user requests, such as creating visual representations of road maps. This approach complements the techniques employed in \cite{blattmann2022semi} for generating images with specific attributes.

Furthermore, the Semi-Parametric Neural Image Synthesis (SPADE) method introduced in \cite{blattmann2022semi} generates new images from existing images and their associated attributes using a neural network architecture. This method produces highly realistic images conditioned on input attributes and can be employed for image-to-image translation, inpainting, and style transfer in mobile networks. The SPADE method shares similarities with the text-to-image generation approach in \cite{gafni2022make}, where both techniques focus on generating high-quality, realistic images based on input data.

However, the development of AI-generated image technology also raises concerns around deep fake technology, which uses AI-based techniques to generate realistic photos, movies, or audio depicting nonexistent events or individuals, as discussed in \cite{westerlund2019emergence}. Deep fakes can interfere with system performance and affect mobile user tasks, leading to ethical and legal concerns that require more study and legislation.

\subsubsection{AI-generated Videos}
AI-generated videos, like AI-generated images, can be utilized in mobile networks for various applications, such as video compression, enhancement, summarization, and synthesis~\cite{clark2019adversarial}. AI-generated videos offer several advantages over AI-generated images in mobile networks. They provide a more immersive and engaging user experience by dynamically conveying more information~\cite{du2022exploring}. Moreover, AI-generated videos can be tailored to specific characteristics, such as style, resolution, or frame rate, to improve user experience or create videos for specific purposes, such as advertising, entertainment, or educational content~\cite{ho2022video}. Furthermore, AI-generated videos can generate new content from existing videos or other types of data, such as images, text, or audio, offering new storytelling methods~\cite{ho2022video}.

Various models can be employed to achieve AI-generated videos in mobile networks. One such model is Imagen Video, presented in~\cite{ho2022imagen}, which is a text-conditioned video generation system based on a cascade of video diffusion models. Imagen Video generates high-definition videos from text input using a base video generation model and an interleaved sequence of spatial and temporal video super-resolution models. The authors describe the process of scaling up the system as a high-definition text-to-video model, including design choices such as selecting fully-convolutional temporal and spatial super-resolution models at specific resolutions and opting for v-parameterization for diffusion models. They also apply progressive distillation with classifier-free guidance to video models for rapid, high-quality sampling~\cite{ho2022imagen,ho2022video}. Imagen Video not only produces high-quality videos but also boasts a high level of controllability and world knowledge, enabling the generation of diverse videos and text animations in various artistic styles and with 3D object comprehension.

\subsubsection{AI-generated 3D}
AI-generated 3D content is becoming increasingly promising for various wireless mobile network applications, including AR and VR~\cite{jin2022dr, chen2023introduction}. It also enhances network efficiency and reduces latency through optimal base station placement~\cite{chou2022diffusionsdf, nichol2022point}. Researchers have proposed several techniques for generating high-quality and diverse 3D content using deep learning (DL) models, some of which complement one another in terms of their applications and capabilities.

One such technique is the Latent-NeRF model, proposed in~\cite{metzer2022latent}, which generates 3D shapes and textures from 2D images using the NeRF architecture. This model is highly versatile and can be used for various applications, such as 3D object reconstruction, 3D scene understanding, and 3D shape editing for wireless VR services. Another technique, the Latent Point Diffusion (LPD) model presented in~\cite{zeng2022lion}, generates 3D shapes with fine-grained details while controlling the overall structure. LPD has been shown to create more diverse shapes than other state-of-the-art models, making it suitable for 3D shape synthesis, 3D shape completion, and 3D shape interpolation. The LPD model complements the latent-NeRF approach by offering more diverse shapes and finer details.

Moreover, researchers in~\cite{li2022diffusion} proposed the Diffusion-SDF model, which generates 3D shapes from natural language descriptions. This model utilizes a combination of voxelized signed distance functions and diffusion-based generative models, producing high-quality 3D shapes with fine-grained details while controlling the overall structure. This technique accurately generates 3D shapes from natural language descriptions, making it useful for applications such as 3D shape synthesis, completion, and interpolation. It shares similarities with the Latent-NeRF and LPD models in terms of generating high-quality 3D content~\cite{lin2022magic3d}.
{\color{black}
\subsection{Advantages of Mobile AIGC}
We then discuss several advantages of generative AI in mobile networks.
\subsubsection{Efficiency}
% Generative AI models offer several efficiency benefits in mobile networks. One of the primary advantages is automation. Generative AI models can automate creating text, images, and other types of media, reducing the need for human labor and significantly boosting productivity~\cite{wu2022generative}. The outputs of generative models can be generated quickly and with minimal human intervention. This is particularly beneficial for tasks such as data augmentation in mobile networks, where a substantial amount of synthetic data is required to train ML models for applications like object recognition or network optimization.
% Moreover, generative AI models can be implemented at the edge of mobile networks~\cite{du2023enabling, li2020noma}, allowing them to produce data locally on devices like smartphones and IoT sensors. This is especially advantageous for tasks that demand generating a large volume of data, such as image and video synthesis for AR applications. 
Generative AI models offer several efficiency benefits in mobile networks. As demonstrated in the applications of AI-generated text models like ALBERT \cite{lan2019albert} and MobileBERT \cite{sun2020mobilebert}, these models can automate the process of creating text, reducing the need for human labor and significantly boosting productivity~\cite{wu2022generative}. Moreover, as shown in the applications of AI-generated audio models like Audio Albert \cite{chi2021audio}, these models can be implemented at the edge of mobile networks~\cite{du2023enabling, li2020noma}, allowing them to produce data locally on devices like smartphones and IoT sensors. This results in improved user experiences and reduced latency in mobile applications that rely on real-time data generation and processing~\cite{du2023enabling}.

\subsubsection{Reconfigurability}
The reconfigurability of AIGC in mobile networks is a significant advantage. As demonstrated in the ChatGPT application, AIGC can produce a vast array of content, which can be seamlessly adjusted to suit evolving network demands and user preferences~\cite{fan2022frido}. Furthermore, as shown in the applications of AI-generated image models like Make-a-Scene \cite{gafni2022make} and SPADE \cite{gafni2022make}, AIGC can contribute to reconfigurability in mobile networks by utilizing image and audio-generative models. These models can be trained to generate new visuals and auditory content based on specific parameters, such as user preferences or contextual information.

\subsubsection{Accuracy}
Employing generative AI models in mobile networks provides significant benefits in terms of accuracy, leading to more precise predictions and well-informed decision-making~\cite{dhariwal2021diffusion}.
% Employing generative AI models in mobile networks provides significant benefits in terms of accuracy, leading to more precise predictions and well-informed decision-making~\cite{dhariwal2021diffusion}. Enhanced accuracy in AI-generated content can substantially improve the overall user experience across various applications within the mobile network ecosystem.
% For example, AI-generated text can automate responses to mobile user inquiries, augmenting the efficiency and precision of mobile user support. This application not only reduces response times but also ensures accurate and contextually relevant information is provided to users, leading to better customer satisfaction and streamlined support services~\cite{oppenlaender2022prompt}.
Similarly, AI-generated visuals and audio can be employed to improve the quality and accuracy of network-provided content, encompassing domains such as advertising, entertainment, and accessibility services. By using generative AI models, tailored and engaging content can be produced, resulting in a more impactful and personalized user experience. In the context of mobile networks, this can mean generating high-quality images or videos adapted to various devices and network conditions, improving the user perception of the provided services.
By harnessing the power of generative AI models, mobile networks can offer more accurate and efficient services, ultimately fostering a superior user experience and enabling innovative solutions tailored to the diverse needs of mobile users~\cite{oppenlaender2022prompt}.

\subsubsection{Scalability and Sustainability}
Utilizing AIGC in mobile networks offers significant scalability and sustainability benefits~\cite{dhariwal2021diffusion}. AIGC can produce a wide range of content~\cite{ho2022imagen}, enhancing mobile networks' overall scalability and sustainability in numerous ways. Specifically, AIGC facilitates scalability in mobile networks by reducing the reliance on human labor and resources.
% Utilizing AIGC in mobile networks offers significant scalability and sustainability benefits~\cite{dhariwal2021diffusion}. AIGC can produce a wide range of content, including text, images, and audio, enhancing mobile networks' overall scalability and sustainability in numerous ways. Specifically, AIGC facilitates scalability in mobile networks by reducing the reliance on human labor and resources.
% For instance, AIGC can generate automated responses to customer inquiries, alleviating the need for human customer support staff. This approach decreases the energy consumption associated with operating human-staffed contact centers and reduces the carbon footprint linked to human labor~\cite{cao2022survey}.
Furthermore, AIGC streamlines the entire content production process, encapsulating activities from initial capture to retouching, and from synergistic designer collaboration to large-scale production. This process efficiency leads to a substantial time saving, which not only results in diminished energy consumption, but also contributes to a reduced carbon footprint associated with maintaining physical storage infrastructures~\cite{ma2023towards}.
Despite the challenges associated with {\color{black} generative AI models}, such as large model sizes and complex training processes, leveraging edge servers in mobile networks can help mitigate these issues by adopting an ``AIGC-as-a-Service" approach~\cite{du2023enabling}. Users can interact with the system by submitting requests through their mobile devices and subsequently receiving computational results from edge servers. This strategy eliminates the need to deploy {\color{black} generative AI models} on devices with constrained computing resources, optimizing overall efficiency and improving scalability and sustainability within the mobile network infrastructure~\cite{cao2022survey}.

\subsubsection{Security and Privacy}
AIGC can offer potential security and privacy advantages by embedding sensitive information within AI-generated content. This approach can serve as a form of steganography, a technique that conceals data within other types of data, making it difficult for unauthorized parties to detect the hidden information. However, it is essential to be aware of potential security and privacy risks associated with AIGC, such as adversarial attacks on AI models or the misuse of AI-generated content for malicious purposes, like deepfakes~\cite{westerlund2019emergence}. To ensure the secure and privacy-preserving use of AIGC in mobile networks, robust security measures and encryption techniques must be in place, along with ongoing research to counter potential threats~\cite{lin2023blockchain}.}
% AIGC can offer potential security and privacy advantages by embedding sensitive information within AI-generated content. This approach can serve as a form of steganography, a technique that conceals data within other types of data, making it difficult for unauthorized parties to detect the hidden information.
% For instance, AI-generated images or audio can be used to encode confidential information in imperceptible ways. This technique can improve privacy in mobile networks, as sensitive data can be transmitted without being explicitly discernible. In addition, AI-generated content can be employed as a security measure, such as AI-generated audio for voice biometrics or AI-generated facial images for authentication purposes, adding an extra layer of security to mobile network services~\cite{cao2022survey}.
% However, it is essential to be aware of potential security and privacy risks associated with AIGC, such as adversarial attacks on AI models or the misuse of AI-generated content for malicious purposes, like deepfakes~\cite{westerlund2019emergence}. To ensure the secure and privacy-preserving use of AIGC in mobile networks, robust security measures and encryption techniques must be in place, along with ongoing research to counter potential threats~\cite{lin2023blockchain}.

\section{Case Studies of AIGC in Mobile Network}\label{sec:case}
Many case studies have been done for achieving effective and efficient mobile AIGC networks as shown in Table.~\ref{tab:sfae}. In this section, we review several representative cases, e.g., the AIGC service provider (ASP) selection, generative AI-empowered traffic and driving simulation, AI-generated incentive mechanism, and blockchain-powered lifecycle management for AIGC.
\begin{table*}[!t]
\centering
    {\color{black} {\small \begin{tabular}{m{1.6cm}|m{8cm}|m{7cm}}
        \toprule[1pt]
        \hline
        \textbf{Reference} & \textbf{System Model} & \textbf{Method Used} \\
        \hline
        \cite{du2023beyond} & A comprehensive tutorial on generative diffusion models in various network optimization problems & Integration of diffusion models with DRL, incentive design, semantic communications, and IoV networks \\
        \hline
        \cite{du2023ai} & Users sharing information through full-duplex device-to-device semantic communications & Diffusion model-based incentive mechanism generation to maximize the users' utilities \\
        \hline
        \cite{du2023generative} & Selection of AIGC service providers (ASPs) capable of effectively executing user tasks & Generative diffusion model for optimal decision generation in ASP selection problem \\
        \hline
        \cite{du2023exploring} & Distributed diffusion model where the user transmits the results after several shared denoising steps to other users & A collaborative distributed diffusion-based AIGC framework \\
        \hline
        \cite{du2023enabling} & Large-scale deployment of AaaS with 20 AIGC service providers (ASPs) and 1000 edge users & Deep reinforcement learning (DRL)-enabled solution to maximize a utility function \\
        \hline
        \cite{liu2023blockchain} & AIGC lifecycle management framework with three ESPs and three producers, supported by the Draw Things application & Blockchain technology to protect the ownership and copyright of AIGC, along with a reputation-based service provider selection strategy \\
        \hline
        \cite{liu2023deep} & Deep generative model-empowered wireless network management and use cases, e.g., network routing, resource allocation, and network economics & Diffusion model to generate effective contracts for incentivizing mobile AIGC services \\
        \hline
        \cite{wang2023guiding} & Wireless sensing platform based on the 801.11ac protocol with a signal transmitter and five receivers & Multi-scale wireless perception for AIGC services \\
        \hline
        \cite{du2023spear} & A user requests a specific number of images from a service provider that is attacked by data poisoning, while diffusion models provide the defense & Generative diffusion model-aided optimization to identify the optimal diffusion steps to minimize the total energy cost \\
        \hline
        \cite{zhang2023generative} & A multi-modality semantic-aware framework for generative AI-enabled vehicular networks & A double deep Q-network-based approach to address the resource allocation problem in generative AI-enabled V2V communication \\
        \hline
        \cite{lin2023unified} & An integrated semantic communication and AIGC (ISCA) framework for Metaverse services & Diffusion model-based joint resource allocation in ISCA systems\\
        \hline
        \cite{du2023yolo} & A semantic communication framework based on You Only Look Once (YOLO) to construct a virtual apple orchard & Semantic communications with generative diffusion model-aided resource optimization \\
        \hline
        \cite{xu2023joint} & A foundation model caching and inference framework to balance the tradeoff among inference latency, accuracy, and resource consumption & Managing cached foundation models and user requests during the provisioning of generative AI services \\
        \hline
        \cite{xu2023sparks} &  A framework of joint model caching and inference for managing models and allocating resources &  A least context algorithm for managing cached models at edge servers  \\
        \hline
        \cite{xu2023generative} & An autonomous driving architecture, where generative AI is leveraged to synthesize conditioned traffic and driving data & A multi-task digital twin offloading model and  a multi-task enhanced auction-based mechanism \\
        \hline
        \cite{chen2023revolution} & A framework that used mobile AIGC to drive Human Digital Twin (HDT) applications, focusing on personalized healthcare solutions & Using generative diffusion model for the resource allocation in mobile AIGC-driven HDT system\\
        \hline
        \cite{huang2023federated} & The model combines Federated Learning (FL) with AIGC to improve AIGC creation and privacy in wireless networks & Using FL techniques to fine-tune AIGC, yielding reduced communication cost and training latency\\
        \hline
        \cite{wang2023generative} & Exploring the application of Generative Artificial Intelligence (GAI) in the physical layer of Integrated Sensing and Communications (ISAC) systems & Using a diffusion model-based method for signal direction estimation demonstrates GAI's efficacy in near-field ISAC\\
        \hline
        \cite{du2023generativemul} & GAI-aided Semantic Communication (SemCom) system that uses multi-model prompts for accurate content decoding and incorporates security measures & Using a diffusion model to ensure secure and accurate message transmission\\
        \hline
        \cite{liu2023optimizing} & Using Pretrained Foundation Models (PFMs) and prompt engineering to expand the applications of AIGC in edge networks & Using ChatGPT to train an effective prompt optimizer, measuring its impact on user experience\\
        \hline
        \cite{zheng2023flexible} & Flexible-position multiple-input multiple-output (MIMO) systems & Using generative diffusion model to generate optimal antenna trajectories to maximize system efficiency\\
        \hline
        \cite{lin2023blockchain} & A blockchain-aided semantic communication framework for AIGC services in virtual transportation networks & A training-based targeted semantic attack scheme and counters it with a blockchain and zero-knowledge proof-based defense mechanism\\
        \hline
        \cite{lin2023blockchain} & A blockchain-aided semantic communication framework for AIGC services in virtual transportation networks & A training-based targeted semantic attack scheme and counters it with a blockchain and zero-knowledge proof-based defense mechanism\\
        \hline
        \cite{wang2023unified} & A framework that uses wireless perception to guide generative AI in producing digital content & A Sequential Multi-Scale Perception algorithm for user skeleton prediction and a diffusion model-based approach to generate an optimal pricing strategy\\
        \hline
    \bottomrule[1pt]
    \end{tabular}}}
    \caption{Key Literature Considering AIGC within Wireless Network.}
    \label{tab:sfae}
\end{table*}

\subsection{AI-Generated Incentive Mechanism}
In this case study, we present the idea of using AI-generated optimization solutions with a focus on the use of diffusion models and their ability to optimize the utility function.

\begin{figure*}
    \centering
    \includegraphics[width=1\linewidth]{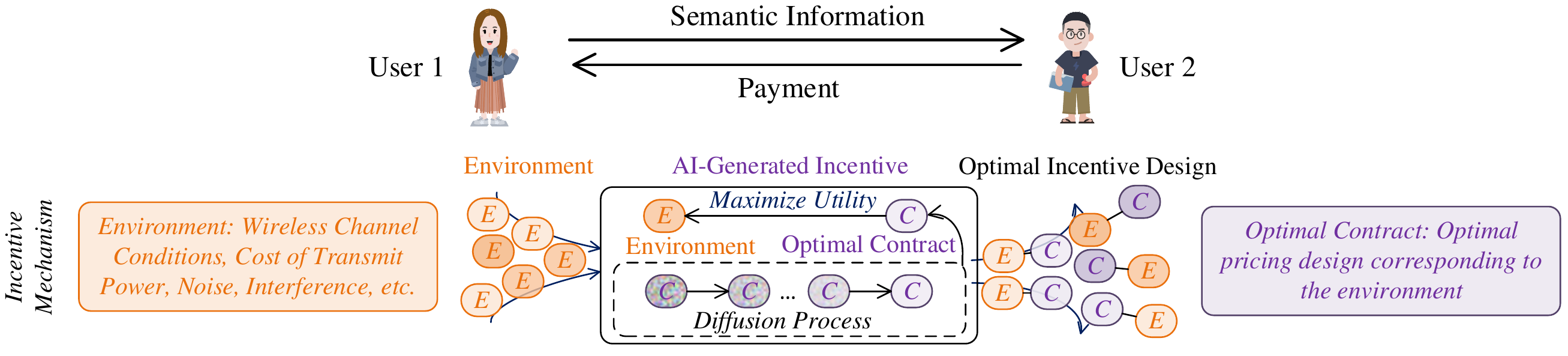}
    \caption{{\color{black}System model of contract design in semantic information sharing network, and the AI-generated contract algorithm. The diffusion models generate different optimal contract designs under different environmental variables.}}
    \label{figsemm}
\end{figure*}
\begin{figure}[!t]
\centering
\subfigure[Training process, with diffusion step $N = 10$~\cite{du2023ai}.]
% , batch size $N_b = 512$, discount factor $\gamma = 0.95$, soft target update parameter $\tau = 0.005$, exploration noise $\epsilon = 0.01$, contract generation network ${\varepsilon}_\theta$ learning rate is $10^{-5}$, and contract quality network $Q_\upsilon$ learning rate is $10^{-5}$
{\includegraphics[width=0.4\textwidth]{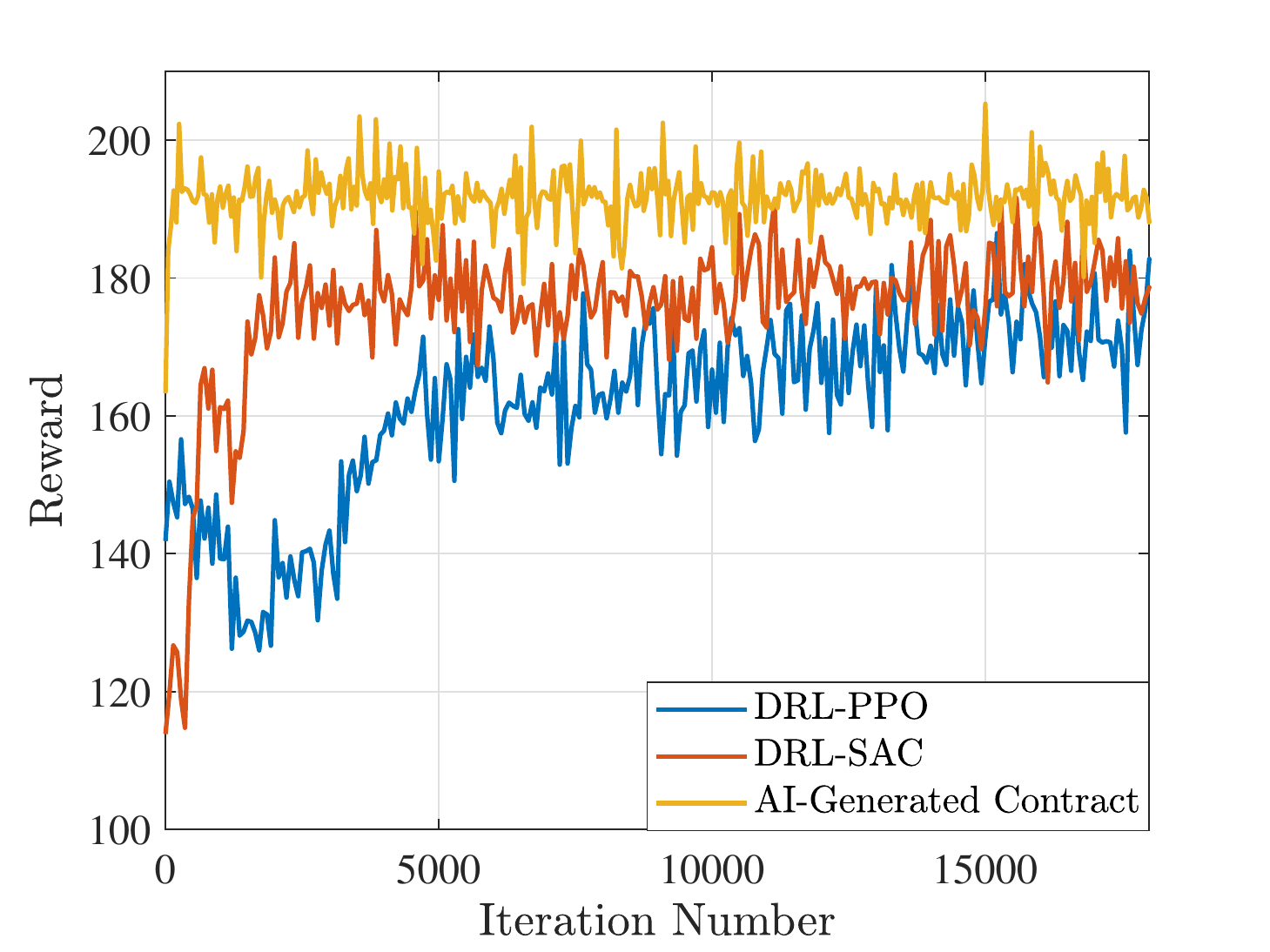}}
\subfigure[{\color{black} The designed contracts.}]{\includegraphics[width=0.4\textwidth]{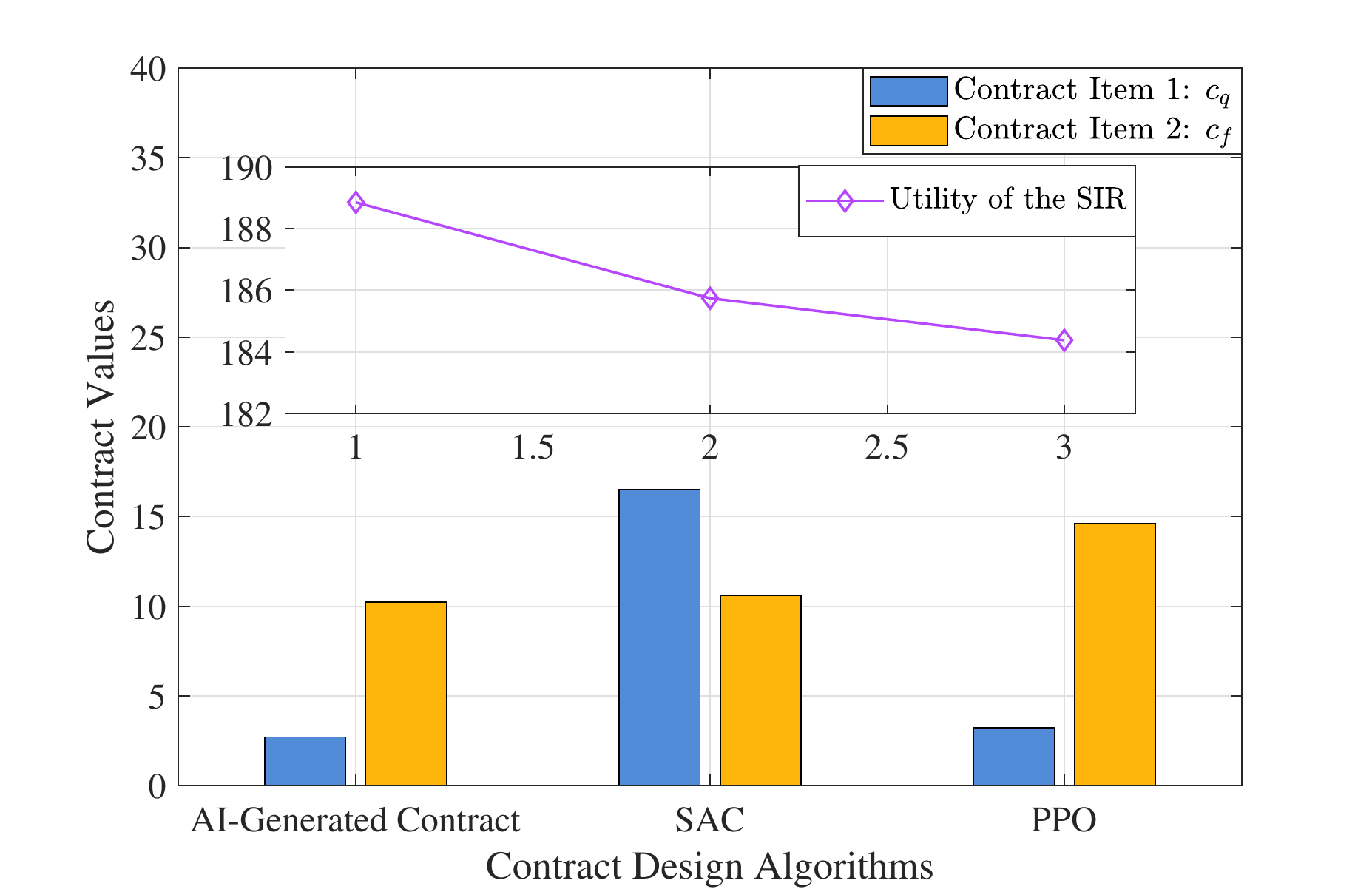}}
\caption{{\color{black} The effect of different incentive design schemes, e.g., PPO, SAC, and AI-generated contract~\cite{du2023ai}.}}
% , with $c_{s} = 1000$, $c_p = 0.03$, $D_{jk} = 6$ m, $\alpha = 4$, $\mu = 4$, $\beta_{k} = 12.9$,  $\sigma _N^2 = \sigma _S^2 = 0.2$, $P_I = 3$ ${\rm dBW}$, ${\eta _{k}} = 0.3$, ${\upsilon _k}=0.2$, $B_{jk} = 10$ ${\rm MHz}$, and $N =3$
\label{free}
\end{figure}
In today's world of advanced internet services, including the Metaverse, MR technology is essential for delivering captivating and immersive user experiences~\cite{wang2023semantic,wang2022wireless}. Nevertheless, the restricted processing power of head-mounted displays (HMDs) used in MR environments poses a significant challenge to the implementation of these services.
To tackle this problem, the researchers in~\cite{du2023ai} introduce an innovative information-sharing strategy that employs full-duplex device-to-device semantic communication~\cite{yang2022semantic}. This method enables users to circumvent computationally demanding and redundant processes, such as producing AIGC in-view images for all MR participants. By allowing users to transmit generated content and semantic data derived from their view image to nearby users, these individuals can subsequently utilize the shared information to achieve spatial matching of computational outcomes within their view images.
In their work, the authors of~\cite{du2023ai} primarily concentrate on developing a contract theoretic incentive mechanism to promote semantic information exchange among users. Their goal is to create an optimal contract that, while adhering to the utility threshold constraints of the semantic information provider, simultaneously maximizes the utility of the semantic information recipient. Consequently, they devised a diffusion model-based AI-generated contract algorithm~\cite{du2023beyond}, as illustrated in Fig.~\ref{figsemm}.

Specifically, the researchers developed a cutting-edge algorithm for creating AI-generated incentive mechanisms~\cite{du2023beyond}, which tackle the challenge of utility maximization by devising optimal contract designs~\cite{du2023ai}. This approach is distinct from traditional neural network backpropagation algorithms or DRL methods, as it primarily focuses on enhancing contract design through iterative denoising of the initial distribution instead of optimizing model parameters.
The policy for contract design is defined by the reverse process of a conditional diffusion model, linking environmental states to contract arrangements. The primary goal of this policy is to produce a deterministic contract design that maximizes the expected total reward over a series of time steps.
% The contract design policy, represented by the reverse process of a conditional diffusion model, is given by
% \begin{equation}
% {\pi _\theta }\left( {\left. {\bf{c}}_0 \right|{\bf{e}}} \right) = {p_\theta }\left( {\left. {{{\bf c}_{0:N}}} \right|{\bf e}} \right) = \mathcal{N}\left( {{{\bf c}_N};0,I} \right)\prod\limits_{i = 1}^N {{p_\theta }\left( {\left. {{{\bf c}_{i - 1}}} \right|{{\bf c}_i},{\bf e}} \right)},
% \end{equation}
% where ${\bf e}$ is the environment vector, ${\bf c}_0$ is the generated contract design, ${\bf c}_N$ is the Gaussian noise.
% The task of training the contract design policy in complex and high-dimensional environments is effectively transferred to training the contract generation network with parameter $\theta$. With a well-trained diffusion model $\theta$, we can generate the optimal contract according to
% \begin{equation}\label{denoise}
% {{\mathbf{c}}_{i - 1}}\mid {{\mathbf{c}}_i} = \frac{{{{\mathbf{c}}_i}}}{{\sqrt {{\alpha _i}} }} - \frac{{{\beta _i}}}{{\sqrt {{\alpha _i}\left( {1 - {{\bar \alpha }_i}} \right)} }}{{\bm{\varepsilon}} _\theta }\left( {{{\mathbf{c}}_i},{\mathbf{e}},i} \right) + \sqrt {{\beta _i}} {\bm{\varepsilon}},
% \end{equation}
% where $\beta_i$ is the variance schedule parameter, $\alpha_i = 1 - \beta_i$, and ${\bar \alpha}_i = \prod\limits_{s = 1}^i a_s$.
To optimize system utility through contract design, the researchers in~\cite{du2023ai} create a contract quality network that associates an environment-contract pair with a value representing the expected total reward when an agent implements a particular contract design policy from the current state and adheres to it in the future. The optimal contract design policy maximizes the system's predicted cumulative utility. {\color{black}The researchers then carried out an} extensive comparison between their suggested AI-powered contract algorithm and two DRL algorithms, specifically SAC and PPO.
As illustrated in the training process in~\cite{du2023ai} (see Fig.~\ref{free}), PPO requires more iteration steps to achieve convergence, while SAC converges more quickly but with a lower final reward value in comparison to the AI-driven contract algorithm.

The enhanced performance of the suggested AI-driven contract algorithm can be ascribed to two main aspects:
\begin{itemize}
    \item Improved sampling quality: By configuring the diffusion step to 10 and applying multiple refinement steps, the diffusion models generate higher quality samples, mitigating the influence of uncertainty and augmenting sampling precision~\cite{dhariwal2021diffusion}.
    \item Enhanced long-term dependence processing capability: Unlike conventional neural network generation models that take into account only the current time step input, the diffusion model creates samples with additional time steps through numerous refinement iterations, thereby bolstering its long-term dependence processing capability~\cite{nichol2021glide}.
\end{itemize}
As demonstrated in Fig.~\ref{free}, the authors in~\cite{du2023ai} examine the optimal contract design capacities of the trained models. For a specific environmental state, the AI-driven contract algorithm provides a contract design that attains a utility value of 189.1, markedly outperforming SAC's 185.9 and PPO's 184.3. These results highlight the practical advantages of the proposed AI-based contract algorithm in contrast to traditional DRL techniques.

{\textit{{\textbf{Lesson Learned:}}}} 
The case study in this research highlights the potential of AI-generated optimization solutions, particularly diffusion models, for addressing complex utility maximization problems within incentive mechanism design. The authors in~\cite{du2023ai} present an innovative approach that employs full-duplex device-to-device semantic communication for information-sharing in mixed reality environments, overcoming the limitations of HMDs.
The diffusion model-based AI-generated contract algorithm proposed in this study demonstrates superior performance compared to traditional DRL algorithms, such as SAC and PPO. The superior performance of the AI-generated contract algorithm can be attributed to improved sampling quality and enhanced long-term dependence processing capability.
This study underscores the effectiveness of employing AI-generated optimization solutions in complex, high-dimensional environments, particularly in the context of incentive mechanism design. Some promising directions for future research include:
\begin{itemize}
    \item Expanding the application of diffusion models: Investigate the application of diffusion models in other domains, such as finance, healthcare, transportation, and logistics, where complex utility maximization problems often arise.
    \item Developing novel incentive mechanisms: Explore the development of new incentive mechanisms that combine AI-generated optimization solutions with other approaches, such as game theory or multi-agent reinforcement learning, to create even more effective incentive designs.
    \item Exploring the role of human-AI collaboration: Investigate how AI-generated optimization solutions can be combined with human decision-making to create hybrid incentive mechanisms that capitalize on the strengths of both human intuition and AI-driven optimization.
\end{itemize}

\subsection{AIGC Service Provider Selection}
\begin{figure}[t]
    \centering
    \includegraphics[width=1\linewidth]{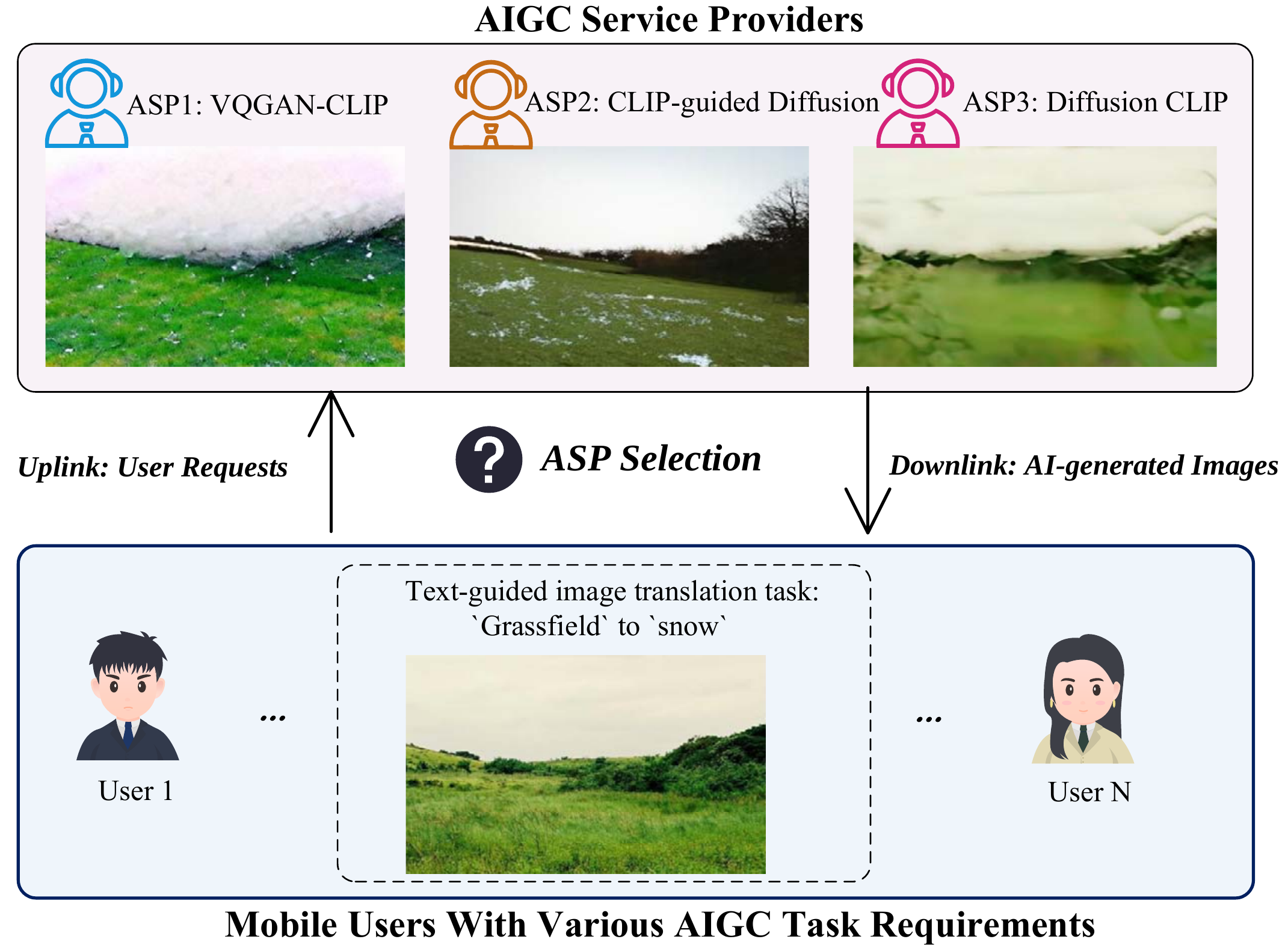}
    \caption{{\color{black}The system model of AIGC service provider selection. Different ASPs performing user tasks can bring different results and different user utilities. Considering that different mobile users have different task requirements and different ASP's AI models have different capabilities and computation capacities, a proper ASP selection algorithm is needed to maximize the total utilities of network users.}}
    \label{fig:model1}
\end{figure}
The integration of {\color{black} generative AI models} within wireless networks offers significant potential, as these state-of-the-art technologies have exhibited exceptional capabilities in generating a wide range of high-quality content. By harnessing the power of artificial intelligence, {\color{black} generative AI models} can astutely analyze user inputs and produce tailored, contextually relevant content in real-time~\cite{dhariwal2021diffusion}. This stands to considerably enhance user experience and foster the creation of innovative applications across various domains, such as entertainment, education, and communication. Nonetheless, the deployment and application of these advanced models give rise to challenges, including extensive model sizes, complex training processes, and resource constraints. Consequently, deploying large-scale AI models on every network edge device poses considerable difficulties.

To address this challenge, the authors in~\cite{du2023enabling} introduce the ``AIGC-as-a-service" architecture. This approach entails ASPs deploying AI models on edge servers, which facilitates the provision of instantaneous services to users via wireless networks, thereby ensuring a more convenient and adaptable experience. By enabling users to effortlessly access and engage with AIGC, the proposed solution minimizes latency and resource consumption. Consequently, edge-based AIGC-as-a-service holds the potential to transform the creation and delivery of AIGC across wireless networks.

However, one problem is that the effectiveness of ASP in meeting user needs displays significant variability due to a variety of factors. Certain ASPs may concentrate on generating specific content types, while others boast more extensive content generation capabilities. For instance, some providers may specialize in producing particular content categories, whereas others offer a wider range of content generation options. Moreover, several ASPs may have access to advanced computing and communication resources, empowering them to develop and deploy more sophisticated {\color{black} generative AI models} within the mobile network. As depicted in Fig.~\ref{fig:model1}, users uploading images and requirement texts to different ASPs encounter diverse results owing to the discrepancies in models employed. For example, a user attempting to add snow to grass in an image may experience varying outcomes depending on the ASP chosen.

With a large number of mobile users and increasing demand for accessing requests, it is crucial to analyze and select ASPs with the necessary capability, skill, and resources to offer high-quality AIGC services. This requires a rigorous selection process considering the provider's {\color{black} generative AI model} capabilities and computation resources. By selecting a provider with the appropriate abilities and resources, organizations can ensure that they have effective AIGC services to increase the QoE for mobile users. 
Motivated by the aforementioned reasons, the authors in~\cite{du2023enabling} examine the viability of large-scale deployment of AIGC-as-a-Service in wireless edge networks. Specifically, in the ASP selection problem, which can be framed as a resource-constrained task assignment problem, the system consists of a series of sequential user tasks, a set of available ASPs, and the unique utility function for each ASP. The objective is to find an assignment of tasks to ASPs, such that the overall utility is maximized. Note that the utility of the task assigned to the ASP is a function of the required resource. Without loss of generality, the authors in~\cite{du2023enabling} consider that is in the form of the diffusion step of the diffusion model, which is positively correlated to the energy cost. The reason is that each step of the diffusion model has energy consumption as it involves running a neural network to remove Gaussian noise. Finally, the total availability of resources for each ASP is taken into account to ensure that the resource constraints are satisfied.

% Mathematically, the ASP selection problem can be formulated as an integer programming problem with decision variables $x_n \left( \forall n\in\mathcal{T}\right) $ in $\mathcal{X}$, which represent the time series of task assignments to available ASPs. Additionally, $\hat{\mathcal{T}}a$ denotes the set of running tasks on the $a^{\rm th}$ ASP at the time of assigning the current task. Thus, the problem can be formulated as
% \begin{align}\label{eq:utility-target}
% \underset{\mathcal{X}}{\max}\quad &\mathcal{U}=\sum{n\in \mathcal{T}}{u_a\left( R_n \right)},\\
% \mathrm{s}.\mathrm{t}.\quad &a=x_n,\\
% &R_n+\sum_{n'\in \hat{\mathcal{T}}a}{R{n'}} \leqslant \mathcal{R}_a\quad \left( \forall a\in \mathcal{A} \right) , \label{eq:system-model-resource-constraint}\\
% & a = 1, \ldots, M, {\text{and}} \:\: n = 1, \ldots, N.
% \end{align}
In this formulation of AIGC service provisioning, the resource constraints are incorporated through the resource constraint, which specifies the limitations on the available resources. Note that failing to satisfy the resource constraint can result in the crash of ASP, causing the termination and restart of its running tasks.

Several baseline policies are used for comparison:
\begin{itemize}
	\item {\textit{\textbf{Random Allocation Policy.}}} This strategy distributes tasks to ASPs in a haphazard manner, without accounting for available resources, task duration, or any restrictions. The random allocation serves as a minimum benchmark for evaluating scheduling efficiency.
	\item {\textit{\textbf{Round-Robin Policy.}}} The round-robin policy allocates tasks to ASPs sequentially in a repeated pattern. This approach can generate effective schedules when tasks are evenly distributed. However, its performance may be suboptimal when there are significant disparities among them.
	\item {\textit{\textbf{Crash-Avoid Policy.}}} The crash-avoid policy prioritizes ASPs with greater available resources when assigning tasks. The goal is to prevent overburdening and maintain system stability.
	\item {\textit{\textbf{Upper Bound Policy.}}} In this hypothetical scenario, the scheduler has complete knowledge of the utility each ASP offers to every user before task distribution. The omniscient allocation strategy sets an upper limit on the performance of user-centric services by allocating tasks to ASPs with the highest utility and avoiding system failures. However, this approach relies on prior information about the unknown utility function, which is unrealistic in practice.
\end{itemize}
The authors in~\cite{du2023enabling} employed a Deep Reinforcement Learning (DRL) technique to optimize Application Service Provider (ASP) selection. In particular, they implemented the Soft Actor-Critic (SAC) method, which alternates between evaluating and improving the policy. Unlike traditional actor-critic frameworks, the SAC approach maximizes a balance between expected returns and entropy, allowing it to optimize both exploitation and exploration for efficient decision-making in dynamic ASP selection scenarios.
To conduct the simulation, the authors consider 20 ASPs and 1000 edge users. Each ASP offered AaaS with a maximum resource capacity, measured by total diffusion timesteps in a given time frame, varying randomly between 600 and 1,500. Each user submits multiple AIGC task requests to ASPs at varying times. These requests detailed the necessary AIGC resources in terms of diffusion timesteps, randomly set between 100 and 250. Task arrivals from users adhered to a Poisson distribution, with a rate of 0.288 requests per hour over a 288-hour duration, amounting to 1,000 tasks in total.
As shown in Fig.~\ref{fig:modeldrl}, simulation results indicate that the proposed DRL-based algorithm outperforms three benchmark policies, i.e., overloading-avoidance, random, and round-robin, by producing higher-quality content for users and achieving fewer crashed tasks.
\begin{figure}[t]
    \centering
    \includegraphics[width=0.9\linewidth]{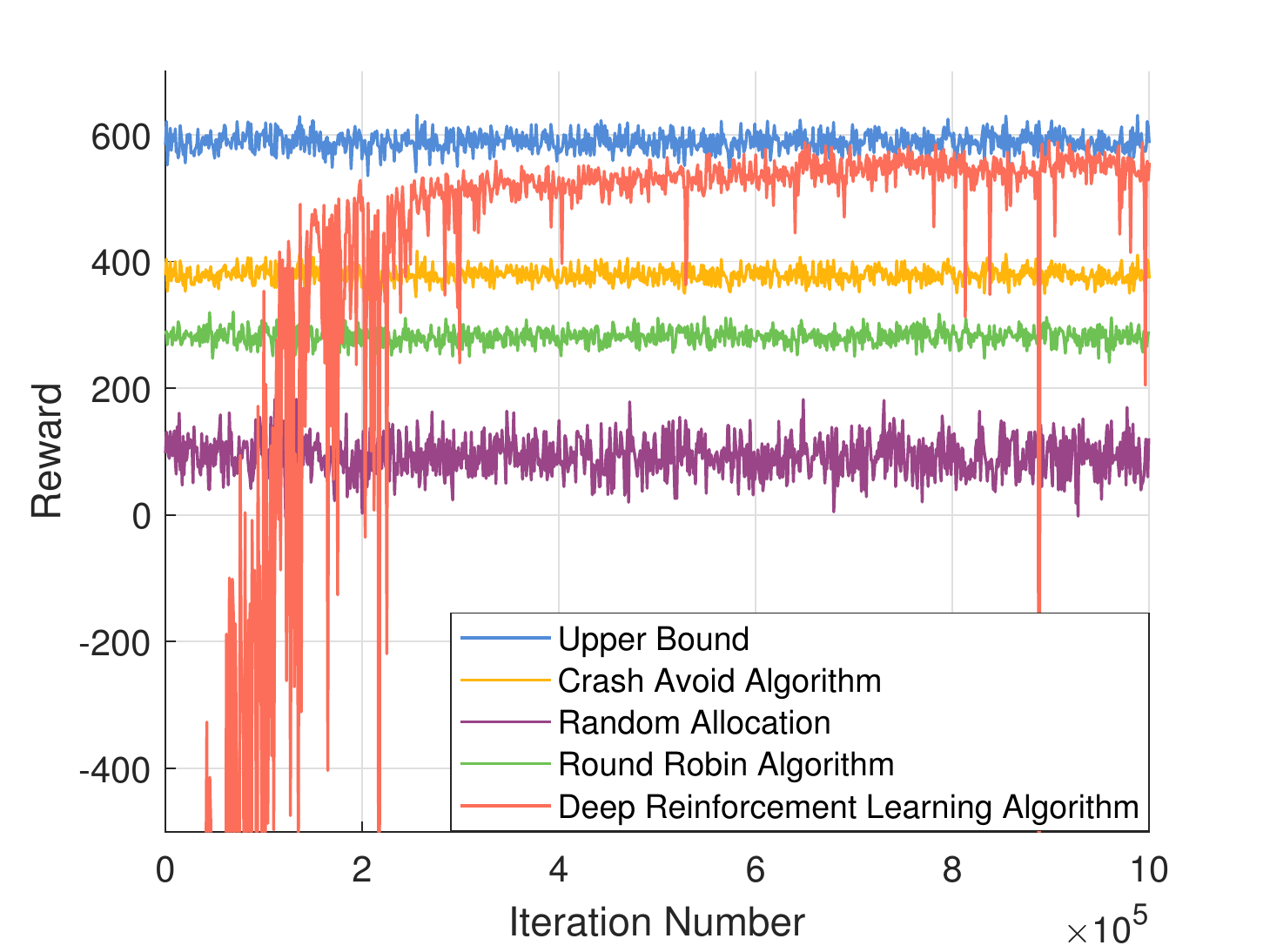}
    \caption{{\color{black}The cumulative rewards under different ASP selection algorithms~\cite{du2023enabling}. DRL-based algorithms can outperform multiple baseline policies, i.e., overloading-avoidance, random, and round-robin, and approximate the optimal policy.}}
    \label{fig:modeldrl}
\end{figure}
{\textit{{\textbf{Lesson Learned:}}}} 
The lesson learned from this study is that the proper selection of ASPs is crucial for maximizing the total utilities of network users and enhancing their experience. The authors in~\cite{du2023enabling} introduced a DRL-based algorithm for ASP selection, which outperforms other baseline policies, such as overloading-avoidance, random, and round-robin. By leveraging the SAC approach, the algorithm strikes a balance between exploitation and exploration in decision-making for dynamic ASP selection scenarios. Consequently, this method can provide higher-quality content for users and lead to fewer crashed tasks, ultimately improving the quality of service in wireless edge networks.
To further enhance research in the area of AIGC service provider selection, future studies could have:
\begin{itemize}
    \item Investigate the integration of FL and distributed training methods to improve the efficiency of {\color{black} generative AI model} updates and reduce the communication overhead among ASPs.
    \item Explore advanced DRL algorithms and meta-learning techniques to adaptively adjust the ASP selection strategy in response to changing network conditions and user requirements.
    \item Assess the impact of real-world constraints, such as network latency, data privacy, and security concerns, on the ASP selection process and devise strategies to address these challenges.
    \item Develop multi-objective optimization techniques for ASP selection that consider additional factors, such as energy consumption, cost, and the trade-off between content quality and computational resources.
\end{itemize}
}

\subsection{Generative AI-empowered Traffic and Driving Simulation}

\begin{figure}[t]
    \centering
    \includegraphics[width=1\linewidth]{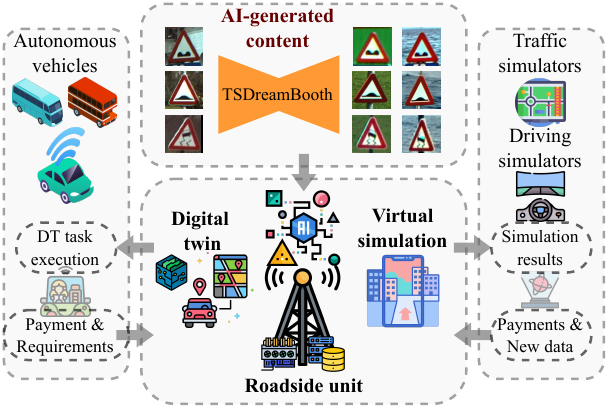}
    \caption{Generative AI-empowered simulations for autonomous driving in vehicular Metaverse, which consists of AVs, virtual simulators, and roadside units.}
    \label{fig:vehicles}
\end{figure}

In autonomous driving systems, traffic and driving simulation can affect the performance of connected autonomous vehicles (AVs). Existing simulation platforms are established based on historical road data and real-time traffic information. However, these data collection processes are difficult and costly, which hinders the development of fully automated transportation systems. Fortunately, generative AI-empowered simulations can largely reduce the cost of data collection and labeling by synthesizing traffic and driving data via generative AI models. Therefore, as illustrated in Fig.~\ref{fig:vehicles}, the authors in~\cite{xu2023generative} design a specialized generative AI model, namely TSDreambooth, for conditional traffic sign generation in the proposed vehicular mixed reality Metaverse architecture. In detail, TSDreambooth is a variation of stable diffusion~\cite{rombach2022high} fine-tuned based on the Belgium traffic sign (BelgiumTS) dataset~\cite{mathias2013traffic}. The performance of TSDreambooth is validated via the pre-trained traffic sign classification model as generative scores. In addition, the newly generated datasets are leveraged to improve the performance of original traffic sign classification models.

In the vehicular Metaverse, connected AVs, roadside units, and virtual simulators can develop simulation platforms in the virtual space collaboratively. Specifically, AVs maintain their representations in the virtual space via digital twin (DT) technologies. Therefore, AVs need to continuously generate multiple DT tasks and execute them to update the representations. To offload these DT tasks to roadside units for remote execution in real-time, AVs need to pay for the communication and computing resources of roadside units. Therefore, to provide fine-grained incentives for RSUs in executing DT tasks with heterogeneous resource demands and various required deadlines, the authors in~\cite{xu2023generative} propose a multi-task enhanced physical-virtual synchronization auction-based mechanism, namely MTEPViSA, to determine and price the resources of RSUs. There are two stage of this mechanism the online submarket for provisioning DT services and the offline submarket for provisioning traffic and driving simulation services. In the online simulation submarket, the multi-task DT scoring rule is proposed to resolve the externalities from the offline submarket. In the meanwhile, the price scaling factor is leveraged to reduce the effect of asymmetric information among driving simulators and traffic simulators in the offline submarket.
The simulation experiments are performed in a vehicular Metaverse system with 30 AVs, 30 virtual traffic simulators, 1 virtual driving simulator, and 1 RSU. The experimental results demonstrate that the proposed mechanism can improve 150\% social surplus compared with other baseline mechanisms. Finally, they develop a simulation testbed of generative AI-empowered simulation systems in the vehicular Metaverse.

\begin{figure}[t]
    \centering
    \includegraphics[width=0.78\linewidth]{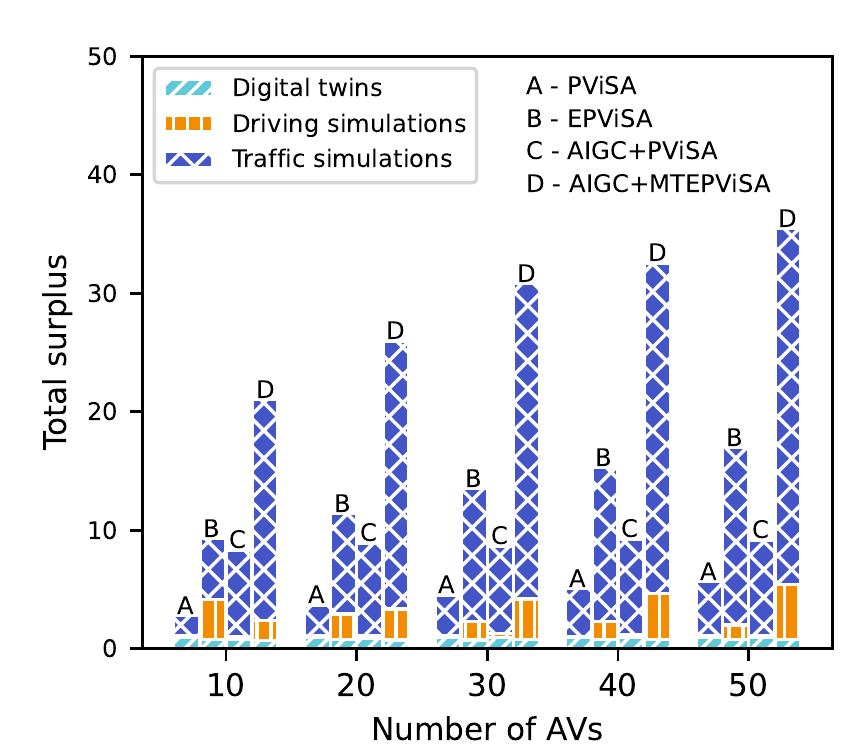}
    \caption{Performance evaluation of the MTEPViSA under different sizes of the market.}
    \label{fig:MTEPViSA}
\end{figure}

The vehicular mixed-reality (MR) Metaverse simulation environment was constructed employing a 3D model representing several city blocks within New York City. Geopipe, Inc. developed this model by leveraging artificial intelligence to generate a digital replica based on photographs taken throughout the city. The simulation encompasses an autonomous vehicle navigating a road, accompanied by strategically positioned highway advertisements. Eye-tracking data were gathered from human participants immersed in the simulation, utilizing the HMD Eyes addon provided by Pupil Labs. Subsequent to the simulation, participants completed a survey aimed at evaluating their subjective level of interest in each simulated scenario. As the experimental results shown in Fig.~\ref{fig:MTEPViSA}, According to the study, as the number of AVs continues to increase, the supply and demand mechanisms in the market are changing. Therefore, to improve market efficiency and total surplus, some mechanisms need to be adopted to coordinate supply and demand. We investigate the market mechanism and propose a mechanism based on AIGC technology to enhance market efficiency. Compared with the existing Physical-virtual Synchronization auction (PViSA) and Enhanced Physical-virtual Synchronization auction (EPViSA) mechanisms~\cite{xu2022epvisa,xu2023generative1}, the AIGC-empowered mechanism can double the total surplus under different numbers of AVs.

{\textit{{\textbf{Lesson Learned:}}}} This case study on generative AI-empowered autonomous driving opens a new paradigm for the vehicular Metaverse, where data and resources can be utilized more efficiently. The authors demonstrate the potential of generative AI models in synthesizing traffic and driving data to reduce the cost of data collection and labeling. The proposed MTEPViSA mechanism also provides a solution to determine and price the resources of roadside units for remote execution of digital twin tasks, improving market efficiency and total surplus. However, there are still several open issues that need to be addressed in this field. Firstly, it is necessary to investigate the potential negative impacts of generative AI models in synthesizing traffic and driving data, such as biases and inaccuracies. Secondly, more research is needed to develop robust and trustworthy mechanisms for determining and pricing the resources of RSUs to ensure fair and efficient allocation of resources. Thirdly, the proposed mechanism needs to be tested and evaluated in more complex and varied scenarios to ensure its scalability and applicability in real-world situations.

{\color{black}
\subsection{Blockchain-Powered Lifecycle Management for AI-Generated Content Products}
This case study delves into the application of a blockchain-based framework for managing the lifecycle of AIGC products within edge networks. The framework, proposed by the authors in~\cite{liu2023blockchain}, addresses concerns related to stakeholders, the blockchain platform, and on-chain mechanisms. We explore the roles and interactions of the stakeholders, discuss the blockchain platform's functions, and elaborate on the framework's on-chain mechanisms.
Within edge networks, the AIGC product lifecycle encompasses four main stakeholders: content creators, Edge Service Providers (ESPs), end-users, and adversaries. The following describes their roles and interplay within the system:
\begin{itemize}
\item {\textbf{Producers:}} Initiate the AIGC product lifecycle by proposing prompts for ESPs to generate content. They retain ownership rights and can publish and sell the generated products.
\item {\textbf{ESPs:}} Possess the resources to generate content for producers, charging fees based on the time and computing power used for the tasks.
\item {\textbf{Consumers:}} View and potentially purchase AIGC products, participating in multiple trading transactions throughout the product lifecycle.
\item {\textbf{Attackers:}} Seek to disrupt normal operations of AIGC products for profit through ownership tampering and plagiarism.
\end{itemize}
Considering the roles of these stakeholders, the blockchain platform fulfills two primary functions: providing a traceable and immutable ledger and supporting on-chain mechanisms. Transactions are recorded in the ledger and validated by full nodes using a consensus mechanism, ensuring security and traceability. ESPs act as full nodes, while producers and consumers serve as clients.

To address the concerns arising from stakeholder interactions, the framework employs three on-chain mechanisms~\cite{liu2023blockchain}:
\begin{itemize}
\item {\textbf{Proof-of-AIGC:}} A mechanism that defends against plagiarism by registering AIGC products on the blockchain. It comprises two phases: proof generation and challenge.
\item {\textbf{Incentive Mechanism:}} Safeguards the exchange of funds and AIGC ownership using Hashed Timelock Contracts (HTLCs).
\item {\textbf{Reputation-based ESP Selection:}} Efficiently schedules AIGC generation tasks among ESPs based on their reputation scores.
\end{itemize}
The Proof-of-AIGC mechanism plays a vital role in maintaining the integrity of AIGC products. It encompasses two stages: proof generation and challenge. The objective of proof generation is to record AIGC products on the blockchain, while the challenge phase allows content creators to raise objections against any on-chain AIGC product they deem infringing upon their creations. If the challenge is successful, the duplicate product can be removed from the registry, thus protecting the original creator's intellectual property rights.

To further strengthen the security of the AIGC ecosystem, a pledged deposit is necessary to initiate a challenge, preventing arbitrary challenges that could burden the blockchain. This process comprises four steps: fetching the proofs, verifying the challenger's identity, measuring the similarity between the original product and the duplicate, and checking the results.

The AIGC economic system necessitates an incentive mechanism to motivate stakeholders and ensure legitimate exchanges of funds and ownership. The Incentive Mechanism rewards ESPs for maintaining the ledger and providing blockchain services. There are no transaction fees, and block generators follow a first-come-first-serve strategy. A two-way guarantee protocol using Hash Time Lock (HTL) is designed to build mutual trust and facilitate AIGC circulation during both the generation and trading phases.

The Proof-of-AIGC mechanism tackles issues like ownership manipulation and AIGC plagiarism, while the incentive mechanism ensures compliance with pre-established contracts. Furthermore, a reputation-based ESP selection accommodates ESP heterogeneity, which is crucial for efficient AIGC lifecycle management.
Specifically, within the AIGC lifecycle management architecture, producers can concurrently interact with multiple heterogeneous ESPs, necessitating the identification of a trustworthy ESP for a specific task. Conventional approaches involve selecting the most familiar ESP to minimize potential risks, which may result in unbalanced workload distribution and increased service latency among ESPs.
To address this challenge, a reputation-based ESP selection strategy is incorporated into the framework. This strategy ranks all accessible ESPs according to their reputation, which is computed using Multi-weight Subjective Logic (MWSL). The primary objectives are to assist producers in choosing the most reliable ESP, distribute the workload evenly across multiple ESPs, and motivate ESPs to accomplish tasks promptly and honestly, as a negative reputation impacts their earnings.

Producers identify suitable ESPs by computing the reputation of all potential ESPs, ranking them based on their current reputation, and allocating the AIGC generation task to the ESP with the highest standing. In MWSL, the concept of "opinion" serves as the fundamental element for reputation calculation. Local opinions represent the assessments of a specific producer who has directly interacted with the ESPs, while recommended opinions are derived from other producers who have also engaged with the ESPs.
To mitigate the effect of subjectivity, an overall opinion is generated for each producer by averaging all the acquired recommended opinions. As producers possess varying degrees of familiarity with ESPs, the weight of their recommended opinions differs. Reputation is determined by combining a producer's local opinion with the overall opinion.
The reputation scheme accomplishes its design objectives by quantifying the trustworthiness of ESPs, aiding producers in selecting the most dependable ESP, reducing service bottlenecks, and incentivizing ESPs to deliver high-quality AIGC services to maximize their profits.

A demonstration of the AIGC lifecycle management framework is conducted to verify the proposed reputation-based ESP selection approach~\cite{liu2023blockchain}. The experimental setup comprises three ESPs and three producers, with the AIGC services facilitated by the Draw Things application. Several parameters are configured, and producers can employ the Softmax function to ascertain the probability of choosing each ESP.
\begin{figure}[t]
    \centering
    \includegraphics[width=0.9\columnwidth]{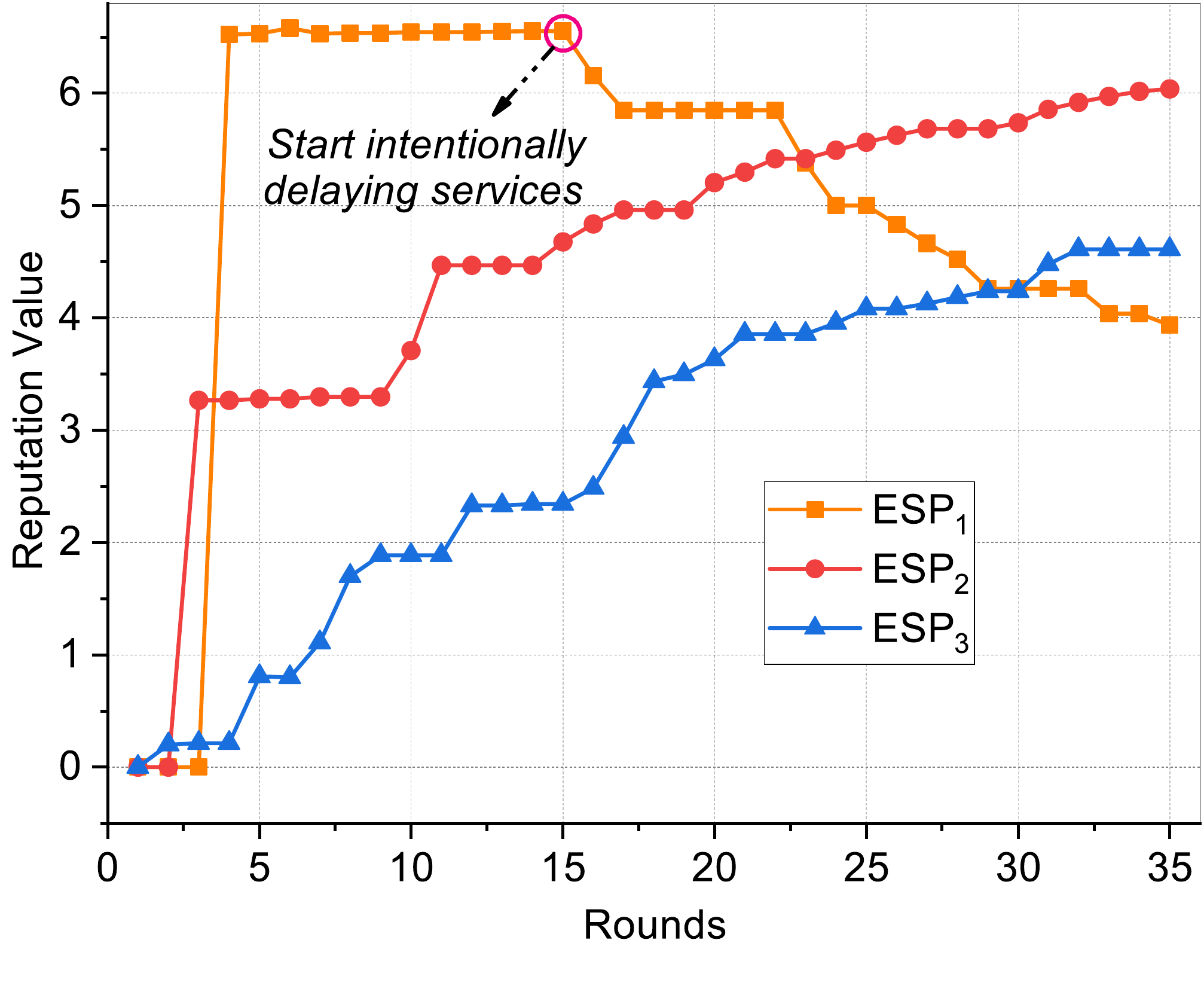}
    \caption{The reputation trends of three ESPs (from the perspective of a random producer)~\cite{liu2023blockchain}.}
    \label{figsemme}
\end{figure}
The reputation trends of the three ESPs are shown in Fig.~\ref{figsemme}, with ESP1 attaining the highest rank and remaining stable owing to its superior service quality. When ESP1 deliberately postpones AIGC services, its reputation declines sharply, while the reputations of ESP2 and ESP3 continue to rise. The proposed reputation strategy effectively measures the trustworthiness of ESPs, enabling producers to effortlessly discern the most reliable ESPs and motivating ESPs to operate with integrity. {\color{black} In reality, the dynamics of ESP selection would become more complex with an increase in the number of ESPs and producers. This underlines the potential challenges and importance of effective reputation management strategies in such expanded scenarios. The reputation-based selection method's robustness and scalability in a larger network is a subject for future work.}
\begin{figure}[t]
\centering
\includegraphics[width=0.9\columnwidth]{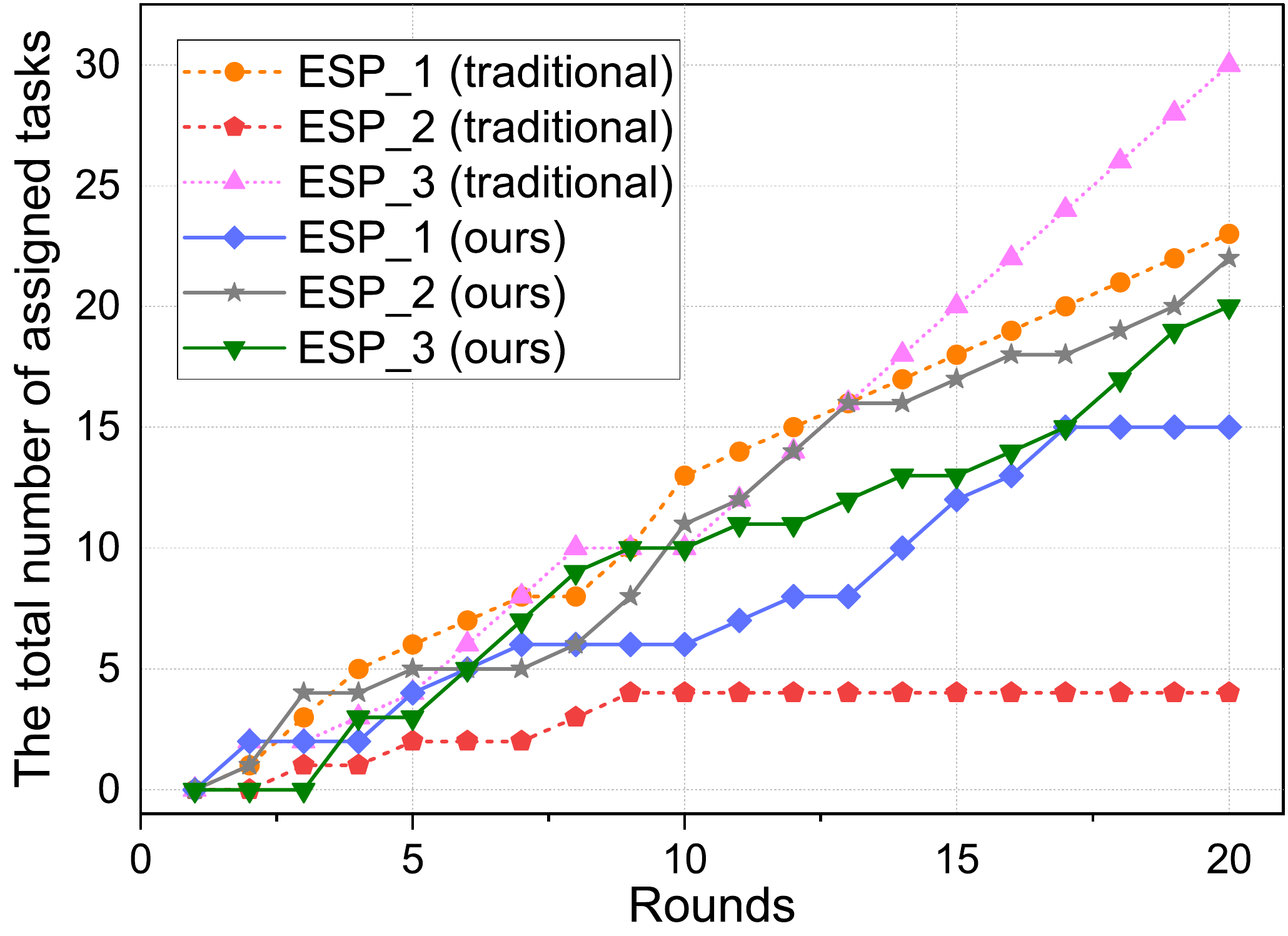}
\caption{The total number of assigned tasks of three ESPs~\cite{liu2023blockchain}.}
\label{faefaef}
\end{figure}
The workload of ESPs under different ESP selection methods is also demonstrated in Fig.~\ref{faefaef}. Traditional methods lead to uneven workloads and extended service latencies. {\color{black}Conversely, the proposed reputation-based method effectively balances the workload among ESPs. This is achieved by enabling producers to quantitatively assess the trustworthiness of ESPs without solely relying on their experiential judgment. The effectiveness of this approach in a network with a larger number of ESPs is an aspect that invites further exploration.}

{\textit{{\textbf{Lesson Learned:}}}} 
The case study on blockchain-powered lifecycle management for AI-generated content products highlights the potential of a blockchain-based framework in addressing key concerns like stakeholder interactions, platform functionality, and on-chain mechanisms. The primary lessons learned emphasize the importance of defining clear stakeholder roles, implementing robust mechanisms such as Proof-of-AIGC and Incentive Mechanism to ensure system integrity, and employing a reputation-based ESP selection scheme to balance workload and encourage honest performance. These insights collectively contribute to the effective management of the AIGC product lifecycle within edge networks. Future research in blockchain-powered lifecycle management for AI-generated content products can explore several promising directions:
\begin{itemize}
    \item Enhancing the efficiency and scalability of the blockchain platform to handle an increased number of transactions and support a growing AIGC ecosystem might be critical.
    \item Refining the reputation-based ESP selection scheme to account for more sophisticated factors, such as task complexity, completion time, and user feedback, could lead to more accurate and dynamic trustworthiness evaluations.
    \item Incorporating privacy-preserving techniques to protect sensitive data in AIGC products and user information without compromising the transparency and traceability of blockchain technology would be valuable. 
\end{itemize}
}

\section{Implementation Challenges in Mobile AIGC Networks}\label{sec:challenges}

When providing AIGC services, a significant amount of computational and storage resources are required to run the {\color{black} generative AI model}. These computation and storage-intensive services pose new challenges to existing mobile edge computing infrastructure. As discussed in Section~\ref{sec:infrastructure}, a cloud-edge-mobile collaborative computing architecture can be implemented to provide AIGC services. However, several critical implementation challenges must be addressed to improve resource utilization and the user experience.

\subsection{Edge Resource Allocation}

AIGC service provisioning based on edge intelligence is computationally and communication-intensive for resource-constrained edge servers and mobile devices~\cite{hu2019dynamic, zhang2021deep}. Specifically, AIGC users send service allocation requests to edge services. Upon receiving these AIGC requests, edge servers perform the AIGC tasks and deliver the output to users~\cite{zhang2021joint}. During this AIGC service provisioning interaction, model accuracy and resource consumption are the most common metrics. Consequently, significant efforts are being made to coordinate mobile devices and edge servers for deploying generative AI at mobile edge networks. As summarized in Table \ref{table:resourceallocation}, several Key Performance Indicators (KPIs) for edge resource allocation in AIGC networks are presented below.
\begin{table*}[htbp]
\small\centering
\caption{Summary of scenarios, problems, benefits/challenges, and mathematical tools of edge resource allocation.}
\begin{tabular}{|m{.05\textwidth}<{\centering}|m{.2\textwidth}<{\centering}|m{.2\textwidth}<{\centering}|m{.2\textwidth}<{\centering}|m{.2\textwidth}<{\centering}|}
    \hline
    % \rowcolor[HTML]{C0C0C0} 
    \textbf{Ref.} &
    \textbf{Scenarios} &
    \textbf{Performance Metrics/Decision Variables} &
    \textbf{Benefits/Challenges} &
    \textbf{Mathematical Tools} \\ \hline
    \cite{wang2018edge} &
    Adaptive control for distributed edge learning &
    Model loss/Steps of local updates, the total number of iterations &
    Provisioning AIGC services in resource-constrained edge environments &
    Control theory \\ \hline
    \cite{hsieh2017gaia} &
    Geo-distributed ML &
    Execution time/Selective barrier, mirror clock &
    Provisioning Localized AIGC services &
    Convergence analysis \\ \hline
    \cite{lin2021optimizing} &  AI service placement in mobile edge intelligence  &
    Total time and energy consumption/Service placement decision, local CPU frequencies, uplink bandwidth, edge CPU frequency &
    Fully utilize scarce wireless spectrum and edge computing resources in provisioning AIGC services &
    ADMM \\ \hline
    \cite{li2020optimizing} &  Joint model training and task inference &
    Energy consumption and execution latency/Model download decision and task splitting ratio &
    Integrated fine-tuning and inference for {\color{black} generative AI models} with heterogeneous computing resources  &
    ADMM \\ \hline
    \cite{zhao2022edgeadaptor} &  Serving edge DNN inference for multiple applications and multiple models &
    Inference accuracy, latency, resource cost/Application configuration, DNN model selection, and edge resources &
    Provision rich AIGC services for long-term utility maximization  &
    Regularization-based online optimization \\ \hline
    \cite{tang2020joint} &  Multi-user collaborative DNN partitioning  &
    Execution latency/Partitioning, computation resources &
    Providing insights for partitioning {\color{black} generative AI models} under edge-mobile collaboration &
    Iterative alternating optimization \\ \hline
    \cite{lim2021decentralized} &   Hierarchical federated edge learning  &
    Data convergence and revenue/Cluster selection and payment &
    Provisioning privacy-preserving AIGC services in edge networks &
    Evolutionary game and auction \\ \hline
\end{tabular}%
\label{table:resourceallocation}
\end{table*}
Here are several KPIs for edge resource allocation in AIGC networks.
\begin{itemize}
\item Model accuracy: In a resource-constrained edge computing network, a key issue when allocating edge resources is optimizing the accuracy of AI services while fully utilizing network resources~\cite{yang2023over}. Besides objective image recognition and classification tasks, AI models are also based on the content's degree of personalization and adaptation. Thus, optimizing AIGC content networks may be more complex than traditional optimization since personalization and customization make evaluating model accuracy more unpredictable.
\item Bandwidth utilization: While providing AIGC services, the edge server must maximize its channel utilization to ensure reliable service in a high-density edge network. To allocate its bandwidth resources more efficiently, the edge server must control channel access to reduce interference between user requests and maximize the quality of its AIGC service to attract more users.
\item Edge resource consumption: Deploying AIGC services in edge networks requires computationally intensive AI training and inference tasks that consume substantial resources. Due to the heterogeneous nature of edge devices, edge services consume resources in generating appropriate AIGC while processing users' requests~\cite{zhang2023energy}. Deployment of AIGC services necessitates continuous iteration to meet actual user needs, as generation results of {\color{black} generative AI models} are typically unstable. This constant AIGC service provisioning at edge servers leads to significant resource consumption.
\end{itemize}

\begin{figure}
    \centering
    \includegraphics[width=1\linewidth]{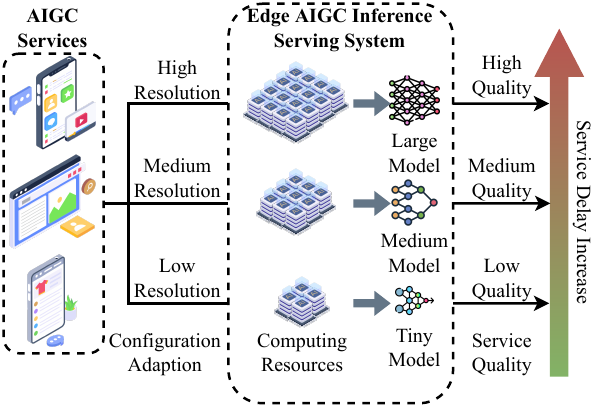}
    \caption{Dynamic AIGC application configuration and generative AI model compression for serving AIGC services in mobile AIGC networks.}
    \label{fig:adaptor}
\end{figure}

{\color{black}Obtaining a balance between model accuracy and resource consumption can be challenging in resource-constrained edge computing networks. One potential strategy is to adjust the trade-off between model accuracy and resource consumption according to the needs of the users. For example, in some cases, a lower level of model accuracy may be acceptable if it results in faster response times or lower resource consumption. Another approach is to use transfer learning, which involves training an existing model on new data to improve accuracy while requiring fewer computational resources. Model compression techniques can also be used to reduce the size of the AI model without significantly impacting accuracy. However, it is important to note that these techniques may not be applicable in all scenarios, as personalization and customization can make evaluating model accuracy more unpredictable. Deployment of AIGC services necessitates continuous iteration to meet actual user needs, as generation results of {\color{black} generative AI models} are typically unstable. Due to the heterogeneous nature of edge devices, edge services consume resources in generating appropriate AIGC while processing users' requests. This constant AIGC service provisioning at edge servers leads to significant resource consumption.}

To provide intelligent applications at mobile edge networks, considerable effort should focus on the relationship between model accuracy, networking, communication, and computation resources at the edge. Simultaneously, offering AIGC services is challenging due to the dynamic network environment and user requirements at mobile edge networks. The authors in \cite{hsieh2017gaia} propose a threshold-based approach for reducing traffic at edge networks during collaborative learning. By considering computation resources, the authors in~\cite{wang2018edge} examine the distributed ML problem under communication, computation, storage, and privacy constraints. Based on the theoretical results obtained from the distributed gradient descent convergence rate, they propose an adaptive control algorithm for distributed edge learning to balance the trade-off between local updates and global parameter aggregations. The experimental results demonstrate the effectiveness of their algorithm under various system settings and data distributions.

{\color{black}Generative AI models} often require frequent fine-tuning and retraining for newly generated data and dynamic requests in non-stationary mobile edge networks~\cite{ditzler2015learning}. Due to limited storage resources at edge servers and the different customization demands of AIGC providers, the AIGC service placement problem is investigated in~\cite{lin2021optimizing}. To minimize total time and energy consumption in edge AI systems, the AI service placement and resource allocation problem is formulated as an MINLP. In the optimization problem, AI service placement and channel allocation are discrete decision variables, while device and edge frequencies are continuous variables. However, solving this problem is not trivial, particularly in large-scale network environments. Thus, the authors propose an alternating direction method of multipliers (ADMM) to reduce the complexity of solving this problem. The experimental results demonstrate that this method achieves near-optimal system performance while the computational complexity grows linearly as the number of users increases. Moreover, when edge intelligence systems jointly consider AI model training and inference~\cite{li2020optimizing}, the ADMM method can optimize edge resources. Additionally, the authors~\cite{zhao2022edgeadaptor} explore how to serve multiple AI applications and AI models at the edge. They propose EdgeAdapter, as illustrated in Fig.~\ref{fig:adaptor}, to balance the triple trade-off between inference accuracy, latency, and resource consumption. To provide inference services with long-term profit maximization, they first analyze the problem as an NP-hard problem and then solve it with a regularization-based online algorithm.

In mobile AIGC networks, an effective architecture for providing AIGC services is to partition a large {\color{black} generative AI model} into multiple smaller models for local execution~\cite{kang2017neurosurgeon}. In~\cite{tang2020joint}, the authors consider a multi-user scenario with massive IoT~\cite{zhang2022inverse} devices that cooperate to support an intelligent application collaboratively. Although partitioning large ML models and distributing smaller models to mobile devices for collaborative execution is feasible, the model distribution and result aggregation might incur extra latency during model training and inference. Additionally, the formulated optimization problem is complex due to its numerous constraints and vast solution space. To address these issues, the authors propose an alternative iterative optimization to obtain solutions in polynomial time. Furthermore, AIGC services allow users to input their preferences into {\color{black} generative AI models}. Therefore, to preserve user privacy among multiple users during collaborative model training and inference~\cite{wu2022collaborative}, the authors in~\cite{lim2021decentralized} investigate the communication efficiency issues of decentralized edge intelligence enabled by FL. In the FL network, thousands of mobile devices participate in model training. However, selecting appropriate cluster heads for aggregating intermediate models can be challenging. Decentralized learning approaches can improve reliability while sacrificing some communication performance, unlike centralized learning with a global controller. A two-stage approach can be adopted in decentralized learning scenarios to improve the participation rate. In this approach, evolutionary game-based allocation can be used for cluster head selection, and DL-based auction effectively rewards model owners.

\subsection{Task and Computation Offloading}
\begin{table*}[htbp]
\small\centering
\caption{Summary of scenarios, problems, benefits/challenges, and mathematical tools of task and computation offloading.}
\begin{tabular}{|m{.05\textwidth}<{\centering}|m{.2\textwidth}<{\centering}|m{.2\textwidth}<{\centering}|m{.2\textwidth}<{\centering}|m{.2\textwidth}<{\centering}|}
    \hline
    % \rowcolor[HTML]{C0C0C0} 
    \textbf{Ref.} &
    \textbf{Scenarios} &
    \textbf{Performance Metrics/Decision variables} &
    \textbf{Benefits/Challenges} &
    \textbf{Mathematical Tools} \\ \hline
    \cite{fan2021accuracy} &
    Edge intelligence in IoT &
    Processing delay/Task offloading decisions&
    Offload AIGC tasks for improving inference accuracy&
    Optimization theory \\ \hline
    \cite{chen2021dnnoff} &
    Intelligent IoT applications &
    Processing time/Offloading decisions&
    Support on-demand changes for AIGC applications &
    Random forest regression \\ \hline
    \cite{kang2017neurosurgeon} &
    Collaborative intelligence between the cloud and mobile edge &
    Latency and energy consumption/DNN computation partitioning&
    Cloud and mobile edge collaborative intelligence for {\color{black} generative AI models} &
    Greedy algorithm\\ \hline
    \cite{chen2020computation} &
    Cloud-edge intelligence &
    Service response time/Task processing node&
    Reduce the average response time for multi-task parallel AIGC services &
    Genetic algorithm \\ \hline
    \cite{lin2019cost} &
    Cost-driven offloading for DNN-Based applications &
    System costs/Number of layers &
    Minimize costs of AIGC services in a cloud-edge-end collaborative environment &
    Genetic algorithm based on particle swarm optimization \\ \hline
    \cite{ren2019coding} &
    Industrial edge intelligence &
    A weighted sum of task execution time and energy consumption/Task assignment&
    Multi-objective optimization of large-scale AIGC tasks with multiple connected devices &
    Generative coding evolutionary algorithm\\ \hline
    \cite{jeong2018computation} &
    Computation offloading for ML web apps &
    Inference time/Pre-sending decisions&
    Reduce execution overheads of AIGC tasks with pre-sending snapshots &
    Hill climbing algorithm\\ \hline
    \cite{li2022multi} &
    Cooperative edge intelligence &
    Quality of experience/Offloading decisions&
    Enhance vertical-horizontal cooperation in multi-user AIGC co-inference scenarios&
    Federated multi-agent reinforcement learning\\ \hline
\end{tabular}%
\label{table:offloading}
\end{table*}

In general, executing {\color{black} generative AI models} that generate creative and valuable content necessitates substantial computational resources, which is impractical for mobile devices with limited resources~\cite{cao2022survey, zhan2020deep}. Offering high-quality and low-latency AIGC services is challenging for mobile devices with low processing power and limited battery life. Fortunately, AIGC users can offload the tasks and computations of {\color{black} generative AI models} over the RAN to edge servers located in proximity to the users. This alleviates the computational burden on mobile devices.

As listed in Table~\ref{table:offloading}, several KPIs are specifically relevant to computation offloading in mobile AIGC networks:
\begin{itemize}
\item Service latency: Service latency refers to the delay associated with data input and retrieval as well as the model inference computations that users perform to generate AIGC~\cite{lin2023efficient}. By offloading AIGC tasks from mobile devices, such as fine-tuning and inference, to edge servers for execution, the total latency in mobile AIGC networks can be reduced. Unlike local execution of the {\color{black} generative AI model}, offloading AI tasks to the edge server for execution introduces additional latency when transmitting personalized instructions and downloading AIGC content.
\item Reliability: Reliability evaluates users' success rate in obtaining personalized data accurately. On the one hand, when connecting to the edge server, users may experience difficulty uploading the requested data to edge servers or downloading the results from servers due to dynamic channel conditions and wireless network instability. On the other hand, the content generated by the {\color{black} generative AI model} may not fully meet the needs of AIGC users in terms of personalization and customization features. Unsuccessful content reception and invalid content affect the AIGC network's reliability.
\end{itemize}

\begin{figure}
    \centering
    \includegraphics[width=1\linewidth]{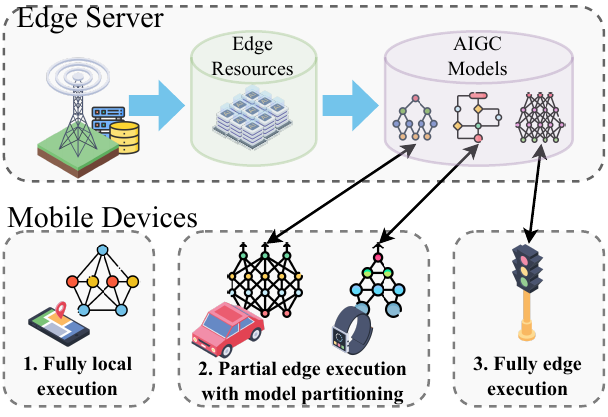}
    \caption{Model partitioning in mobile AIGC networks. The {\color{black} generative AI models} of mobile devices can be split and full or partial of them can be offloaded to edge servers for remote execution.}
    \label{fig:c1}
\end{figure}

{\color{black} When implementing cloud-edge collaborative training and fine-tuning for {\color{black} generative AI models}~\cite{wang2023overview}, it is important to consider specific algorithms or techniques that enable effective collaboration between cloud and edge servers~\cite{wu2020accuracy, zhang2021deep}. For example, FL and distributed training approaches can facilitate the collaboration process by allowing edge servers to train models locally and then send the updated weights to the cloud server for aggregation~\cite{yang2022federated}. The division of responsibilities between cloud and edge servers can also greatly affect the overall efficiency and performance of the {\color{black} generative AI models}. Therefore, it is crucial to discuss and implement appropriate schemes for determining which tasks are offloaded to the edge servers and which are performed on the cloud server.}
To provide AIGC services in edge intelligence-empowered IoT, offloading ML tasks to edge servers for remote execution is a promising approach for computation-intensive AI model inference~\cite{tian2022jmsnas}. For instance, in Fig~\ref{fig:c1}, multiple lightweight ML models can be loaded into IoT devices, while large-scale ML models can be installed and executed on edge servers~\cite{zhou2019edge}. Heterogeneous {\color{black} generative AI models} can be deployed on mobile devices and edge servers according to their resource demands and service requirements~\cite{zhang2020q}. However, the multiple attributes of ML tasks, such as accuracy, inference latency, and reliability, render the offloading problem of AIGC highly complex. Therefore, the authors in~\cite{fan2021accuracy} propose an ML task offloading scheme to minimize task execution latency while guaranteeing inference accuracy. Considering error inference leading to extra delays in task processing, they initially model the inference process as M/M/1 queues, which are also applicable to the AIGC service process. Furthermore, the optimization problem of ML task execution is formulated as a Mixed-Integer Nonlinear Programming (MINLP) to minimize provisioning delay, which can be adopted in the inference process of AIGC services. To extend the deterministic environment in~\cite{fan2021accuracy} into a more general environment, the authors in \cite{chen2021dnnoff} first propose an adaptive translation mechanism to automatically and dynamically offload intelligent IoT applications. Then, they make predictive offloading decisions using a random forest regression model. Their experiments demonstrate that the proposed framework reduces response times for complex applications by half. Such ML methods can also be used to analyze AIGC network traffic to improve service delivery efficiency and reliability.

{\color{black} The success of edge-mobile collaboration for AIGC services is dependent on several factors, including the type of service, user characteristics, computational resources, and network conditions~\cite{wu2022ai, wen2023task, koda2020communication}. For instance, a real-time AIGC service may have different latency requirements compared to an offline service. Similarly, the required computational resources may vary depending on the model's complexity~\cite{wu2023split}. Additionally, the user profile, including location and device type, may affect the selection of edge servers for task offloading. Furthermore, network conditions such as bandwidth and packet loss rate can impact the reliability and latency of the service. Therefore, it is necessary to implement effective resource allocation and task offloading schemes to ensure high-quality and low-latency AIGC services in dynamic and diverse environments.} Cloud-edge collaborative intelligence enables local tasks to be offloaded to edge and cloud servers. AIGC can benefit from cloud-edge intelligence, as edge servers can provide low-latency AIGC services while cloud servers can offer high-quality AIGC services. The authors in~\cite{kang2017neurosurgeon} develop a scheme called Neurosurgeon to select the optimal partitioning point based on model architectures, hardware platforms, network conditions, and load information at the servers to automatically partition the computation of tensors of DNNs between cloud and edge servers. Furthermore, the authors in~\cite{saguil2020layer} find that the layered approach can reduce the number of messages transmitted between devices by up to 97\% while only decreasing the accuracy of models by a mere 3\%. However, multiple AIGC services should be considered in cloud-edge collaborative intelligence that differs in types (e.g., text, images, and videos) and their diverse quality of service (QoS) requirements~\cite{kang2022personalized}. In multi-task parallel scheduling~\cite{chen2020computation}, the genetic algorithm can also be used to make real-time model partitioning decisions. The authors in~\cite{lin2019cost} propose a cost-driven strategy for AI application offloading through a self-adaptive genetic algorithm based on particle swarm optimization.

In industrial edge intelligence, where edge intelligence is embedded in the industrial IoT~\cite{yang2021local, ren2019coding, zhang2020deephealth, zhang2021optimizing}, offloading computation tasks to edge servers is an efficient solution for self-organizing, autonomous decision-making, and rapid response throughout the manufacturing lifecycle, which is similarly required by mobile AIGC networks. Therefore, efficiently solving task assignment problems is crucial for effective {\color{black} generative AI model} inference. However, the coexistence of multiple tasks among devices makes system response slow for various tasks. For example, text-based and image-based AIGC may coexist on the same edge device. As one solution, in \cite{ren2019coding}, the authors propose a coding group evolution algorithm to solve large-scale task assignment problems, where tasks span the entire lifecycle of various products, including real-time monitoring, complex control, product structure computation, multidisciplinary cooperation optimization, and production process computation. Likewise, the AIGC lifecycle includes data collection, labeling, model training and optimization, and inference. Furthermore, a simple grouping strategy is introduced to parallel partition the solution space and accelerate the evolutionary optimization process. In contrast to VM-level adaptation to specific edge servers\cite{matsubara2019distilled}, the authors propose application-level adaptation for generic servers. The lighter adaptation framework in~\cite{jeong2018computation} further improves transmission time and user data privacy performance, including offloading and data/code recovery to generic edge servers.

{\color{black} Ensuring dependable task offloading is crucial in providing superior AIGC services with minimal latency in edge computing. For instance, data transmission redundancy can enhance dependability by transmitting data via multiple pathways to mitigate network congestion or failures. By incorporating these techniques, task offloading dependability in edge computing can be enhanced, thereby leading to more efficient and effective AIGC services.} Most intelligent computing offloading solutions converge slowly, consume significant resources, and raise user privacy concerns~\cite{jiang2021intelligence,sun2021cooperative}. The situation is similar when leveraging learning-based approaches to make AIGC service offloading decisions. Consequently, the authors enhance multi-user QoE~\cite{he2020edge} for cooperative edge intelligence in~\cite{li2022multi} with federated multi-agent reinforcement learning. They formulate the cooperative offloading problem as a Markov Decision Process (MDP). The state is composed of current tasks, local loads, and edge loads. Learning agents select task processing positions to maximize multi-user QoE, which simultaneously considers service latency, energy consumption, task drop rate, and privacy protection. Similarly, AIGC service provisioning systems can easily adopt the proposed solution for maximizing QoE in AIGC services.

\subsection{Edge Caching}

\begin{figure}[t]
    \centering
    \includegraphics[width=1\linewidth]{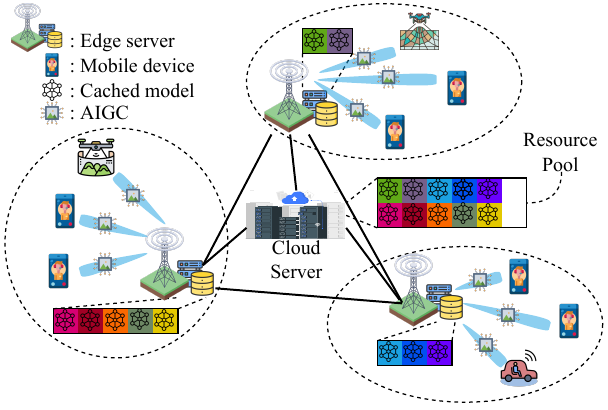}
    \caption{An overview of edge caching in mobile AIGC networks. By caching the {\color{black} generative AI model} on the edge servers, the latency of AIGC services can be reduced and the network congestion in the core network can be reduced.}
    \label{fig:caching}
\end{figure}
\begin{table*}[htbp]
\small\centering
\caption{Summary of scenarios, problems, performance metrics, and mathematical tools for edge caching in AIGC networks.}
\begin{tabular}{|m{.05\textwidth}<{\centering}|m{.2\textwidth}<{\centering}|m{.2\textwidth}<{\centering}|m{.2\textwidth}<{\centering}|m{.2\textwidth}<{\centering}|}
    \hline
    % \rowcolor[HTML]{C0C0C0} 
    \textbf{Ref.} &
    \textbf{Scenarios} &
    \textbf{Performance Metrics/Decision Variables} &
    \textbf{Benefits/Challenges} &
    \textbf{Mathematical Tools} \\ \hline
    \cite{guo2019edgeserve} &
    DL Model caching at the edge &
    Runtime memory consumption and loading time/Model preload policy &
    Manage and utilize GPU memories of edge servers for caching {\color{black} generative AI models} &
    Cache replacement algorithms \\ \hline
    \cite{ogden2021many} &
    Caching many models at the edge &
    Model load and execution latency and monetary cost /Caching eviction policy &
    Improve scalability of mobile AIGC networks via model-level caching deployment and replacement&
    Model utility calculation \\ \hline
    \cite{xu2018deepcache} &
    Cache for mobile deep vision applications&
    Latency, accuracy loss, energy saving/Caching policy, user selection, transmit power, bandwidth ratio &
    Caching for users' requests for multimodal AIGC services &
    Greedy algorithm\\ \hline
    \cite{fuerst2021faascache} &
    Cache for functions in serverless computing&
    Execution time, cold start proportion/Function keep-alive policy &
    Keep {\color{black} generative AI models} alive and warm for in-contextual inference &
    Greedy-dual based approach\\ \hline
    \cite{zheng2021knowledge} &
    Knowledge caching for FL &
    Transmission latency and energy consumption/Caching policy, user selection, transmit power, bandwidth ratio &
    Privacy-preserving model caching via knowledge of AIGC requests &
    Optimization theory \\ \hline
\end{tabular}%
\label{table:edgecaching}
\end{table*}
Edge caching is the delivery of low-latency content and computing services using the storage capacity of edge base stations and mobile devices~\cite{wang2020attention, mu2023communication}. As illustrated in Fig.~\ref{fig:caching}, in mobile AIGC networks, users can request AIGC services without accessing cloud data centers by caching {\color{black} generative AI models} in edge servers and mobile devices. Unlike the cache in traditional content distribution networks, the {\color{black} generative AI model} cache also requires computing resources to support its execution. Additionally, the {\color{black} generative AI model} needs to gather user historical requests and profiles in context to provide personalized services during the AIGC service process. As shown in Table~\ref{table:edgecaching}, here are several KPIs for edge caching in AIGC networks:
\begin{itemize}
\item Model access delay: Model access latency is an important indicator of AIGC service quality. The latency is lowest when the {\color{black} generative AI model} is cached in the mobile device~\cite{yao2022loading}. The model access latency must also be calculated considering the delay in the wireless communication network when the edge server provides the {\color{black} generative AI model}. Finally, the core network latency must be considered when the cloud provides the AIGC service.
\item Backhaul traffic load: The load on the backhaul traffic is significantly reduced, as the requests and results of AIGC services do not need to go through the core network when the {\color{black} generative AI model} is cached in the mobile edge network.
\item Model hit rate: Similar to content hit rate, the model hit rate is an important metric for {\color{black} generative AI models} in the edge cache. It can be used for future model exits and loading during model replacement.
\end{itemize}

As there is sufficient infrastructure and resources in the cloud computing infrastructure, the {\color{black} generative AI model} can be fully loaded into the GPU memory for real-time service requests. In contrast, the proposed EdgeServe in~\cite{guo2019edgeserve} keeps models in main memory or GPU memory so that they can be effectively managed or used at the edge. Similar to traditional CDNs, the authors use model execution caches at edge servers to provide immediate AI delivery. In detail, there are mainly three challenges in {\color{black} generative AI model} caching:
\begin{itemize}
\item Resource-constraint edge servers: Compared to the resource-rich cloud, the resources of servers in the edge network, such as GPU memory, are limited~\cite{shi2020communication}. Therefore, caching all {\color{black} generative AI models} on one edge server is infeasible.
\item Model-missing cost: When the mobile device user requests AIGC, the corresponding model is missed if the {\color{black} generative AI model} used to generate the AIGC is not cached in the current edge server~\cite{ogden2021many}. In contrast to the instantly available AIGC service, if the {\color{black} generative AI model} is missing, the edge server needs to send a model request to the cloud server and download the model, which causes additional overhead in terms of bandwidth and latency.
\item Functionally equivalent models: The number of {\color{black} generative AI models} is large and increases depending on the number of detailed tasks~\cite{xie2022robust}. Meanwhile, AI models have similar functions in different applications, i.e., functionally equivalent. For example, for image recognition tasks, a large number of models with different architectures are proposed to recognize features in images, which have different model architectures and computation requirements.
\end{itemize}
To address these challenges, the authors in~\cite{guo2019edgeserve} formulate the problem of edge modeling as determining which DL models should be preloaded into memory and which should be discarded when the memory is full while satisfying the requirements of inferential response times. Fortunately, this edge model caching problem can be solved using existing cache replacement policies for edge content caching. The accuracies and computation complexities of DL models make this optimization problem more complicated than conventional edge caching problems. Similarly, for resource-constrained edge servers, the {\color{black} generative AI model} can be dynamically deployed and replaced. However, an effective caching algorithm for loading and unloading the {\color{black} generative AI models} to maximize the hit rate has not yet been investigated.

As the capabilities of AI services continue to grow and diversify, multiple models need to be deployed simultaneously at the edge to achieve various tasks, including classification, recognition, text/image/video generation~\cite{ogden2020mdinference}. Especially in mobile AIGC networks, multiple base models need to work together to generate a large amount of multimodal synthetic data. Many models play a synergistic role in the AIGC services at the edge of the network, while the support of multiple models also poses a challenge to the limited GPU memory of the edge servers. Therefore, the authors in~\cite{ogden2021many} propose a model-level caching system with an eviction policy according to model characteristics and workloads. The model eviction policy is based on model utility calculation from cache miss penalty and the number of requests. This model-aware caching approach introduces a new direction for providing AIGC services at mobile edge networks with heterogeneous requests. Experimental results show that compared to the non-penalty-aware eviction policy, the model load delay can be reduced by 1/3. This eviction policy can also be adopted in the problem of which unpopular {\color{black} generative AI models} should be unloaded.

At mobile AIGC networks, not only the {\color{black} generative AI model} needs to be cached, but also the AIGC requests and results can be cached to reduce the latency of service requests in AIGC networks. To this end, the authors devise a principled cache design to accelerate the execution of CNN models by exploiting the temporal locality of video for continuous vision tasks to support mobile vision applications~\cite{buckler2018eva2}. The authors in~\cite{xu2018deepcache} propose a principled cache scheme, named DeepCache, to retrieve reusable results and reuse them within a fine-grained CNN by exploiting the temporal locality of the mobile video stream. In DeepCache, mobile devices do not need to offload any data to the cloud and can support the most popular models. Additionally, without requiring developers to retrain models or tune parameters, DeepCache caches inference results for unmodified CNN models. Overall, DeepCache can reduce energy consumption by caching content to reduce model inference latency while sacrificing a small fraction of model accuracy.

In serverless computing for edge intelligence, mobile devices can call functions of AIGC services at edge servers, which is more resource-efficient compared to container and virtual machine (VM)-based AIGC services. Nevertheless, such functions suffer from the cold-start problem of initializing their code and data dependencies at edge servers. Although the execution time of each function is usually short, initialization, i.e., fetching and installing prerequisite libraries and dependencies before execution, is time-consuming~\cite{oakes2018sock}. Fortunately, the authors in~\cite{fuerst2021faascache} show that the caching-based keep-alive policy can be used to address the cold-start problem by demonstrating that the keep-alive function is equivalent to caching. Finally, to balance the trade-off between server memory utilization and cold-start overhead, a greedy dual-based caching algorithm is proposed.

Frequently, a large-scale {\color{black} generative AI model} can be partitioned into multiple computing functions that can be efficiently managed and accessed during training, fine-tuning, and inference. FL models can be cached on edge servers to facilitate user access to instances and updates, thus addressing user privacy concerns~\cite{chen2020joint, xu2022privacy}. For example, the authors in~\cite{zheng2021knowledge} propose a knowledge cache scheme for FL in which participants can simultaneously minimize training delay and training loss according to their preference. Their insight is that there are two stimulations for caching knowledge for FL~\cite{chen2021communication}: i) training data sufficiency and ii) connectivity stability. Experimental results show that the proposed preference-driven caching policy, based on the preferences (i.e., demands or desires for global models) of participants in FL, can outperform the random policy when user preferences are intense. Therefore, preference-based {\color{black} generative AI model} caching should be extensively investigated for providing personalized and customized AIGC services at edge servers.

\subsection{Mobility Management}
\begin{figure}[t]
    \centering
    \includegraphics[width=1\linewidth]{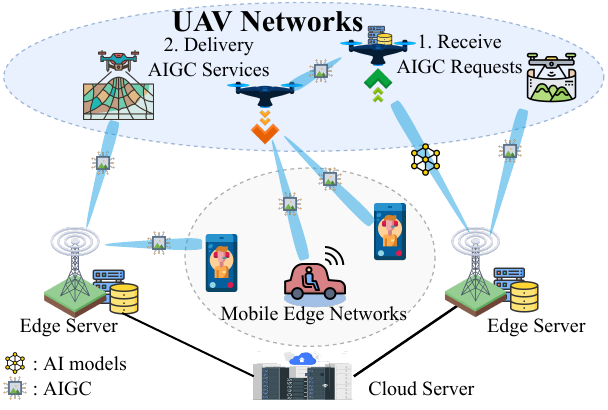}
    \caption{An overview of mobility management in mobile AIGC networks. The coverage of the mobile AIGC network will be significantly enhanced by UAV processing the user's server request and providing AIGC services.}
    \label{fig:uav}
\end{figure}
\begin{table*}[!]
\small\centering
\caption{Summary of scenarios, problems, benefits/challenges, and mathematical tools for mobility management.}
\begin{tabular}{|m{.05\textwidth}<{\centering}|m{.2\textwidth}<{\centering}|m{.2\textwidth}<{\centering}|m{.2\textwidth}<{\centering}|m{.2\textwidth}<{\centering}|}
    \hline
    % \rowcolor[HTML]{C0C0C0} 
    \textbf{Ref.} &
    \textbf{Scenarios} &
    \textbf{Performance Metrics/Problems} &
    \textbf{Benefits/Challenges} &
    \textbf{Mathematical Tools} \\ \hline
    \cite{wang2021dynamic} &
    Jointing vehicle-edge deep neural network inference &
    Latency, failure rate/CPU frequency &
    Robust AIGC service provisioning via layer-level offloading &
    Chemical reaction optimization \\ \hline
    \cite{yang2022novel} &
    Vehicular edge intelligence &
    Weighted average completion time and task acceptance ratio/Task dispatching policy &
    Provisioning AIGC service in multi-vehicle environments with motion prediction&
    Greedy algorithm \\ \hline
    \cite{sun2022meet} &
    Mobility-enhanced edge intelligence&
    Task completion ratio and model accuracy/Offloading redundancy, task assignment, beam selection &
    Sustainable AIGC service provisioning with mobility management &
    FL \\ \hline

    \cite{wang2021resource} &
    Edge intelligence-assisted IoV&
    Average delay and energy consumption/Transmission decision, task offloading decision, bandwidth, and computation resource allocation &
    Flexible network model selection for AIGC services for balancing the tradeoff adaptively &
    Quantum-inspired reinforcement learning \\ \hline
    \cite{balasubramanian2022venet} &
    Cooperative edge intelligence in IoV&
    Average delay and energy consumption/Trajectory prediction accuracy &
    Optimize AIGC service with spatial and temporal correlations of users' requests &
    Hybrid stacked autoencoder learning \\ \hline
    \cite{dong2021uavs} &
    UAVs as an intelligent service&
    Model accuracy and energy consumption/Number of local iterations &
    Provision AIGC services via a network of UAVs &
    Greedy algorithm \\ \hline
    \cite{luo2022keepedge} &
    Knowledge distillation-empowered edge intelligence&
    Accuracy and inference delay/Size of model parameters &
    Visual information-aided {\color{black} generative AI model} deployment and inference scheduling &
    Knowledge distillation \\ \hline
    
\end{tabular}%
\label{table:mobility}
\end{table*}

Mobile edge intelligence for the Internet of Vehicles and Unmanned Aerial Vehicle (UAV) networks relies on effective mobility management solutions~\cite{zhou2019exploiting, zhu2020millimeter, du2022performance, 9959884} to provide mobile AIGC services. Furthermore, UAV-based AIGC service distribution offers advantages such as ease of deployment, flexibility, and extensive coverage for enhanced edge intelligence~\cite{huynh2022uav, alsamhi2021drones}. Specifically, UAVs, with their line-of-sight communication links, can extend the reach of edge intelligence~\cite{wang2022interference}. For example, flexible UAVs equipped with AIGC servers enable users to access AIGC services with ultra-low latency and high reliability, especially when fixed-edge servers are often overloaded in hotspot areas or expensive to deploy in remote areas, as illustrated in Fig.~\ref{fig:uav}. In addition, UAV-enabled edge intelligence can be utilized to implement mobile AIGC content and service delivery.

As summarized in Table~\ref{table:mobility}, here are several KPIs for mobility management in AIGC networks:
\begin{itemize}
\item Task accomplishment ratio: The provisioning of AIGC services at mobile edge networks must consider the dynamic nature of users~\cite{yang2022multi}. As a result, services must be completed before users leave the base station. To measure the effectiveness of mobility management in AIGC networks, the task completion rate can be used.
\item Coverage enhancement: Vehicles and UAVs can serve as reconfigurable base stations to enhance the coverage of mobile AIGC networks~\cite{wang2021federated}, providing {\color{black} generative AI models} and content to users anywhere and anytime.
\end{itemize}

In vehicular networks, intelligent applications, such as AIGC-empowered navigation systems, are reshaping existing transportation systems. In~\cite{wang2021dynamic}, the authors propose a joint vehicle-edge inference framework to optimize energy consumption while reducing the execution latency of DNNs. In detail, vehicles and edge servers determine an optimal partition point for DNNs and dynamically allocate resources for DNN execution. They propose a chemical reaction optimization-based algorithm to accelerate convergence when solving the resource allocation problem. This framework offers insights for implementing mobile AIGC networks, where vehicles can collaborate with base stations to provide real-time AIGC services based on DNNs during their movement.

AIGC applications require sufficient processing and memory resources to perform extensive AIGC services~\cite{quan2018adaptive, misra2019soft, sun2020edge, wang2021green}. However, resource-constrained vehicles cannot meet the QoS requirements of the tasks. The authors in~\cite{yang2022novel} propose a distributed scheduling framework that develops a priority-driven transmission scheduling policy to address the dynamic network topologies of vehicle networks and promote vehicle edge intelligence. To meet the various QoS requirements of intelligent tasks, large-volume tasks can be partitioned and sequentially uploaded. Additionally, the impact of vehicle motion on task completion time and edge server load balancing can be independently handled by intelligent task processing requests. The effectiveness of the proposed framework is demonstrated in single-vehicle and multi-vehicle environments through simulation and deployment experiments. To facilitate smart and green vehicle networks~\cite{sun2022meet}, the real-time accuracy of AI tasks, such as {\color{black} generative AI model} inference, can be monitored through on-demand model training using infrastructure vehicles and opportunity vehicles.

The heterogeneous communication and computation requirements of AIGC services in highly dynamic, time-varying Internet of Vehicles (IoV) warrant further investigation~\cite{huang2021fedparking, xu2022secure, li2020deep, wu2022delay}. To dynamically make transmission and offload decisions, the authors in~\cite{wang2021resource} formulate a Markov decision process for time-varying environments in their joint communication and computation resource allocation strategy. Finally, they develop a quantum-inspired reinforcement learning algorithm, in which quantum mechanisms can enhance learning convergence and performance. The authors in~\cite{balasubramanian2022venet} propose a stacked autoencoder to capture spatial and temporal correlations to combine road traffic management and data network traffic management. To reduce vehicle energy consumption and learning delay, the proposed learning model can minimize the required signal traffic and prediction errors. Consequently, the accuracy of AIGC services based on autoencoder techniques can be improved through this management framework.

With UAV-enhanced edge intelligence, UAVs can serve as aerial wireless base stations, edge computing servers, and edge caching providers in mobile AIGC networks~\cite{wu2023splitmag, yao2023split}. To demonstrate the performance of UAV-enhanced edge intelligence while preserving user privacy at mobile edge networks, the authors in~\cite{dong2021uavs} use UAV-enabled FL as a use case. Moreover, the authors suggest that flexible switching between compute and cache services using adaptive scheduling UAVs is a topic for future research. Therefore, flexible AIGC service provisioning and UAV-based AIGC delivery are essential for satisfying real-time service requirements and reliable generation. In this regard, the authors in~\cite{luo2022keepedge} propose a visually assisted positioning solution for UAV-based AIGC delivery services where GPS signals are weak or unstable. Specifically, knowledge distillation is leveraged to accelerate inference speed and reduce resource consumption while ensuring satisfactory model accuracy.

\subsection{Incentive Mechanism}

\begin{table*}[htbp]
\small\centering
\caption{Summary of scenarios, problems, benefits/challenges, and mathematical tools of incentive mechanism.}

\begin{tabular}{|m{.05\textwidth}<{\centering}|m{.2\textwidth}<{\centering}|m{.2\textwidth}<{\centering}|m{.2\textwidth}<{\centering}|m{.2\textwidth}<{\centering}|}
    \hline
    % \rowcolor[HTML]{C0C0C0} 
    \textbf{Ref.} &
    \textbf{Scenarios} &
    \textbf{Problems} &
    \textbf{Benefits/Challenges} &
    \textbf{Mathematical Tools} \\ \hline
    \cite{zhan2020incentive} &
    Efficient edge learning &
    A weighted sum of training time and payment/Total payment and training time &
    Incentivize AIGC service providers with heterogeneous resources under the uncertainty of edge network bandwidth &
    Deep reinforcement learning \\ \hline
    \cite{liu2021incentive} &
    Efficient edge learning &
    Model accuracy, number of training rounds, time efficiency/The total price &
    Long-term incentive mechanism for AIGC services with long-term and short-term pricing strategies  &
    Hierarchical deep reinforcement learning \\ \hline
    \cite{deng2021fair} &
    Quality-aware FL &
    Model accuracy and loss reduction/Learning quality estimation and quality-aware incentive mechanism &
    Estimate the performance of AIGC services with privacy-preserving methods for distributing proper incentives &
    Reverse auction \\ \hline
    \cite{ren2022ai} &
    Cloud-Edge computing power trading for ubiquitous AI services &
    Profits, resource utilization, security/Computing-power unit price &
    Trustworthy edge-cloud resource trading framework for AIGC services &
    Stackelberg game and multi-agent reinforcement learning \\ \hline
\end{tabular}%
\label{table:incentive}
\end{table*}

As suitable incentive mechanisms are designed, more edge nodes participate in and contribute to the AIGC services~\cite{liu2023blockchain, wang2022infedge, zhan2021survey,du2022reconfigurable}. This increases the computational capacity of the system. In addition, the nodes are motivated to earn rewards by providing high-quality services. Thus, the overall quality of AIGC services is improved. Finally, nodes are encouraged to engage in secure operations without security concerns by recording resource transactions through the blockchain.

As listed in Table~\ref{table:incentive}, here are several KPIs for incentive mechanisms in AIGC networks:
\begin{itemize}
\item Social welfare: AIGC's social welfare is the sum of the value of AIGC's services to the participants of the current network. Higher social welfare means that more AIGC users and AIGC service providers are participating in the AIGC network and providing high-value AIGC services within the network.
\item Revenue: Providers of AIGC use a large amount of computing and energy resources to provide AIGC, which may be offset by revenue from AIGC users. The higher the revenue, the more the AIGC service provider can be motivated to improve the AIGC service to a higher quality.
\item Economic properties: In AIGC networks, AIGC providers and users should be risk-neutral, which indicates the incentive mechanisms should satisfy economic properties, e.g., individually rational, incentive compatible, and budget balance~\cite{chen2022resource}.
\end{itemize}

While edge learning has several promising benefits, the learning time for satisfactory performance and appropriate monetary incentives for resource providers are nontrivial challenges for AIGC. In~\cite{zhan2020incentive, wu2022sustainable, zhan2020learning}, where mobile devices are connected to the edge server, the authors design the incentive mechanism for efficient edge learning. Specifically, mobile devices collect data and train private models locally with computational resources based on the price of edge servers in each training round. Then, the updated models are uploaded to the edge server and aggregated to minimize the global loss function. Furthermore, the authors in~\cite{du2021resource} not only analyze the optimal pricing strategy but also use Deep Reinforcement Learning to learn the pricing strategy to obtain the optimal solution in each round in a dynamic environment and with incomplete information. In the absence of prior knowledge, the DRL agent can learn from experience to find the optimal pricing strategy that balances payment and training time. To extend~\cite{zhan2020incentive} to long-term incentive provisioning, the authors in \cite{liu2021incentive} propose a long-term incentive mechanism for edge learning frameworks. To obtain the optimal short-term and long-term pricing strategies, the hierarchical deep reinforcement learning algorithm is used in the framework to improve the model accuracy with budget constraints.

In the process of fine-tuning the AIGC edge, the incentives described above can be used to balance the time and adaptability of the fine-tuned {\color{black} generative AI model}.
In providing incentives to AIGC service providers, the quality of AIGC services also needs to be considered in the incentive mechanism. The authors in~\cite{deng2021fair} propose a quality-aware FL framework to prevent inferior model updates from degrading the global model quality. Specifically, based on an AI model trained from historical learning results, the authors estimate the learning quality of mobile devices. To motivate participants to contribute high-quality services, the authors propose a reverse auction-based incentive mechanism under the recruitment budget of edge servers, taking into account the model quality. Finally, the authors propose an algorithm for integrating the model quality into the aggregation process and for filtering non-optimal model updates to further optimize the global learning model.

Traditionally, resource utilization is inefficient, and trading mechanisms are unfair in cloud-edge computing power trading~\cite{ren2019collaborative} for AIGC services. To address this issue, the authors in~\cite{ren2022ai} develop a general trading framework for computing power grids. As illustrated in Fig.~\ref{fig:blockchain}, the authors solve the problem of the under-utilization of computing power with AI consumers in this framework. The computing-power trading problem is first formulated as a Stackelberg game and then solved with a profit-driven multi-agent reinforcement learning algorithm. Finally, a blockchain is designed for transaction security in the trading framework. In mobile AIGC networks with multiple AIGC service providers and multiple AIGC users, the Stackelberg game and its extension can still provide a valid framework for equilibrium analysis. In addition, multi-agent reinforcement learning also learns the equilibrium solution of the game by exploration and exploitation in the presence of incomplete information about the game.

\subsection{Security and Privacy}

Mobile AIGC networks leverage a collaborative computing framework on the cloud side to provide AIGC services, utilizing a large amount of heterogeneous data and computing power~\cite{tian2022comprehensive, liu2018survey, xue2023blockchain, chen2023challenges}. When mobile users are kind, AIGC can greatly enhance their creativity and efficiency. However, malicious users can also utilize AIGC for destructive purposes, posing a threat to users in mobile edge networks. For example, AI-generated text can be used by malicious users to complete phishing emails, thus compromising the security and privacy of normal users~\cite{crothers2022machine}. To ensure secure AIGC services, providers must choose trusted AIGC solutions and securely train AI models while providing secure hints and answers to AIGC service users~\cite{kang2023adversarial}.

\begin{figure}
    \centering
    \includegraphics[width=1\linewidth]{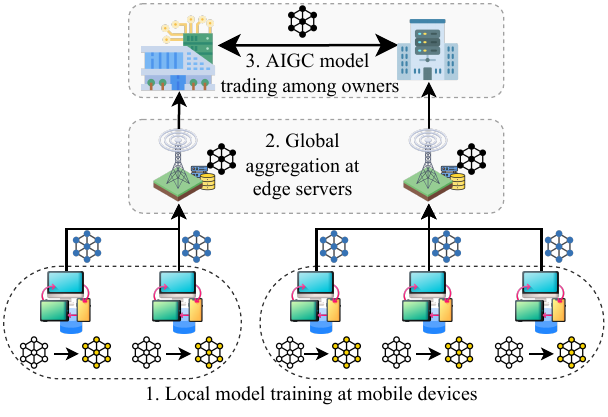}
    \caption{Federated Learning in mobile AIGC networks, including the local model training at mobile devices, global aggregation at edge servers, and cross-server model trading.}
    \label{fig:FL}
\end{figure}

\subsubsection{Privacy-preserving AIGC Service Provisioning}\label{sec:fl}
During the lifecycle of providing AIGC services, privacy information in large-scale datasets and user requests needs to be kept secure to prevent privacy breaches. In mobile AIGC networks, the generation and storage of data for {\color{black} generative AI model} training occur at edge servers and mobile devices~\cite{zhang2023split, li2021federated, wang2023privacy}. Unlike resourceful cloud data centers, edge and mobile layers have limited defense capacities against various attacks. Fortunately, several privacy-preserving distributed learning frameworks, such as FL\cite{liu2022wireless, kang2022blockchain}, have been proposed to empower privacy-preserving {\color{black} generative AI model} fine-tuning and inference at mobile AIGC networks. In preserving user privacy in AIGC networks, FL is a distributed ML approach that allows users to transmit local models instead of data during model training \cite{9785702, zhang2021optimizing, cui2021security}. Specifically, as illustrated in Fig.~\ref{fig:FL}, there are two major approaches to employing FL in AIGC networks
\begin{itemize}
    \item Secure aggregation: While FL is being learned, the mobile devices send local updates to edge servers for global aggregation. During global aggregation, authenticated encryption allows the use of secret sharing mechanisms.
    \item Differential privacy: Differential privacy can prevent FL servers from identifying the owners of a local update. Differential privacy is similar to secure aggregation in that it prevents FL servers from identifying owners of local updates.
\end{itemize} 

Therefore, in~\cite{augenstein2019generative}, the authors propose a differential private federated generative model to synthesize representative examples of private data. With guaranteed privacy, the proposed model can solve many common data problems without human intervention. Moreover, in~\cite{fan2020federated}, the authors propose an FL-based generative learning scheme to improve the efficiency and robustness of GAN models. The proposed scheme is particularly effective in the presence of varying parallelism and highly skewed data distributions. To find an inherent cluster structure in users' data and unlabeled datasets, the authors propose in~\cite{chung2022federated} the unsupervised Iterative Federated Clustering algorithm, which uses generative models to deal with the statistical heterogeneity that may exist among the participants of FL. Since the centralized FL frameworks in~\cite{fan2020federated, chung2022federated} might raise security concerns and risk single-point failure, the authors propose in~\cite{wang2021efficient} a decentralized FL framework based on a ring topology and deeply generated models. On the one hand, a method for synchronizing the ring topology can improve the communication efficiency and reliability of the system. On the other hand, generative models can solve data-related problems, such as incompleteness, low quality, insufficient quantity, and sensitivity. Finally, an InterPlanetary File System (IPFS)-based data-sharing system is developed to reduce data transmission costs and traffic congestion.

\subsubsection{Secure AIGC Service Provisioning}\label{sec:blockchain}
Given the numerous benefits of provisioning AIGC services in mobile and edge layers, multi-tier collaboration among cloud servers, edge servers, and mobile devices enables ubiquitous AIGC service provision by heterogeneous stakeholders~\cite{shen2022edgematrix, gai2020blockchain, lin2023unified, lin2022blockchain}. A trustworthy collaborative AIGC service provisioning framework must be established to provide reliable and secure AIGC services. Compared to central cloud AIGC providers, mobile and edge AIGC providers can customize AIGC services by collaborating with many user nodes while distributing data to different devices~\cite{9711561}. Therefore, a secure access control mechanism is required for multi-party content streaming to ensure privacy and security. However, the security of AIGC transmission cannot be ensured due to various attacks on mobile AIGC networks~\cite{huang2022blockchain}. Fortunately, blockchain~\cite{huang2022blockchain, shen2019privacy, shen2022secure, shen2020blockchain}, based on distributed ledger technologies, can be utilized to explore a secure and reliable AIGC service provisioning framework and record resource and service transactions to encourage data sharing among nodes, forming a trustworthy and active mobile AIGC ecosystem~\cite{xu2022quantum}. As illustrated in Fig.~\ref{fig:blockchain}, there are several benefits that blockchain brings to mobile AIGC networks\cite{wang2022integrating}:
\begin{itemize}
\item Computing and Communication Management: Blockchain enables heterogeneous computing and communication resources to be managed securely, adaptively, and efficiently in mobile AIGC networks~\cite{9969941}.
\item Data Administration: By recording AIGC resource and service transactions in blockchain with smart contracts, data administration in mobile AIGC networks is made profitable, collaborative, and credible.
\item Optimization: During optimization in AIGC services, the blockchain always provides available, complete, and secure historical data for input to optimization algorithms.
\end{itemize}
For instance, the authors in~\cite{dirgantoro2020generative} propose an edge intelligence framework based on deep generative models and blockchain. To overcome the accuracy issue of the limited dataset, GAN is leveraged in the framework to synthesize training samples. Then, the output of this framework is confirmed and incentivized by smart contracts based on the proof-of-work consensus algorithm. Furthermore, the multimodal outputs of AIGC can be minted as NFTs and then recorded on the blockchain. The authors in~\cite{tann2022predicting} develop a conditional generative model to synthesize new digital asset collections based on the historical transaction results of previous collections. First, the context information of NFT collections is extracted based on unsupervised learning. Based on the historical context, the newly minted collections are generated based on future token transactions. The proposed generative model can synthesize new NFT collections based on the contexts, i.e., the extracted features of previous transactions.
\begin{figure}
    \centering
    \includegraphics[width=1\linewidth]{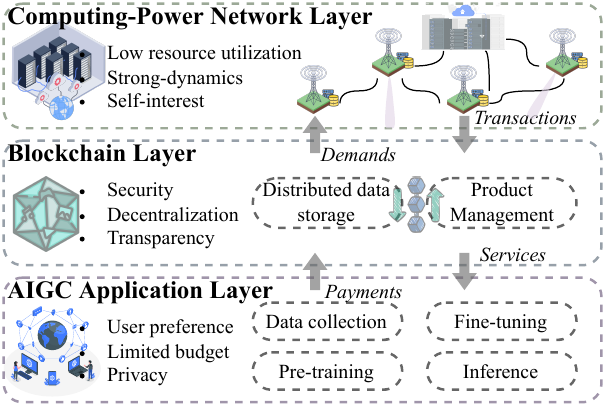}
    \caption{Blockchain in mobile AIGC networks~\cite{ren2022ai}, including the AIGC application layer, blockchain layer, and computing-power network layers, for provisioning AIGC services.}
    \label{fig:blockchain}
\end{figure}
\subsection{Lessons Learned}

\subsubsection{Multi-Objective Quality of AIGC Services}
In mobile AIGC networks, the quality of AIGC services is determined by several factors, including model accuracy, service latency, energy consumption, and revenue. Consequently, AIGC service providers must optimally allocate edge resources to satisfy users' multidimensional quality requirements for AIGC services~\cite{zhao2022edgeadaptor}. Moreover, the migration of AIGC tasks and computations can enhance the reliability and efficiency of AIGC services. Notably, dynamic network conditions in the edge network necessitate users to make online decisions to achieve load balancing and efficient use of computing resources. {\color {black}A variety of methodologies are proposed, enhancing the multi-objective quality of AIGC services within mobile edge networks~\cite{xu2023joint}. The techniques encompass multi-objective optimization among QoS, QoE, latency, and resource consumption. The primary objective of designing these strategies is to optimize key parameters such as accuracy, latency, resource consumption, and user satisfaction. The benefits including heightened performance and superior user experience, are attained, albeit at the potential cost of an increase in complexity, resource consumption, and potential privacy issues.} Attaining high-quality AIGC services requires proper considerations and practices to address the challenges discussed above, meet the quality requirements of multiple objectives, and improve user satisfaction and service quality.

\subsubsection{Edge Caching for Efficient Delivery of AIGC Services}
Edge caching plays a pivotal role in the efficient delivery of AIGC services in mobile AIGC networks. Tackling the challenges of constrained-memory edge servers, model-missing costs, and functionally equivalent models is essential for optimizing caching policies. Developing model-aware caching approaches, investigating preference-driven caching policies, and implementing principled cache designs to reduce latency and energy consumption are promising directions for enhancing the performance of mobile AIGC networks. {\color{black} In the quest for the efficient delivery of AIGC services via edge caching in mobile edge networks, the need for well-designed edge caching algorithms is emphasized~\cite{yao2022loading}. The benefits associated with these algorithms include enhanced efficiency, decreased latency, and improved dependability. Conversely, the challenges that may arise from these strategies include escalated complexity, heightened costs, and potential privacy concerns.} As AI services continue to evolve, further research in caching strategies is crucial for providing effective, personalized, and low-latency AIGC services for mobile users.

\subsubsection{Preference-aware AIGC Service Provisioning}
Offering AIGC services based on user preferences not only improves user satisfaction but also reduces service latency and resource consumption in mobile edge networks. To implement preference-based AIGC service delivery, AIGC service providers must first collect historical user data and analyze it thoroughly. In providing AIGC services, the service provider makes personalized recommendations and adjusts its strategy according to user feedback. {\color{black} The exploration of preference-aware AIGC service provisioning is conducted considering several techniques, which include collaborative filtering, DRL, context awareness, user profiling, and multi-objective optimization.} Although user preferences play a significant role in AIGC service provision, it is essential to use and manage this information properly to protect user privacy.
% \begin{table*}[t]
% \small\centering
% \caption{A summary of future directions in mobile AIGC networks.}
% \resizebox{\textwidth}{!}{%
% \begin{tabular}{|c|c|c|}
% \hline
% \textbf{Future Directions} & \textbf{Problems} & \textbf{Potential Techniques} \\ \hline
% \multirow{3}{*}{Networking and Computing Issues} & Decentralized Mobile AIGC Networks & Blockchain \\ \cline{2-3} 
%  & Sustainability in Mobile AIGC Networks & Green computing and communication \\ \cline{2-3} 
%  & {\color{black}Wireless Communication in Mobile AIGC Networks} & {\color{black}Robustness, reliability, latency}\\ \hline
% \multirow{2}{*}{Machine Learning Issues} & {\color{black} Generative AI Model} Compression & Pruning, quantization, and knowledge distillation \\ \cline{2-3} 
%  & Privacy-preserving AIGC Services & Differential privacy, secure multi-party computation, and homomorphic encryption \\ \cline{2-3} 
%  & {\color{black}AI-generated Network Design} & {\color{black}Design, analysis, control, monitoring, and traffic prediction} \\ \hline
% \multirow{3}{*}{Practical Implementation Issues} & Integrating AIGC and Digital Twins & Monitoring, analyzing, and predictions \\ \cline{2-3} 
%  & Immersive Streaming & AR and VR
%  \\ \cline{2-3} 
%  & {\color{black}Alignment} & {\color{black}Digital humans and avatars}\\ \hline
% \end{tabular}%
% }\label{tab:future}
% \end{table*}
\subsubsection{Life-cycle Incentive Mechanism throughout AIGC Services}

In mobile AIGC networks, the entire life cycle of AIGC services necessitates appropriate incentives for participants. A single AIGC service provider cannot provide AIGC services alone. Throughout the data collection, pre-training, fine-tuning, and inference of AIGC services, stakeholders with heterogeneous resources require reasonable incentives and must share the benefits according to their contributions. Conversely, from the users' perspective, evaluation mechanisms must be introduced. For instance, users can assess the reputation of AIGC service providers based on their transaction history to promote service optimization and improvement. Ultimately, the provisioning and transmission logs of AIGC services can also be recorded in a tamper-proof distributed ledger. {\color{black} Incentive strategies for participants in the life cycle of AIGC services in mobile edge networks are also examined. The use of smart contracts, distributed ledger technology, evaluation mechanisms, and incentive design is proposed as a means to strengthen collaboration and enhance the overall quality of AIGC services~\cite{ren2022ai}. These methodologies introduce automation, transparency, and improved reputation, which are seen as distinct advantages.}

\subsubsection{Blockchain-based System Management of Mobile AIGC Networks}
Furthermore, mobile AIGC networks connect heterogeneous user devices to edge servers and cloud data centers. This uncontrolled demand for content generation introduces uncertainty and security risks into the system.  Therefore, secure management and auditing methods are required to manage devices in edge environments, such as dynamically accessing, departing, and identifying IoT devices.  In the traditional centralized management architecture, the risk of central node failure is unavoidable. Thus, a secure and reliable monitoring and equipment auditing system should be developed. {\color{black} Lastly, we analyze a suite of techniques aimed at improving blockchain-based system management of mobile AIGC networks. Such techniques include blockchain-based data administration, secure management and auditing methods, collaborative infrastructure, decentralized management architecture, and blockchain-based optimization~\cite{liu2023blockchain}. }

\section{Future Research Directions and Open Issues}~\label{sec:future} 

In this section, we discuss future research directions and open issues from the perspectives of networking and computing, ML, and practical implementation.
% \usepackage{multirow}
% \usepackage{graphicx}
% \cite{wu2020investigating}
% \subsection{Quantum-inspired AIGC}

% \subsection{Generative Modeling for Networking and Communication System Designs}
% 
\subsection{Networking and Computing Issues}

\subsubsection{Decentralized Mobile AIGC Networks}
With the advancement of blockchain technologies \cite{li2020blockchain}, decentralized mobile AIGC networks can be realized based on distributed data storage, the convergence of computing and networking, and proof-of-ownership of data~\cite{xu2022quantum}. Such a decentralized network structure, enabled by digital identities and smart contracts, can protect AIGC users' privacy and data security. Furthermore, based on blockchain technologies, mobile AIGC networks can achieve decentralized management of the entire lifecycle of AIGC services. Therefore, future research should investigate specific consensus mechanisms~\cite{li2020blockchain, liu2021proof}, off-chain storage systems, and token structures for the deployment of decentralized mobile AIGC networks~\cite{du2023exploring}.

\subsubsection{Sustainability in Mobile AIGC Networks}

In mobile AIGC networks, the pre-training, fine-tuning, and inference of generative AI models typically consume a substantial amount of computing and networking resources~\cite{mao2017survey, zhang2023sustainable}. Hence, future research can focus on the green operations of mobile AIGC networks that provide AIGC services with minimal energy consumption and carbon emissions. To this end, effective algorithms and frameworks should be developed to operate mobile AIGC networks under dynamic service configurations, operating modes of edge nodes, and communication links. Moreover, intelligent resource management and scheduling techniques can also be proposed to balance the tradeoff between service quality and resource consumption~\cite{ma2023reliability}.

\subsubsection{Wireless Communications in Mobile AIGC Networks}
{\color{black}The influence of wireless communications on AIGC services is a critical area for future research. A key aspect to investigate is the robustness of AIGC services to the challenges posed by wireless communications~\cite{du2023ai}. This includes understanding how factors such as transmit power, fading, and device mobility within an edge network can affect the performance of distributed diffusion model-based AIGC computing~\cite{wang2021dynamic}. Initial research in this area, such as the study in~\cite{du2023exploring}, has shown that despite the increase in bit error probability, distributed AIGC computing exhibits relatively high robustness. Further exploration of this robustness, as well as the development of strategies to enhance it, could significantly improve the performance and reliability of AIGC services in wireless networks. This can involve, for example, the development of adaptive physical layer transmission strategies~\cite{wang2022uplink} that take into account the current state of the wireless channel or the design of error correction mechanisms that can recover from bit errors introduced during wireless transmission~\cite{wang2023uplink,zhu2021multi}. 
In addition, the use of AI-generated optimization solutions, particularly diffusion models, to overcome the challenges posed by the wireless environment and generate optimal solutions for network design is a promising avenue for future research. This can involve the development of AI-generated incentive mechanisms to promote semantic information exchange among users, as demonstrated by the authors~\cite{du2023ai}. Such mechanisms can help to create an optimal contract that adheres to the utility threshold constraints of the semantic information provider while maximizing the utility of the semantic information recipient.}

High-quality data resources are also critical for the sustainability of mobile AIGC networks~\cite{du2023generative}. The performance of generative models depends not only on effective network architectures but also on the quality of training datasets~\cite{shi2023task}. However, as AIGC becomes pervasive, training datasets are gradually replaced by synthesized data that might be irrelevant to real data. Therefore, improving the quality and reliability of data in mobile AIGC networks, such as through multimodal data fusion and incremental learning technology, can further enhance the accuracy and performance of the models.

\subsection{Machine Learning Issues}

\subsubsection{{\color{black} Generative AI Model} Compression}

As {\color{black} generative AI models} become increasingly complex, model compression techniques are becoming more important to reduce service latency and resource consumption in provisioning AIGC services~\cite{cheng2017survey}. Fortunately, several techniques have been developed for {\color{black} generative AI model} compressions, such as pruning, quantization, and knowledge distillation. First, pruning involves removing unimportant weights from the model, while quantization reduces the precision of the weights~\cite{li2022compact}. Then, knowledge distillation involves training a smaller model to mimic the larger model's behavior. Future research on {\color{black} generative AI model} compression might continue to focus on developing and refining these techniques to improve their efficiency and effectiveness for deploying {\color{black} generative AI models} in edge nodes and mobile devices. It is necessary to consider the limited resources of such devices and develop specialized compression techniques that can balance model size and accuracy.

{\color{black} \subsubsection{AI-generated Network Design}
Generative AI models have various potential applications in mobile networks, including design, analysis, control, monitoring, and traffic prediction~\cite{du2023beyond,huang2023ai}. They can be utilized to create efficient network architectures, understand network behavior, predict network loads, develop network control algorithms, detect anomalies, and predict future network traffic patterns and demands~\cite{du2023beyond}. Future research directions in machine learning for mobile AIGC networks can focus on improving the efficiency and effectiveness of existing applications, exploring new applications and use cases, and addressing the challenges posed by the unique characteristics of mobile networks, such as mobility, limited resources, and privacy concerns.}

\subsubsection{Privacy-preserving AIGC Services} 
To provide privacy-preserving AIGC services, it is necessary to consider privacy computing techniques in both {\color{black} generative AI model} training and inference~\cite{lin2023blockchain,lim2020federated}. Techniques such as differential privacy, secure multi-party computation, and homomorphic encryption can be used to protect sensitive data and prevent unauthorized access. Differential privacy involves adding noise to the data to protect individual privacy, while secure multi-party computation allows multiple parties to compute a function without revealing their inputs to one another. Homomorphic encryption enables computations to be performed on encrypted data without decryption. To successfully deploy {\color{black} generative AI models} in edge nodes and mobile devices, the limited resources of such devices should be considered and specialized techniques that can balance privacy and performance should be developed~\cite{du2023generativemul}. Additionally, concerns such as data ownership and user privacy leakage should be taken into account.

% \subsection{AIGC Algorithms}

% \subsubsection{Fast Deployment via Prompt Fune-tuning}

% \subsubsection{Parallel Execution of {\color{black} generative AI models}}

\subsection{Practical Implementation Issues}

\subsubsection{Integrating AIGC and Digital Twins}
Digital twins enable the maintenance of representations to monitor, analyze, and predict the status of physical entities~\cite{8424832}. On one hand, the integration of AIGC and digital twin technologies has the potential to significantly improve the performance of mobile AIGC networks. By creating virtual representations of physical mobile AIGC networks, service latency, and quality can be optimized through the analysis of historical data and online predictions. Furthermore, AIGC can also enhance digital twin applications by reducing the time required for designers to create simulation entities. However, several issues need to be considered during the integration of AIGC and DTs, such as efficient and secure synchronization.

% Digital twins can maintenance of representations to monitor, analyze, and predict the status of physical entities. Since the digital representation contains the lifecycle data of the physical entities, the digital twin provides a large number of prompts for generating AIGC requests that can be used to generate new data. Therefore, users in mobile AIGC networks can use leverage digital twins to send service requests to AIGC service providers in advance to reduce waiting time. On the other hand, AIGC can also provide a large amount of new synthetic content for the simulation platform in the digital twin. In this way, the cost of producing content for the simulation platform can be significantly reduced. Therefore, the AIGC-supported simulation platform can achieve better simulation performance with lower resource consumption.

\subsubsection{Immersive Streaming}
AIGC can create immersive streaming content, such as AR and VR, that can transport viewers to virtual worlds~\cite{clemm2020toward}, which can be used in various applications such as education, entertainment, and social media. Immersive streaming can enhance the AIGC delivery process by providing a platform for viewers to interact with the generated content in real-time. However, combining AIGC and immersive streaming raises some concerns. Future research should focus on addressing the potential for biased content generation by the AIGC algorithms and the high bandwidth requirements of immersive streaming, which can cause latency issues, resulting in the degradation of the viewer's experience.

{\color{black}\subsubsection{Alignment}

In human-oriented applications that involve digital humans and avatars, the alignment of generative AI models~\cite{ouyang2022training, chen2023multi, chen2023multiple} in mobile AIGC networks should be well-investigated for safety and ethnicity. There are several potential research directions for AI alignment, such as personalized AI alignment, ethical guidelines for AI-generated content, trust and transparency, emotional alignment, cultural alignment, and robustness to adversarial attacks. By focusing on these areas, future AI alignment research in mobile AIGC networks can help maintain a user-centric, respectful, and ethically responsible approach for mobile AIGC networks and their applications.}

\section{Conclusions}\label{sec:conclusion}
In this paper, we have focused on the deployment of mobile AIGC networks,  {\color{black}which serve} generative AI models, services, and applications at mobile edge networks. We have discussed the background and fundamentals of generative models and the lifecycle of AIGC services at mobile AIGC networks. We have also explored AIGC-driven creative applications and use cases for mobile AIGC networks, as well as the implementation, security, and privacy challenges of deploying mobile AIGC networks. Finally, we have highlighted some future research directions and open issues for the full realization of mobile AIGC networks.

\bibliographystyle{IEEEtran}
\bibliography{survey}

\begin{thebibliography}{100}
\providecommand{\url}[1]{#1}
\csname url@samestyle\endcsname
\providecommand{\newblock}{\relax}
\providecommand{\bibinfo}[2]{#2}
\providecommand{\BIBentrySTDinterwordspacing}{\spaceskip=0pt\relax}
\providecommand{\BIBentryALTinterwordstretchfactor}{4}
\providecommand{\BIBentryALTinterwordspacing}{\spaceskip=\fontdimen2\font plus
\BIBentryALTinterwordstretchfactor\fontdimen3\font minus
  \fontdimen4\font\relax}
\providecommand{\BIBforeignlanguage}[2]{{%
\expandafter\ifx\csname l@#1\endcsname\relax
\typeout{** WARNING: IEEEtran.bst: No hyphenation pattern has been}%
\typeout{** loaded for the language `#1'. Using the pattern for}%
\typeout{** the default language instead.}%
\else
\language=\csname l@#1\endcsname
\fi
#2}}
\providecommand{\BIBdecl}{\relax}
\BIBdecl

\bibitem{du2023beyond}
H.~Du, R.~Zhang, Y.~Liu, J.~Wang, Y.~Lin, Z.~Li, D.~Niyato, J.~Kang, Z.~Xiong,
  S.~Cui \emph{et~al.}, ``Beyond deep reinforcement learning: {A} tutorial on
  generative diffusion models in network optimization,'' \emph{{\rm{arXiv
  preprint arXiv:2308.05384}}}, 2023.

\bibitem{cetinic2022understanding}
E.~Cetinic and J.~She, ``Understanding and creating art with {AI}: Review and
  outlook,'' \emph{ACM Transactions on Multimedia Computing, Communications,
  and Applications}, vol.~18, no.~2, pp. 1--22, Feb. 2022.

\bibitem{lee2021creators}
L.-H. Lee, Z.~Lin, R.~Hu, Z.~Gong, A.~Kumar, T.~Li, S.~Li, and P.~Hui, ``When
  creators meet the metaverse: A survey on computational arts,''
  \emph{{\rm{arXiv preprint arXiv:2111.13486}}}, Apr. 2021.

\bibitem{wu2022ai}
W.~Wu, C.~Zhou, M.~Li, H.~Wu, H.~Zhou, N.~Zhang, X.~S. Shen, and W.~Zhuang,
  ``{AI}-native network slicing for {6G} networks,'' \emph{IEEE Wireless
  Communications}, vol.~29, no.~1, pp. 96--103, Apr. 2022.

\bibitem{wang2023survey}
Y.~Wang, Y.~Pan, M.~Yan, Z.~Su, and T.~H. Luan, ``A survey on {ChatGPT}:
  {AI}-generated contents, challenges, and solutions,'' \emph{{\rm{arXiv
  preprint arXiv:2305.18339}}}, Feb. 2023.

\bibitem{bond2021deep}
S.~Bond-Taylor, A.~Leach, Y.~Long, and C.~G. Willcocks, ``Deep generative
  modelling: A comparative review of {VAE}s, {GAN}s, normalizing flows,
  energy-based and autoregressive models,'' \emph{IEEE Transactions on Pattern
  Analysis and Machine Intelligence}, vol.~44, no.~11, pp. 7327--7347, Sep.
  2021.

\bibitem{radford2021learning}
A.~Radford, J.~W. Kim, C.~Hallacy, A.~Ramesh, G.~Goh, S.~Agarwal, G.~Sastry,
  A.~Askell, P.~Mishkin, J.~Clark \emph{et~al.}, ``Learning transferable visual
  models from natural language supervision,'' in \emph{Proceedings of the
  International Conference on Machine Learning}, Virtual Conference, Jul. 2021,
  pp. 8748--8763.

\bibitem{ramesh2021zero}
A.~Ramesh, M.~Pavlov, G.~Goh, S.~Gray, C.~Voss, A.~Radford, M.~Chen, and
  I.~Sutskever, ``Zero-shot text-to-image generation,'' in \emph{Proceedings of
  the International Conference on Machine Learning}, Virtual Conference, Jul.
  2021, pp. 8821--8831.

\bibitem{ramesh2022hierarchical}
A.~Ramesh, P.~Dhariwal, A.~Nichol, C.~Chu, and M.~Chen, ``Hierarchical
  text-conditional image generation with {CLIP} latents,'' \emph{{\rm{arXiv
  preprint arXiv:2204.06125}}}, Apr. 2022.

\bibitem{GenerativeAI2022}
S.~Huang, P.~Grady, and GPT-3, ``Generative{AI}: A creative new world,''
  ``Accessed Feb. 4, 2023", [Online]. Available:
  \url{https://www.sequoiacap.com/article/generative-{AI}-a-creative-new-world/}.

\bibitem{crothers2022machine}
E.~Crothers, N.~Japkowicz, and H.~Viktor, ``Machine generated text: A
  comprehensive survey of threat models and detection methods,''
  \emph{{\rm{arXiv preprint arXiv:2210.07321}}}, Oct. 2022.

\bibitem{ChatGPT2022}
O.~AI, ``Chatgpt: Optimizing language models for dialogue,'' ``Accessed Feb. 4,
  2023", [Online]. Available: \url{https://openai.com/blog/chatgpt/}.

\bibitem{ho2022imagen}
J.~Ho, W.~Chan, C.~Saharia, J.~Whang, R.~Gao, A.~Gritsenko, D.~P. Kingma,
  B.~Poole, M.~Norouzi, D.~J. Fleet \emph{et~al.}, ``Imagen video: High
  definition video generation with diffusion models,'' \emph{{\rm{arXiv
  preprint arXiv:2210.02303}}}, Oct. 2022.

\bibitem{kim2023bargaining}
M.~Kim, A.~DeRieux, and W.~Saad, ``A bargaining game for personalized, energy
  efficient split learning over wireless networks,'' in \emph{2023 IEEE
  Wireless Communications and Networking Conference (WCNC)}, Glasgow, United
  Kingdom, May 2023, pp. 1--6.

\bibitem{wang2020convergence}
X.~Wang, Y.~Han, V.~C. Leung, D.~Niyato, X.~Yan, and X.~Chen, ``Convergence of
  edge computing and deep learning: A comprehensive survey,'' \emph{IEEE
  Communications Surveys \& Tutorials}, vol.~22, no.~2, pp. 869--904, Jan.
  2020.

\bibitem{westerlund2019emergence}
M.~Westerlund, ``The emergence of deepfake technology: A review,''
  \emph{Technology Innovation Management Review}, vol.~9, no.~11, pp. 40--53,
  Nov. 2019.

\bibitem{yuan2023computing}
X.~Yuan, L.~Pu, L.~Jiao, X.~Wang, M.~Yang, and J.~Xu, ``When computing power
  network meets distributed machine learning: An efficient federated split
  learning framework,'' \emph{{\rm{arXiv preprint arXiv:2305.12979}}}, Mar.
  2023.

\bibitem{zhang2019mobile}
J.~Zhang and K.~B. Letaief, ``Mobile edge intelligence and computing for the
  {Internet} of vehicles,'' \emph{Proceedings of the IEEE}, vol. 108, no.~2,
  pp. 246--261, Jun. 2019.

\bibitem{lim2020federated}
W.~Y.~B. Lim, N.~C. Luong, D.~T. Hoang, Y.~Jiao, Y.-C. Liang, Q.~Yang,
  D.~Niyato, and C.~Miao, ``Federated learning in mobile edge networks: A
  comprehensive survey,'' \emph{IEEE Communications Surveys \& Tutorials},
  vol.~22, no.~3, pp. 2031--2063, Apr. 2020.

\bibitem{makhmutov2020survey}
M.~Makhmutov, S.~Varouqa, and J.~A. Brow, ``Survey on copyright laws about
  music generated by artificial intelligence,'' in \emph{Proceedings of the
  IEEE Symposium Series on Computational Intelligence}, ACT, Australia, Jan.
  2020, pp. 3003--3009.

\bibitem{chen2021distributed}
M.~Chen, D.~G{\"u}nd{\"u}z, K.~Huang, W.~Saad, M.~Bennis, A.~V. Feljan, and
  H.~V. Poor, ``Distributed learning in wireless networks: Recent progress and
  future challenges,'' \emph{IEEE Journal on Selected Areas in Communications},
  vol.~39, no.~12, pp. 3579--3605, Oct. 2021.

\bibitem{zhan2021multimodal}
F.~Zhan, Y.~Yu, R.~Wu, J.~Zhang, and S.~Lu, ``Multimodal image synthesis and
  editing: A survey,'' \emph{{\rm{arXiv preprint arXiv:2112.13592}}}, Dec.
  2021.

\bibitem{shen2021holistic}
X.~Shen, J.~Gao, W.~Wu, M.~Li, C.~Zhou, and W.~Zhuang, ``Holistic network
  virtualization and pervasive network intelligence for {6G},'' \emph{IEEE
  Communications Surveys \& Tutorials}, vol.~24, no.~1, pp. 1--30, Dec. 2021.

\bibitem{letaief2021edge}
K.~B. Letaief, Y.~Shi, J.~Lu, and J.~Lu, ``Edge artificial intelligence for
  {6G}: Vision, enabling technologies, and applications,'' \emph{IEEE Journal
  on Selected Areas in Communications}, vol.~40, no.~1, pp. 5--36, Nov. 2021.

\bibitem{cao2022survey}
H.~Cao, C.~Tan, Z.~Gao, G.~Chen, P.-A. Heng, and S.~Z. Li, ``A survey on
  generative diffusion model,'' \emph{{\rm{arXiv preprint arXiv:2209.02646}}},
  Sep. 2022.

\bibitem{wang2022integrating}
X.~Wang, X.~Ren, C.~Qiu, Z.~Xiong, H.~Yao, and V.~C. Leung, ``Integrating edge
  intelligence and blockchain: What, why, and how,'' \emph{IEEE Communications
  Surveys \& Tutorials}, vol.~24, no.~4, pp. 2193--2229, Jul. 2022.

\bibitem{xu2022full}
M.~Xu, W.~C. Ng, W.~Y.~B. Lim, J.~Kang, Z.~Xiong, D.~Niyato, Q.~Yang, X.~S.
  Shen, and C.~Miao, ``A full dive into realizing the edge-enabled metaverse:
  Visions, enabling technologies, and challenges,'' \emph{IEEE Communications
  Surveys \& Tutorials}, vol.~25, no.~1, pp. 656--700, Nov. 2023.

\bibitem{nyatsanga2023comprehensive}
S.~Nyatsanga, T.~Kucherenko, C.~Ahuja, G.~E. Henter, and M.~Neff, ``A
  comprehensive review of data-driven co-speech gesture generation,''
  \emph{{\rm{arXiv preprint arXiv:2301.05339}}}, Jan. 2023.

\bibitem{zhou2019edge}
Z.~Zhou, X.~Chen, E.~Li, L.~Zeng, K.~Luo, and J.~Zhang, ``Edge intelligence:
  Paving the last mile of artificial intelligence with edge computing,''
  \emph{Proceedings of the IEEE}, vol. 107, no.~8, pp. 1738--1762, Jun. 2019.

\bibitem{mao2017survey}
Y.~Mao, C.~You, J.~Zhang, K.~Huang, and K.~B. Letaief, ``A survey on mobile
  edge computing: The communication perspective,'' \emph{IEEE Communications
  Surveys \& Tutorials}, vol.~19, no.~4, pp. 2322--2358, Aug. 2017.

\bibitem{chen2020computation}
Z.~Chen, J.~Hu, X.~Chen, J.~Hu, X.~Zheng, and G.~Min, ``Computation offloading
  and task scheduling for {DNN}-based applications in cloud-edge computing,''
  \emph{IEEE Access}, vol.~8, pp. 115\,537--115\,547, Jun. 2020.

\bibitem{kang2017neurosurgeon}
Y.~Kang, J.~Hauswald, C.~Gao, A.~Rovinski, T.~Mudge, J.~Mars, and L.~Tang,
  ``Neurosurgeon: Collaborative intelligence between the cloud and mobile
  edge,'' \emph{ACM SIGARCH Computer Architecture News}, vol.~45, no.~1, pp.
  615--629, Mar. 2017.

\bibitem{zhang2022intelligent}
H.~Zhang and B.~Di, ``Intelligent omni-surfaces: Simultaneous refraction and
  reflection for full-dimensional wireless communications,'' \emph{IEEE
  Communications Surveys \& Tutorials}, vol.~24, no.~4, pp. 1997--2028, Aug.
  2022.

\bibitem{huang2020location}
D.~Huang, P.~Chen, R.~Zeng, Q.~Du, M.~Tan, and C.~{GAN}, ``Location-aware graph
  convolutional networks for video question answering,'' in \emph{Proceedings
  of the AAAI Conference on Artificial Intelligence}, vol.~34, no.~07, New
  York, New York, Feb. 2020, pp. 11\,021--11\,028.

\bibitem{zhang2022toward}
H.~Zhang, B.~Di, K.~Bian, Z.~Han, H.~V. Poor, and L.~Song, ``Toward ubiquitous
  sensing and localization with reconfigurable intelligent surfaces,''
  \emph{Proceedings of the IEEE}, vol. 110, no.~9, pp. 1401--1422, May 2022.

\bibitem{wang2014cache}
X.~Wang, M.~Chen, T.~Taleb, A.~Ksentini, and V.~C. Leung, ``Cache in the air:
  Exploiting content caching and delivery techniques for 5{G} systems,''
  \emph{IEEE Communications Magazine}, vol.~52, no.~2, pp. 131--139, Feb. 2014.

\bibitem{huang2022fine}
S.~Huang, H.~Zhang, X.~Wang, M.~Chen, J.~Li, and V.~C. Leung, ``Fine-grained
  spatio-temporal distribution prediction of mobile content delivery in 5{G}
  ultra-dense networks,'' \emph{IEEE Transactions on Mobile Computing}, pp.
  1--14, Dec. 2022.

\bibitem{liew2023economics}
Z.~Q. Liew, H.~Du, W.~Y.~B. Lim, Z.~Xiong, D.~Niyato, C.~Miao, and D.~I. Kim,
  ``Economics of semantic communication system: An auction approach,''
  \emph{IEEE Transactions on Vehicular Technology}, 2023.

\bibitem{gozalo2023chatgpt}
R.~Gozalo-Brizuela and E.~C. Garrido-Merchan, ``{ChatGPT} is not all you need.
  a state of the art review of large generative {AI} models,'' \emph{{\rm{arXiv
  preprint arXiv:2301.04655}}}, Jan. 2023.

\bibitem{lin2023split}
Z.~Lin, G.~Qu, X.~Chen, and K.~Huang, ``Split learning in {6G} edge networks,''
  \emph{{\rm{arXiv preprint arXiv:2306.12194}}}, Jun. 2023.

\bibitem{du2022attention}
H.~Du, J.~Liu, D.~Niyato, J.~Kang, Z.~Xiong, J.~Zhang, and D.~I. Kim,
  ``Attention-aware resource allocation and qoe analysis for metaverse xurllc
  services,'' \emph{IEEE Journal on Selected Areas in Communications}, to
  appear, 2023.

\bibitem{yang2022fusing}
Q.~Yang, Y.~Zhao, H.~Huang, Z.~Xiong, J.~Kang, and Z.~Zheng, ``Fusing
  blockchain and {AI} with metaverse: A survey,'' \emph{IEEE Open Journal of
  the Computer Society}, vol.~3, pp. 122--136, Jul. 2022.

\bibitem{ren2023building}
X.~Ren, M.~Xu, D.~Niyato, J.~Kang, Z.~Xiong, C.~Qiu, and X.~Wang, ``Building
  resilient web 3.0 with quantum information technologies and blockchain: An
  ambilateral view,'' \emph{{\rm{arXiv preprint arXiv:2303.13050}}}, Mar. 2023.

\bibitem{tiago2019youtube}
F.~Tiago, F.~Moreira, and T.~Borges-Tiago, ``Youtube videos: A destination
  marketing outlook,'' in \emph{Proceedings of the Strategic Innovative
  Marketing and Tourism}, Northern Aegean, Greece, May 2019, pp. 877--884.

\bibitem{krumm2008user}
J.~Krumm, N.~Davies, and C.~Narayanaswami, ``User-generated content,''
  \emph{IEEE Pervasive Computing}, vol.~7, no.~4, pp. 10--11, Oct. 2008.

\bibitem{croitoru2022diffusion}
F.-A. Croitoru, V.~Hondru, R.~T. Ionescu, and M.~Shah, ``Diffusion models in
  vision: A survey,'' \emph{{\rm{arXiv preprint arXiv:2209.04747}}}, Sep. 2022.

\bibitem{oppenlaender2022prompt}
J.~Oppenlaender, ``Prompt engineering for text-based generative art,''
  \emph{{\rm{arXiv preprint arXiv:2204.13988}}}, Apr. 2022.

\bibitem{marcus2022very}
G.~Marcus, E.~Davis, and S.~Aaronson, ``A very preliminary analysis of {DALL-E}
  2,'' \emph{{\rm{arXiv preprint arXiv:2204.13807}}}, Apr. 2022.

\bibitem{chi2021audio}
P.-H. Chi, P.-H. Chung, T.-H. Wu, C.-C. Hsieh, Y.-H. Chen, S.-W. Li, and H.-y.
  Lee, ``Audio {ALBERT}: A lite {BERT} for self-supervised learning of audio
  representation,'' in \emph{Proceedings of the IEEE Spoken Language Technology
  Workshop}, Shenzhen, China, Jan. 2021, pp. 344--350.

\bibitem{chui2018notes}
M.~Chui, J.~Manyika, M.~Miremadi, N.~Henke, R.~Chung, P.~Nel, and S.~Malhotra,
  ``Notes from the {AI} frontier: Insights from hundreds of use cases,''
  \emph{McKinsey Global Institute}, vol.~2, 2018.

\bibitem{brown2020language}
T.~Brown, B.~Mann, N.~Ryder, M.~Subbiah, J.~D. Kaplan, P.~Dhariwal,
  A.~Neelakantan, P.~Shyam, G.~Sastry, A.~Askell \emph{et~al.}, ``Language
  models are few-shot learners,'' \emph{Advances In Neural Information
  Processing Systems}, vol.~33, pp. 1877--1901, Dec. 2020.

\bibitem{ouyang2022training}
L.~Ouyang, J.~Wu, X.~Jiang, D.~Almeida, C.~Wainwright, P.~Mishkin, C.~Zhang,
  S.~Agarwal, K.~Slama, A.~Ray \emph{et~al.}, ``Training language models to
  follow instructions with human feedback,'' \emph{Advances in Neural
  Information Processing Systems}, vol.~35, pp. 27\,730--27\,744, Jul. 2022.

\bibitem{schulman2017proximal}
J.~Schulman, F.~Wolski, P.~Dhariwal, A.~Radford, and O.~Klimov, ``Proximal
  policy optimization algorithms,'' \emph{{\rm{arXiv preprint
  arXiv:1707.06347}}}, Jul. 2017.

\bibitem{dong2022survey}
Q.~Dong, L.~Li, D.~Dai, C.~Zheng, Z.~Wu, B.~Chang, X.~Sun, J.~Xu, and Z.~Sui,
  ``A survey for in-context learning,'' \emph{{\rm{arXiv preprint
  arXiv:2301.00234}}}, Jan. 2022.

\bibitem{Newbing}
Microsoft, ``Introducing the new {Bing},'' ``Accessed Mar. 19, 2023", [Online].
  Available: \url{https://www.bing.com/new}.

\bibitem{OfficeCopilot}
J.~Spataro, ``Introducing {Microsoft} 365 {Copilot} – your copilot for
  work,'' ``Accessed Mar. 19, 2023", [Online]. Available:
  \url{https://blogs.microsoft.com/blog/2023/03/16/introducing-microsoft-365-copilot-your-copilot-for-work/}.

\bibitem{yang20226g}
Y.~Yang, M.~Ma, H.~Wu, Q.~Yu, P.~Zhang, X.~You, J.~Wu, C.~Peng, T.-S.~P. Yum,
  S.~Shen \emph{et~al.}, ``6g network {AI} architecture for everyone-centric
  customized services,'' \emph{IEEE Network}, pp. 1--10, Jul. 2022.

\bibitem{daniel2018quality}
F.~Daniel, P.~Kucherbaev, C.~Cappiello, B.~Benatallah, and M.~Allahbakhsh,
  ``Quality control in crowdsourcing: A survey of quality attributes,
  assessment techniques, and assurance actions,'' \emph{ACM Computing Surveys
  (CSUR)}, vol.~51, no.~1, pp. 1--40, Jan. 2018.

\bibitem{zhang2022holographic}
H.~Zhang, H.~Zhang, B.~Di, M.~Di~Renzo, Z.~Han, H.~V. Poor, and L.~Song,
  ``Holographic integrated sensing and communication,'' \emph{IEEE Journal on
  Selected Areas in Communications}, vol.~40, no.~7, pp. 2114--2130, Mar. 2022.

\bibitem{deng2019data}
X.~Deng, Y.~Jiang, L.~T. Yang, M.~Lin, L.~Yi, and M.~Wang, ``Data fusion based
  coverage optimization in heterogeneous sensor networks: A survey,''
  \emph{Information Fusion}, vol.~52, pp. 90--105, Dec. 2019.

\bibitem{schuhmann2021laion}
C.~Schuhmann, R.~Vencu, R.~Beaumont, R.~Kaczmarczyk, C.~Mullis, A.~Katta,
  T.~Coombes, J.~Jitsev, and A.~Komatsuzaki, ``{LAION-400M}: Open dataset of
  {CLIP}-filtered 400 million image-text pairs,'' \emph{{\rm{arXiv preprint
  arXiv:2111.02114}}}, 2021.

\bibitem{du2022semantic}
H.~Du, J.~Wang, D.~Niyato, J.~Kang, Z.~Xiong, J.~Zhang, and X.~Shen, ``Semantic
  communications for wireless sensing: {RIS}-aided encoding and self-supervised
  decoding,'' \emph{IEEE Journal on Selected Areas in Communications}, 2023.

\bibitem{jain2022hugging}
S.~M. Jain, ``Hugging face,'' in \emph{Introduction to Transformers for NLP:
  With the Hugging Face Library and Models to Solve Problems}, 2022, pp.
  51--67.

\bibitem{kynkaanniemi2019improved}
T.~Kynk{\"a}{\"a}nniemi, T.~Karras, S.~Laine, J.~Lehtinen, and T.~Aila,
  ``Improved precision and recall metric for assessing generative models,''
  \emph{Advances In Neural Information Processing Systems}, vol.~32, p.
  3927–3936, Dec. 2019.

\bibitem{park2021benchmark}
D.~H. Park, S.~Azadi, X.~Liu, T.~Darrell, and A.~Rohrbach, ``Benchmark for
  compositional text-to-image synthesis,'' in \emph{Proceedings of the Neural
  Information Processing Systems Datasets and Benchmarks Track}, Virtual
  Conference, Dec. 2021.

\bibitem{wu2023visual}
C.~Wu, S.~Yin, W.~Qi, X.~Wang, Z.~Tang, and N.~Duan, ``Visual {ChatGPT}:
  Talking, drawing and editing with visual foundation models,''
  \emph{{\rm{arXiv preprint arXiv:2303.04671}}}, Mar. 2023.

\bibitem{benny2021evaluation}
Y.~Benny, T.~Galanti, S.~Benaim, and L.~Wolf, ``Evaluation metrics for
  conditional image generation,'' \emph{International Journal of Computer
  Vision}, vol. 129, no.~5, pp. 1712--1731, May 2021.

\bibitem{xu2018attngan}
T.~Xu, P.~Zhang, Q.~Huang, H.~Zhang, Z.~Gan, X.~Huang, and X.~He, ``Attngan:
  Fine-grained text to image generation with attentional generative adversarial
  networks,'' in \emph{Proceedings of the IEEE Conference on Computer Vision
  and Pattern Recognition}, Salt Lake City, Utah, Jun. 2018, pp. 1316--1324.

\bibitem{naeem2020reliable}
M.~F. Naeem, S.~J. Oh, Y.~Uh, Y.~Choi, and J.~Yoo, ``Reliable fidelity and
  diversity metrics for generative models,'' in \emph{Proceedings of the
  International Conference on Machine Learning}, Virtual Conference, Nov. 2020,
  pp. 7176--7185.

\bibitem{du2023rethinking}
H.~Du, B.~Ma, D.~Niyato, J.~Kang, Z.~Xiong, and Z.~Yang, ``Rethinking quality
  of experience for metaverse services: A consumer-based economics
  perspective,'' \emph{IEEE Network}, 2023.

\bibitem{goodfellow2020generative}
I.~Goodfellow, J.~Pouget-Abadie, M.~Mirza, B.~Xu, D.~Warde-Farley, S.~Ozair,
  A.~Courville, and Y.~Bengio, ``Generative adversarial networks,''
  \emph{Communications of the ACM}, vol.~63, no.~11, pp. 139--144, Oct. 2020.

\bibitem{zhao2016energy}
J.~Zhao, M.~Mathieu, and Y.~LeCun, ``Energy-based generative adversarial
  network,'' \emph{{\rm{arXiv preprint arXiv:1609.03126}}}, 2016.

\bibitem{kingma2019introduction}
D.~P. Kingma, M.~Welling \emph{et~al.}, ``An introduction to variational
  autoencoders,'' \emph{Foundations and Trends{\textregistered} in Machine
  Learning}, vol.~12, no.~4, pp. 307--392, Nov. 2019.

\bibitem{rezende2015variational}
D.~Rezende and S.~Mohamed, ``Variational inference with normalizing flows,'' in
  \emph{Proceedings of the International Conference on Machine Learning},
  Lille, France, Jul. 2015, pp. 1530--1538.

\bibitem{zhao2023survey}
W.~X. Zhao, K.~Zhou, J.~Li, T.~Tang, X.~Wang, Y.~Hou, Y.~Min, B.~Zhang,
  J.~Zhang, Z.~Dong \emph{et~al.}, ``A survey of large language models,''
  \emph{{\rm{arXiv preprint arXiv:2303.18223}}}, 2023.

\bibitem{driess2023palm}
D.~Driess, F.~Xia, M.~S. Sajjadi, C.~Lynch, A.~Chowdhery, B.~Ichter, A.~Wahid,
  J.~Tompson, Q.~Vuong, T.~Yu \emph{et~al.}, ``{PaLM-E}: An embodied multimodal
  language model,'' \emph{{\rm{arXiv preprint arXiv:2303.03378}}}, 2023.

\bibitem{chen2023big}
Z.~Chen, Z.~Zhang, and Z.~Yang, ``Big {AI} models for {6G} wireless networks:
  Opportunities, challenges, and research directions,'' \emph{{\rm{arXiv
  preprint arXiv:2308.06250}}}, 2023.

\bibitem{bariah2023large}
L.~Bariah, Q.~Zhao, H.~Zou, Y.~Tian, F.~Bader, and M.~Debbah, ``Large language
  models for {Telecom}: The next big thing?'' \emph{{\rm{arXiv preprint
  arXiv:2306.10249}}}, 2023.

\bibitem{lin2023pushing}
Z.~Lin, G.~Qu, Q.~Chen, X.~Chen, Z.~Chen, and K.~Huang, ``Pushing large
  language models to the {6G} edge: Vision, challenges, and opportunities,''
  \emph{{\rm{arXiv preprint arXiv:2309.16739}}}, 2023.

\bibitem{xu2023sparks}
M.~Xu, D.~Niyato, H.~Zhang, J.~Kang, Z.~Xiong, S.~Mao, and Z.~Han, ``Sparks of
  {GPTs} in edge intelligence for metaverse: Caching and inference for mobile
  {AIGC} services,'' \emph{{\rm{arXiv preprint arXiv:2304.08782}}}, 2023.

\bibitem{dinh2013survey}
H.~T. Dinh, C.~Lee, D.~Niyato, and P.~Wang, ``A survey of mobile cloud
  computing: Architecture, applications, and approaches,'' \emph{Wireless
  Communications and Mobile Computing}, vol.~13, no.~18, pp. 1587--1611, Oct.
  2013.

\bibitem{xu2022personalized}
C.~Xu, Y.~Ding, C.~Chen, Y.~Ding, W.~Zhou, and S.~Wen, ``Personalized location
  privacy protection for location-based services in vehicular networks,''
  \emph{IEEE Transactions on Intelligent Transportation Systems}, vol.~9,
  no.~10, pp. 1633--1637, Jul. 2022.

\bibitem{du2021millimeter}
H.~Du, J.~Zhang, J.~Cheng, and B.~Ai, ``Millimeter wave communications with
  reconfigurable intelligent surfaces: Performance analysis and optimization,''
  \emph{IEEE Transactions on Communications}, vol.~69, no.~4, pp. 2752--2768,
  2021.

\bibitem{wu2022hitdl}
J.~Wu, L.~Wang, Q.~Pei, X.~Cui, F.~Liu, and T.~Yang, ``Hitdl: High-throughput
  deep learning inference at the hybrid mobile edge,'' \emph{IEEE Transactions
  on Parallel and Distributed Systems}, vol.~33, no.~12, pp. 4499--4514, Aug.
  2022.

\bibitem{gpt3}
M.~Zhang and J.~Li, ``A commentary of gpt-3 in mit technology review 2021,''
  \emph{Fundamental Research}, vol.~1, no.~6, pp. 831--833, Feb. 2021.

\bibitem{devlin2018bert}
J.~D. M.-W.~C. Kenton and L.~K. Toutanova, ``{BERT}: Pre-training of deep
  bidirectional transformers for language understanding,'' in \emph{Proceedings
  of NAACL-HLT}, Minneapolis, Minnesota, Jun. 2019, pp. 4171--4186.

\bibitem{thoppilan2022lamda}
R.~Thoppilan, D.~De~Freitas, J.~Hall, N.~Shazeer, A.~Kulshreshtha, H.-T. Cheng,
  A.~Jin, T.~Bos, L.~Baker, Y.~Du \emph{et~al.}, ``Lamda: Language models for
  dialog applications,'' \emph{{\rm{arXiv preprint arXiv:2201.08239}}}, Jan.
  2022.

\bibitem{vaswani2017attention}
A.~Vaswani, N.~Shazeer, N.~Parmar, J.~Uszkoreit, L.~Jones, A.~N. Gomez,
  {\L}.~Kaiser, and I.~Polosukhin, ``Attention is all you need,''
  \emph{Advances In Neural Information Processing Systems}, p. 6000–6010,
  Dec. 2017.

\bibitem{zhu2015aligning}
Y.~Zhu, R.~Kiros, R.~Zemel, R.~Salakhutdinov, R.~Urtasun, A.~Torralba, and
  S.~Fidler, ``Aligning books and movies: Towards story-like visual
  explanations by watching movies and reading books,'' in \emph{Proceedings of
  the IEEE International Conference on Computer Vision}, Santiago, Chile, Dec.
  2015, pp. 19--27.

\bibitem{papineni2002bleu}
K.~Papineni, S.~Roukos, T.~Ward, and W.-J. Zhu, ``Bleu: A method for automatic
  evaluation of machine translation,'' in \emph{Proceedings of the 40th annual
  meeting of the Association for Computational Linguistics}, Philadelphia,
  Pennsylvania, Jul. 2002, pp. 311--318.

\bibitem{lin2004rouge}
C.-Y. Lin, ``{ROUGE}: A package for automatic evaluation of summaries,'' in
  \emph{Text summarization branches out}, Barcelona, Spain, Jul. 2004, pp.
  74--81.

\bibitem{karras2019style}
T.~Karras, S.~Laine, and T.~Aila, ``A style-based generator architecture for
  generative adversarial networks,'' in \emph{Proceedings of the IEEE/CVF
  conference on computer vision and pattern recognition}, Long Beach, CA, Jun.
  2019, pp. 4401--4410.

\bibitem{brock2018large}
A.~Brock, J.~Donahue, and K.~Simonyan, ``Large scale {GAN} training for high
  fidelity natural image synthesis,'' \emph{{\rm{arXiv preprint
  arXiv:1809.11096}}}, Sep. 2018.

\bibitem{sauer2022stylegan}
A.~Sauer, K.~Schwarz, and A.~Geiger, ``Stylegan-xl: Scaling stylegan to large
  diverse datasets,'' in \emph{Proceedings of the ACM SIGGRAPH}, Virtual
  Conference, Jul. 2022, pp. 1--10.

\bibitem{clark2019adversarial}
A.~Clark, J.~Donahue, and K.~Simonyan, ``Adversarial video generation on
  complex datasets,'' \emph{{\rm{arXiv preprint arXiv:1907.06571}}}, Jul. 2019.

\bibitem{chen2022visualgpt}
J.~Chen, H.~Guo, K.~Yi, B.~Li, and M.~Elhoseiny, ``Visualgpt: Data-efficient
  adaptation of pretrained language models for image captioning,'' in
  \emph{Proceedings of the IEEE/CVF Conference on Computer Vision and Pattern
  Recognition}, Virtual Conference, Jun. 2022, pp. 18\,030--18\,040.

\bibitem{kingma2013auto}
D.~P. Kingma and M.~Welling, ``Auto-encoding variational bayes,''
  \emph{{\rm{arXiv preprint arXiv:1312.6114}}}, 2013.

\bibitem{saharia2022photorealistic}
C.~Saharia, W.~Chan, S.~Saxena, L.~Li, J.~Whang, E.~L. Denton, K.~Ghasemipour,
  R.~Gontijo~Lopes, B.~Karagol~Ayan, T.~Salimans \emph{et~al.},
  ``Photorealistic text-to-image diffusion models with deep language
  understanding,'' \emph{Advances In Neural Information Processing Systems},
  vol.~35, pp. 36\,479--36\,494, Nov. 2022.

\bibitem{sohl2015deep}
J.~Sohl-Dickstein, E.~Weiss, N.~Maheswaranathan, and S.~Ganguli, ``Deep
  unsupervised learning using nonequilibrium thermodynamics,'' in
  \emph{Proceedings of the International Conference on Machine Learning},
  Lille, France, Jul. 2015, pp. 2256--2265.

\bibitem{ho2020denoising}
J.~Ho, A.~Jain, and P.~Abbeel, ``Denoising diffusion probabilistic models,''
  \emph{Advances In Neural Information Processing Systems}, vol.~33, pp.
  6840--6851, Dec. 2020.

\bibitem{song2020denoising}
J.~Song, C.~Meng, and S.~Ermon, ``Denoising diffusion implicit models,''
  \emph{{\rm{arXiv preprint arXiv:2010.02502}}}, 2020.

\bibitem{goodfellow2014generative}
I.~J. Goodfellow, ``On distinguishability criteria for estimating generative
  models,'' \emph{{\rm{arXiv preprint arXiv:1412.6515}}}, Dec. 2014.

\bibitem{oord2017neural}
A.~Van Den~Oord, O.~Vinyals \emph{et~al.}, ``Neural discrete representation
  learning,'' \emph{Advances In Neural Information Processing Systems}, p.
  6309–6318, Dec. 2017.

\bibitem{deng2009imagenet}
L.~Fei-Fei, J.~Deng, and K.~Li, ``Imagenet: Constructing a large-scale image
  database,'' \emph{Journal of Vision}, vol.~9, no.~8, pp. 1037--1037, Jun.
  2009.

\bibitem{liu2015deep}
Z.~Liu, P.~Luo, X.~Wang, and X.~Tang, ``Deep learning face attributes in the
  wild,'' in \emph{Proceedings of the IEEE International Conference on Computer
  Vision}, Santiago, Chile, Dec. 2015, pp. 3730--3738.

\bibitem{lin2014microsoft}
T.-Y. Lin, M.~Maire, S.~Belongie, J.~Hays, P.~Perona, D.~Ramanan,
  P.~Doll{\'a}r, and C.~L. Zitnick, ``Microsoft coco: Common objects in
  context,'' in \emph{Proceedings of the European Conference on Computer
  Vision}, Zurich, Switzerland, Sep. 2014, pp. 740--755.

\bibitem{heusel2017gans}
M.~Heusel, H.~Ramsauer, T.~Unterthiner, B.~Nessler, and S.~Hochreiter, ``{GAN}s
  trained by a two time-scale update rule converge to a local {Nash}
  equilibrium,'' \emph{Advances In Neural Information Processing Systems}, p.
  6629–6640, Dec. 2017.

\bibitem{salimans2016improved}
T.~Salimans, I.~Goodfellow, W.~Zaremba, V.~Cheung, A.~Radford, and X.~Chen,
  ``Improved techniques for training {GAN}s,'' \emph{Advances In Neural
  Information Processing Systems}, p. 2234–2242, Dec. 2016.

\bibitem{zhang2018unreasonable}
R.~Zhang, P.~Isola, A.~A. Efros, E.~Shechtman, and O.~Wang, ``The unreasonable
  effectiveness of deep features as a perceptual metric,'' in \emph{Proceedings
  of the IEEE Conference on Computer Vision and Pattern Recognition}, Los
  Alamitos, CA, Jun. 2018, pp. 586--595.

\bibitem{museNet}
A.~Topirceanu, G.~Barina, and M.~Udrescu, ``{MuSeNet}: Collaboration in the
  music artists industry,'' in \emph{Proceedings of the European Network
  Intelligence Conference}, Wroclaw, Poland, Sep. 2014, pp. 89--94.

\bibitem{oord2016wavenet}
A.~{van den Oord}, S.~Dieleman, H.~Zen, K.~Simonyan, O.~Vinyals, A.~Graves,
  N.~Kalchbrenner, A.~Senior, and K.~Kavukcuoglu, ``{WaveNet: A Generative
  Model for Raw Audio},'' in \emph{Proceedings of the 9th ISCA Workshop on
  Speech Synthesis Workshop}, Sunnyvale, California, Sep. 2016, p. 125.

\bibitem{borsos2022audiolm}
Z.~Borsos, R.~Marinier, D.~Vincent, E.~Kharitonov, O.~Pietquin, M.~Sharifi,
  O.~Teboul, D.~Grangier, M.~Tagliasacchi, and N.~Zeghidour, ``Audiolm: A
  language modeling approach to audio generation,'' \emph{{\rm{arXiv preprint
  arXiv:2209.03143}}}, Sep. 2022.

\bibitem{hawthorne2018enabling}
C.~Hawthorne, A.~Stasyuk, A.~Roberts, I.~Simon, C.-Z.~A. Huang, S.~Dieleman,
  E.~Elsen, J.~Engel, and D.~Eck, ``Enabling factorized piano music modeling
  and generation with the {MAESTRO} dataset,'' \emph{{\rm{arXiv preprint
  arXiv:1810.12247}}}, Oct. 2018.

\bibitem{dhariwal2021diffusion}
P.~Dhariwal and A.~Nichol, ``Diffusion models beat {GAN}s on image synthesis,''
  \emph{Advances In Neural Information Processing Systems}, vol.~34, pp.
  8780--8794, Dec. 2021.

\bibitem{ho2022video}
J.~Ho, T.~Salimans, A.~Gritsenko, W.~Chan, M.~Norouzi, and D.~J. Fleet, ``Video
  diffusion models,'' \emph{{\rm{arXiv preprint arXiv:2204.03458}}}, Apr. 2022,
  [Online]. Available: \url{https://arxiv.org/abs/2204.03458}.

\bibitem{poole2022dreamfusion}
B.~Poole, A.~Jain, J.~T. Barron, and B.~Mildenhall, ``Dreamfusion: Text-to-{3D}
  using {2D} diffusion,'' \emph{{\rm{arXiv preprint arXiv:2209.14988}}}, Sep
  2022.

\bibitem{kay2017kinetics}
W.~Kay, J.~Carreira, K.~Simonyan, B.~Zhang, C.~Hillier, S.~Vijayanarasimhan,
  F.~Viola, T.~Green, T.~Back, P.~Natsev \emph{et~al.}, ``The kinetics human
  action video dataset,'' \emph{{\rm{arXiv preprint arXiv:1705.06950}}}, May
  2017.

\bibitem{mildenhall2021nerf}
B.~Mildenhall, P.~P. Srinivasan, M.~Tancik, J.~T. Barron, R.~Ramamoorthi, and
  R.~Ng, ``Ne{RF}: Representing scenes as neural radiance fields for view
  synthesis,'' \emph{Communications of the ACM}, vol.~65, no.~1, pp. 99--106,
  Jan. 2021.

\bibitem{lan2019albert}
Z.~Lan, M.~Chen, S.~Goodman, K.~Gimpel, P.~Sharma, and R.~Soricut, ``{ALBERT}:
  A lite {BERT} for self-supervised learning of language representations,'' in
  \emph{Proceedings of the International Conference on Learning
  Representations}, Addis Ababa, Ethiopia, Apr. 2019.

\bibitem{sun2020mobilebert}
Z.~Sun, H.~Yu, X.~Song, R.~Liu, Y.~Yang, and D.~Zhou, ``{M}obile{BERT}: a
  compact task-agnostic {BERT} for resource-limited devices,'' in
  \emph{Proceedings of the 58th Annual Meeting of the Association for
  Computational Linguistics}, Virtual Conference, Jul. 2020, pp. 2158--2170.

\bibitem{nichol2021glide}
A.~Q. Nichol, P.~Dhariwal, A.~Ramesh, P.~Shyam, P.~Mishkin, B.~Mcgrew,
  I.~Sutskever, and M.~Chen, ``{GLIDE}: Towards photorealistic image generation
  and editing with text-guided diffusion models,'' in \emph{International
  Conference on Machine Learning}, Baltimore, Maryland, Jun. 2022, pp.
  16\,784--16\,804.

\bibitem{shi2022divae}
J.~Shi, C.~Wu, J.~Liang, X.~Liu, and N.~Duan, ``{DiVAE}: Photorealistic images
  synthesis with denoising diffusion decoder,'' \emph{{\rm{arXiv preprint
  arXiv:2206.00386}}}, 2022.

\bibitem{xu2022wireless}
M.~Xu, D.~Niyato, J.~Kang, Z.~Xiong, C.~Miao, and D.~I. Kim, ``Wireless
  edge-empowered metaverse: A learning-based incentive mechanism for virtual
  reality,'' in \emph{Proceedings of IEEE International Conference on
  Communications (ICC)}, Seoul, South Korea, Aug. 2022, pp. 5220--5225.

\bibitem{zhang2023location}
H.~Zhang, S.~Mao, D.~Niyato, and Z.~Han, ``Location-dependent augmented reality
  services in wireless edge-enabled metaverse systems,'' \emph{IEEE Open
  Journal of the Communications Society}, vol.~4, pp. 171--183, Jan. 2023.

\bibitem{du2020mec}
J.~Du, F.~R. Yu, G.~Lu, J.~Wang, J.~Jiang, and X.~Chu, ``Mec-assisted immersive
  vr video streaming over terahertz wireless networks: A deep reinforcement
  learning approach,'' \emph{IEEE {Internet} of Things Journal}, vol.~7,
  no.~10, pp. 9517--9529, Jun. 2020.

\bibitem{gafni2022make}
O.~Gafni, A.~Polyak, O.~Ashual, S.~Sheynin, D.~Parikh, and Y.~Taigman,
  ``Make-a-scene: Scene-based text-to-image generation with human priors,'' in
  \emph{Proceedings of the 17th European Conference on Computer Vision}, Tel
  Aviv, Israel, 2022, pp. 89--106.

\bibitem{blattmann2022semi}
A.~Blattmann, R.~Rombach, K.~Oktay, J.~M{\"u}ller, and B.~Ommer,
  ``Semi-parametric neural image synthesis,'' in \emph{Advances In Neural
  Information Processing Systems}, Virtual Conference, Nov. 2022.

\bibitem{du2022exploring}
H.~Du, J.~Wang, D.~Niyato, J.~Kang, Z.~Xiong, X.~S. Shen, and D.~I. Kim,
  ``Exploring attention-aware network resource allocation for customized
  metaverse services,'' \emph{IEEE Network}, pp. 1--1, 2022.

\bibitem{jin2022dr}
W.~Jin, N.~Ryu, G.~Kim, S.-H. Baek, and S.~Cho, ``{Dr. 3D}: Adapting {3D}
  {GAN}s to artistic drawings,'' in \emph{Proceedings of the SIGGRAPH Asia},
  2022, pp. 1--8.

\bibitem{chen2023introduction}
A.~Chen, S.~Mao, Z.~Li, M.~Xu, H.~Zhang, D.~Niyato, and Z.~Han, ``An
  introduction to point cloud compression standards,'' \emph{GetMobile: Mobile
  Computing and Communications}, vol.~27, no.~1, pp. 11--17, May 2023.

\bibitem{chou2022diffusionsdf}
G.~Chou, Y.~Bahat, and F.~Heide, ``Diffusion{SDF}: Conditional generative
  modeling of signed distance functions,'' \emph{{\rm{arXiv preprint
  arXiv:2211.13757}}}, Nov. 2022.

\bibitem{nichol2022point}
A.~Nichol, H.~Jun, P.~Dhariwal, P.~Mishkin, and M.~Chen, ``Point-e: A system
  for generating 3{D} point clouds from complex prompts,'' \emph{{\rm{arXiv
  preprint arXiv:2212.08751}}}, Dec. 2022.

\bibitem{metzer2022latent}
G.~Metzer, E.~Richardson, O.~Patashnik, R.~Giryes, and D.~Cohen-Or,
  ``Latent-{NeRF} for shape-guided generation of {3D} shapes and textures,''
  \emph{{\rm{arXiv preprint arXiv:2211.07600}}}, Nov. 2022.

\bibitem{zeng2022lion}
X.~Zeng, A.~Vahdat, F.~Williams, Z.~Gojcic, O.~Litany, S.~Fidler, and K.~Kreis,
  ``{LION}: Latent point diffusion models for {3D} shape generation,'' in
  \emph{Advances in Neural Information Processing Systems}, Virtual Conference,
  Nov. 2022.

\bibitem{li2022diffusion}
M.~Li, Y.~Duan, J.~Zhou, and J.~Lu, ``Diffusion-{SDF}: Text-to-shape via
  voxelized diffusion,'' \emph{{\rm{arXiv preprint arXiv:2212.03293}}}, Dec.
  2022.

\bibitem{lin2022magic3d}
C.-H. Lin, J.~Gao, L.~Tang, T.~Takikawa, X.~Zeng, X.~Huang, K.~Kreis,
  S.~Fidler, M.-Y. Liu, and T.-Y. Lin, ``Magic3d: High-resolution text-to-3d
  content creation,'' \emph{{\rm{arXiv preprint arXiv:2211.10440}}}, Nov. 2022.

\bibitem{wu2022generative}
A.~N. Wu, R.~Stouffs, and F.~Biljecki, ``Generative adversarial networks in the
  built environment: A comprehensive review of the application of {GAN}s across
  data types and scales,'' \emph{Building and Environment}, p. 109477, Sep.
  2022.

\bibitem{du2023enabling}
H.~Du, Z.~Li, D.~Niyato, J.~Kang, Z.~Xiong, D.~I. Kim \emph{et~al.}, ``Enabling
  {AI}-generated content ({AIGC}) services in wireless edge networks,''
  \emph{{\rm{arXiv preprint arXiv:2301.03220}}}, Jan. 2023.

\bibitem{li2020noma}
Z.~Li, M.~Xu, J.~Nie, J.~Kang, W.~Chen, and S.~Xie, ``{NOMA}-enabled
  cooperative computation offloading for blockchain-empowered {Internet} of
  things: A learning approach,'' \emph{IEEE {Internet} of Things Journal},
  vol.~8, no.~4, pp. 2364--2378, Aug. 2020.

\bibitem{fan2022frido}
W.-C. Fan, Y.-C. Chen, D.~Chen, Y.~Cheng, L.~Yuan, and Y.-C.~F. Wang, ``Frido:
  Feature pyramid diffusion for complex scene image synthesis,''
  \emph{{\rm{arXiv preprint arXiv:2208.13753}}}, Aug. 2022.

\bibitem{ma2023towards}
H.~Ma, Z.~Zhou, X.~Zhang, and X.~Chen, ``Towards carbon-neutral edge computing:
  Greening edge {AI} by harnessing spot and future carbon markets,'' \emph{IEEE
  {Internet} of Things Journal}, pp. 1--1, Apr. 2023.

\bibitem{lin2023blockchain}
Y.~Lin, H.~Du, D.~Niyato, J.~Nie, J.~Zhang, Y.~Cheng, and Z.~Yang,
  ``Blockchain-aided secure semantic communication for {AI}-generated content
  in metaverse,'' \emph{{\rm{arXiv preprint arXiv:2301.11289}}}, Jan. 2023.

\bibitem{du2023ai}
H.~Du, J.~Wang, D.~Niyato, J.~Kang, Z.~Xiong, and D.~I. Kim, ``{AI}-generated
  incentive mechanism and full-duplex semantic communications for information
  sharing,'' \emph{{\rm{arXiv preprint arXiv:2303.01896}}}, Mar. 2023.

\bibitem{du2023generative}
H.~Du, Z.~Li, D.~Niyato, J.~Kang, Z.~Xiong, H.~Huang, and S.~Mao, ``Generative
  {AI}-aided optimization for {AI}-generated content ({AIGC}) services in edge
  networks,'' \emph{{\rm{arXiv preprint arXiv:2303.13052}}}, 2023.

\bibitem{du2023exploring}
H.~Du, R.~Zhang, D.~Niyato, J.~Kang, Z.~Xiong, D.~I. Kim, H.~V. Poor
  \emph{et~al.}, ``Exploring collaborative distributed diffusion-based
  {AI}-generated content (aigc) in wireless networks,'' \emph{{\rm{arXiv
  preprint arXiv:2304.03446}}}, 2023.

\bibitem{liu2023blockchain}
Y.~Liu, H.~Du, D.~Niyato, J.~Kang, Z.~Xiong, C.~Miao, A.~Jamalipour
  \emph{et~al.}, ``Blockchain-empowered lifecycle management for {AI}-generated
  content ({AIGC}) products in edge networks,'' \emph{{\rm{arXiv preprint
  arXiv:2303.02836}}}, Mar. 2023.

\bibitem{liu2023deep}
Y.~Liu, H.~Du, D.~Niyato, J.~Kang, Z.~Xiong, D.~I. Kim, and A.~Jamalipour,
  ``Deep generative model and its applications in efficient wireless network
  management: A tutorial and case study,'' \emph{{\rm{arXiv preprint
  arXiv:2303.17114}}}, 2023.

\bibitem{wang2023guiding}
J.~Wang, H.~Du, D.~Niyato, Z.~Xiong, J.~Kang, S.~Mao \emph{et~al.}, ``Guiding
  {AI}-generated digital content with wireless perception,'' \emph{{\rm{arXiv
  preprint arXiv:2303.14624}}}, 2023.

\bibitem{du2023spear}
H.~Du, D.~Niyato, J.~Kang, Z.~Xiong, K.-Y. Lam, Y.~Fang, and Y.~Li, ``Spear or
  shield: Leveraging generative {AI} to tackle security threats of intelligent
  network services,'' \emph{{\rm{arXiv preprint arXiv:2306.02384}}}, 2023.

\bibitem{zhang2023generative}
R.~Zhang, K.~Xiong, H.~Du, D.~Niyato, J.~Kang, X.~Shen, and H.~V. Poor,
  ``Generative {AI}-enabled vehicular networks: Fundamentals, framework, and
  case study,'' \emph{{\rm{arXiv preprint arXiv:2304.11098}}}, 2023.

\bibitem{lin2023unified}
Y.~Lin, Z.~Gao, H.~Du, D.~Niyato, J.~Kang, R.~Deng, and X.~S. Shen, ``A unified
  blockchain-semantic framework for wireless edge intelligence enabled web
  3.0,'' \emph{IEEE Wireless Communications}, pp. 1--1, Mar. 2023.

\bibitem{du2023yolo}
B.~Du, H.~Du, H.~Liu, D.~Niyato, P.~Xin, J.~Yu, M.~Qi, and Y.~Tang,
  ``{YOLO}-based semantic communication with generative {AI}-aided resource
  allocation for digital twins construction,'' \emph{{\rm{arXiv preprint
  arXiv:2306.14138}}}, 2023.

\bibitem{xu2023joint}
M.~Xu, D.~Niyato, H.~Zhang, J.~Kang, Z.~Xiong, S.~Mao, and Z.~Han, ``Joint
  foundation model caching and inference of generative {AI} services for edge
  intelligence,'' \emph{{\rm{arXiv preprint arXiv:2305.12130}}}, 2023.

\bibitem{xu2023generative}
M.~Xu, D.~Niyato, J.~Chen, H.~Zhang, J.~Kang, Z.~Xiong, S.~Mao, and Z.~Han,
  ``Generative {AI}-empowered simulation for autonomous driving in vehicular
  mixed reality metaverses,'' \emph{{\rm{arXiv preprint arXiv:2302.08418}}},
  Feb. 2023.

\bibitem{chen2023revolution}
J.~Chen, C.~Yi, H.~Du, D.~Niyato, J.~Kang, J.~Cai \emph{et~al.}, ``A revolution
  of personalized healthcare: {E}nabling human digital twin with mobile
  {AIGC},'' \emph{{\rm{arXiv preprint arXiv:2307.12115}}}, 2023.

\bibitem{huang2023federated}
X.~Huang, P.~Li, H.~Du, J.~Kang, D.~Niyato, D.~I. Kim, and Y.~Wu, ``Federated
  learning-empowered {AI}-generated content in wireless networks,''
  \emph{{\rm{arXiv preprint arXiv:2307.07146}}}, 2023.

\bibitem{wang2023generative}
J.~Wang, H.~Du, D.~Niyato, J.~Kang, S.~Cui, X.~Shen, and P.~Zhang, ``Generative
  {AI} for integrated sensing and communication: {I}nsights from the physical
  layer perspective,'' \emph{{\rm{arXiv preprint arXiv:2310.01036}}}, 2023.

\bibitem{du2023generativemul}
H.~Du, G.~Liu, D.~Niyato, J.~Zhang, J.~Kang, Z.~Xiong, B.~Ai, and D.~I. Kim,
  ``Generative {AI}-aided joint training-free secure semantic communications
  via multi-modal prompts,'' \emph{{\rm{arXiv preprint arXiv:2309.02616}}},
  2023.

\bibitem{liu2023optimizing}
Y.~Liu, H.~Du, D.~Niyato, J.~Kang, S.~Cui, X.~Shen, and P.~Zhang, ``Optimizing
  mobile-edge {AI}-generated everything ({AIGX}) services by prompt
  engineering: {F}undamental, framework, and case study,'' \emph{{\rm{arXiv
  preprint arXiv:2309.01065}}}, 2023.

\bibitem{zheng2023flexible}
J.~Zheng, J.~Zhang, H.~Du, D.~Niyato, S.~Sun, B.~Ai, and K.~B. Letaief,
  ``Flexible-position {MIMO} for wireless communications: {F}undamentals,
  challenges, and future directions,'' \emph{{\rm{arXiv preprint
  arXiv:2308.14578}}}, 2023.

\bibitem{wang2023unified}
J.~Wang, H.~Du, D.~Niyato, J.~Kang, Z.~Xiong, D.~Rajan, S.~Mao \emph{et~al.},
  ``A unified framework for guiding generative {AI} with wireless perception in
  resource-constrained mobile edge networks,'' \emph{{\rm{arXiv preprint
  arXiv:2309.01426}}}, 2023.

\bibitem{wang2023semantic}
J.~Wang, H.~Du, Z.~Tian, D.~Niyato, J.~Kang, and X.~Shen, ``Semantic-aware
  sensing information transmission for metaverse: A contest theoretic
  approach,'' \emph{IEEE Transactions on Wireless Communications}, pp. 1--1,
  Jan. 2023.

\bibitem{wang2022wireless}
J.~Wang, H.~Du, X.~Yang, D.~Niyato, J.~Kang, and S.~Mao, ``Wireless sensing
  data collection and processing for metaverse avatar construction,''
  \emph{{\rm{arXiv preprint arXiv:2211.12720}}}, Nov. 2022.

\bibitem{yang2022semantic}
W.~Yang, H.~Du, Z.~Q. Liew, W.~Y.~B. Lim, Z.~Xiong, D.~Niyato, X.~Chi, X.~S.
  Shen, and C.~Miao, ``Semantic communications for future {Internet}:
  Fundamentals, applications, and challenges,'' \emph{IEEE Communications
  Surveys \& Tutorials}, vol.~25, no.~1, pp. 213--250, Nov. 2023.

\bibitem{rombach2022high}
R.~Rombach, A.~Blattmann, D.~Lorenz, P.~Esser, and B.~Ommer, ``High-resolution
  image synthesis with latent diffusion models,'' in \emph{Proceedings of the
  IEEE/CVF Conference on Computer Vision and Pattern Recognition}, New Orleans,
  Louisiana, Jun. 2022, pp. 10\,684--10\,695.

\bibitem{mathias2013traffic}
M.~Mathias, R.~Timofte, R.~Benenson, and L.~Van~Gool, ``Traffic sign
  recognition—how far are we from the solution?'' in \emph{Proceedings of the
  International joint conference on Neural networks}, Dallas, Texas, Aug. 2013,
  pp. 1--8.

\bibitem{xu2022epvisa}
M.~Xu, D.~Niyato, B.~Wright, H.~Zhang, J.~Kang, Z.~Xiong, S.~Mao, and Z.~Han,
  ``{Epvisa}: Efficient auction design for real-time physical-virtual
  synchronization in the metaverse,'' \emph{{\rm{arXiv preprint
  arXiv:2211.06838}}}, Nov. 2022.

\bibitem{xu2023generative1}
M.~Xu, D.~Niyato, H.~Zhang, J.~Kang, Z.~Xiong, S.~Mao, and Z.~Han, ``Generative
  {AI}-empowered effective physical-virtual synchronization in the vehicular
  metaverse,'' \emph{{\rm{arXiv preprint arXiv:2301.07636}}}, Jan. 2023.

\bibitem{hu2019dynamic}
C.~Hu, W.~Bao, D.~Wang, and F.~Liu, ``Dynamic adaptive {DNN} surgery for
  inference acceleration on the edge,'' in \emph{Proceedings of the IEEE
  INFOCOM}, Paris, France, Apr. 2019, pp. 1423--1431.

\bibitem{zhang2021deep}
W.~Zhang, D.~Yang, H.~Peng, W.~Wu, W.~Quan, H.~Zhang, and X.~Shen, ``Deep
  reinforcement learning based resource management for {DNN} inference in
  industrial {IoT},'' \emph{IEEE Transactions on Vehicular Technology},
  vol.~70, no.~8, pp. 7605--7618, Mar. 2021.

\bibitem{zhang2021joint}
R.~Zhang, K.~Xiong, Y.~Lu, B.~Gao, P.~Fan, and K.~B. Letaief, ``Joint
  coordinated beamforming and power splitting ratio optimization in mu-miso
  swipt-enabled hetnets: A multi-agent ddqn-based approach,'' \emph{IEEE
  Journal on Selected Areas in Communications}, vol.~40, no.~2, pp. 677--693,
  Oct. 2021.

\bibitem{wang2018edge}
S.~Wang, T.~Tuor, T.~Salonidis, K.~K. Leung, C.~Makaya, T.~He, and K.~Chan,
  ``When edge meets learning: Adaptive control for resource-constrained
  distributed machine learning,'' in \emph{Proceedings of the IEEE INFOCOM},
  Honolulu, HI, Jun. 2018, pp. 63--71.

\bibitem{hsieh2017gaia}
K.~Hsieh, A.~Harlap, N.~Vijaykumar, D.~Konomis, G.~R. Ganger, P.~B. Gibbons,
  and O.~Mutlu, ``Gaia: Geo-distributed machine learning approaching {LAN}
  speeds,'' in \emph{Proceedings of the 14th USENIX Symposium on Networked
  Systems Design and Implementation}, Boston, MA, Mar. 2017, pp. 629--647.

\bibitem{lin2021optimizing}
Z.~Lin, S.~Bi, and Y.-J.~A. Zhang, ``Optimizing {AI} service placement and
  resource allocation in mobile edge intelligence systems,'' \emph{IEEE
  Transactions on Wireless Communications}, vol.~20, no.~11, pp. 7257--7271,
  May 2021.

\bibitem{li2020optimizing}
X.~Li, S.~Bi, and H.~Wang, ``Optimizing resource allocation for joint {AI}
  model training and task inference in edge intelligence systems,'' \emph{IEEE
  Wireless Communications Letters}, vol.~10, no.~3, pp. 532--536, Mar. 2020.

\bibitem{zhao2022edgeadaptor}
K.~Zhao, Z.~Zhou, X.~Chen, R.~Zhou, X.~Zhang, S.~Yu, and D.~Wu,
  ``Edge{A}daptor: Online configuration adaption, model selection and resource
  provisioning for edge {DNN} inference serving at scale,'' \emph{IEEE
  Transactions on Mobile Computing}, pp. 1--16, Jul. 2022.

\bibitem{tang2020joint}
X.~Tang, X.~Chen, L.~Zeng, S.~Yu, and L.~Chen, ``Joint multiuser {DNN}
  partitioning and computational resource allocation for collaborative edge
  intelligence,'' \emph{IEEE {Internet} of Things Journal}, vol.~8, no.~12, pp.
  9511--9522, Jul. 2020.

\bibitem{lim2021decentralized}
W.~Y.~B. Lim, J.~S. Ng, Z.~Xiong, J.~Jin, Y.~Zhang, D.~Niyato, C.~Leung, and
  C.~Miao, ``Decentralized edge intelligence: A dynamic resource allocation
  framework for hierarchical federated learning,'' \emph{IEEE Transactions on
  Parallel and Distributed Systems}, vol.~33, no.~3, pp. 536--550, Jul. 2021.

\bibitem{yang2023over}
Y.~Yang, Z.~Zhang, Y.~Tian, Z.~Yang, C.~Huang, C.~Zhong, and K.-K. Wong,
  ``Over-the-air split machine learning in wireless {MIMO} networks,''
  \emph{IEEE Journal on Selected Areas in Communications}, vol.~41, no.~4, pp.
  1007--1022, Feb. 2023.

\bibitem{zhang2023energy}
R.~Zhang, K.~Xiong, Y.~Lu, P.~Fan, D.~W.~K. Ng, and K.~B. Letaief, ``Energy
  efficiency maximization in ris-assisted swipt networks with rsma: A ppo-based
  approach,'' \emph{IEEE Journal on Selected Areas in Communications}, pp.
  1--1, Jan. 2023.

\bibitem{ditzler2015learning}
G.~Ditzler, M.~Roveri, C.~Alippi, and R.~Polikar, ``Learning in nonstationary
  environments: A survey,'' \emph{IEEE Computational Intelligence Magazine},
  vol.~10, no.~4, pp. 12--25, Nov. 2015.

\bibitem{zhang2022inverse}
R.~Zhang, K.~Xiong, X.~Tian, Y.~Lu, P.~Fan, and K.~B. Letaief, ``Inverse
  reinforcement learning meets power allocation in multi-user cellular
  networks,'' in \emph{IEEE INFOCOM 2022-IEEE Conference on Computer
  Communications Workshops (INFOCOM WKSHPS)}, New York, NY, May 2022, pp. 1--2.

\bibitem{wu2022collaborative}
W.~Wu, Y.~Tang, P.~Yang, W.~Zhang, and N.~Zhang, ``Collaborative deep neural
  network inference via mobile edge computing,'' in \emph{Broadband
  Communications, Computing, and Control for Ubiquitous Intelligence}.\hskip
  1em plus 0.5em minus 0.4em\relax Springer, Mar. 2022, pp. 263--290.

\bibitem{fan2021accuracy}
W.~Fan, Z.~Chen, Y.~Su, F.~Wu, B.~Tang, and Y.~Liu, ``Accuracy-based task
  offloading and resource allocation for edge intelligence in {I}o{T},''
  \emph{IEEE Wireless Communications Letters}, vol.~11, no.~2, pp. 371--375,
  Nov. 2021.

\bibitem{chen2021dnnoff}
X.~Chen, M.~Li, H.~Zhong, Y.~Ma, and C.-H. Hsu, ``{DNNOff}: offloading
  {DNN}-based intelligent {I}o{T} applications in mobile edge computing,''
  \emph{IEEE Transactions on Industrial Informatics}, vol.~18, no.~4, pp.
  2820--2829, Apr. 2021.

\bibitem{lin2019cost}
B.~Lin, Y.~Huang, J.~Zhang, J.~Hu, X.~Chen, and J.~Li, ``Cost-driven
  off-loading for {DNN}-based applications over cloud, edge, and end devices,''
  \emph{IEEE Transactions on Industrial Informatics}, vol.~16, no.~8, pp.
  5456--5466, Aug. 2019.

\bibitem{ren2019coding}
L.~Ren, Y.~Laili, X.~Li, and X.~Wang, ``Coding-based large-scale task
  assignment for industrial edge intelligence,'' \emph{IEEE Transactions on
  Network Science and Engineering}, vol.~7, no.~4, pp. 2286--2297, Sep. 2019.

\bibitem{jeong2018computation}
H.-J. Jeong, I.~Jeong, H.-J. Lee, and S.-M. Moon, ``Computation offloading for
  machine learning web apps in the edge server environment,'' in
  \emph{Proceedings of the IEEE 38th International Conference on Distributed
  Computing Systems}, Vienna, Austria, Jul. 2018, pp. 1492--1499.

\bibitem{li2022multi}
X.~Li, C.~Sun, J.~Wen, X.~Wang, M.~Guizani, and V.~C. Leung, ``Multi-user qoe
  enhancement: Federated multi-agent reinforcement learning for cooperative
  edge intelligence,'' \emph{IEEE Network}, vol.~36, no.~5, pp. 144--151, Nov.
  2022.

\bibitem{zhan2020deep}
Y.~Zhan, S.~Guo, P.~Li, and J.~Zhang, ``A deep reinforcement learning-based
  offloading game in edge computing,'' \emph{IEEE Transactions on Computers},
  vol.~69, no.~6, pp. 883--893, Jan. 2020.

\bibitem{lin2023efficient}
Z.~Lin, G.~Zhu, Y.~Deng, X.~Chen, Y.~Gao, K.~Huang, and Y.~Fang, ``Efficient
  parallel split learning over resource-constrained wireless edge networks,''
  \emph{{\rm{arXiv preprint arXiv:2303.15991}}}, Mar. 2023.

\bibitem{wang2023overview}
Y.-C. Wang, J.~Xue, C.~Wei, and C.-C.~J. Kuo, ``An overview on generative {AI}
  at scale with edge-cloud computing,'' \emph{{\rm{arXiv preprint
  arXiv:2306.17170}}}, Jun. 2023.

\bibitem{wu2020accuracy}
W.~Wu, P.~Yang, W.~Zhang, C.~Zhou, and X.~Shen, ``Accuracy-guaranteed
  collaborative {DNN} inference in industrial {IoT} via deep reinforcement
  learning,'' \emph{IEEE Transactions on Industrial Informatics}, vol.~17,
  no.~7, pp. 4988--4998, Aug. 2020.

\bibitem{yang2022federated}
Z.~Yang, M.~Chen, K.-K. Wong, H.~V. Poor, and S.~Cui, ``Federated learning for
  {6G}: Applications, challenges, and opportunities,'' \emph{Engineering},
  vol.~8, pp. 33--41, Jan. 2022.

\bibitem{tian2022jmsnas}
Y.~Tian, Z.~Zhang, Z.~Yang, and Q.~Yang, ``Jmsnas: Joint model split and neural
  architecture search for learning over mobile edge networks,'' in \emph{2022
  IEEE International Conference on Communications Workshops (ICC Workshops)},
  Seoul, South Korea, May 2022, pp. 103--108.

\bibitem{zhang2020q}
R.~Zhang, K.~Xiong, W.~Guo, X.~Yang, P.~Fan, and K.~B. Letaief,
  ``Q-learning-based adaptive power control in wireless {RF} energy harvesting
  heterogeneous networks,'' \emph{IEEE Systems Journal}, vol.~15, no.~2, pp.
  1861--1872, Sep. 2020.

\bibitem{wen2023task}
D.~Wen, X.~Jiao, P.~Liu, G.~Zhu, Y.~Shi, and K.~Huang, ``Task-oriented
  over-the-air computation for multi-device edge split inference,'' in
  \emph{2023 IEEE Wireless Communications and Networking Conference (WCNC)},
  Glasgow, United Kingdom, Mar. 2023, pp. 1--6.

\bibitem{koda2020communication}
Y.~Koda, J.~Park, M.~Bennis, K.~Yamamoto, T.~Nishio, M.~Morikura, and
  K.~Nakashima, ``Communication-efficient multimodal split learning for mmwave
  received power prediction,'' \emph{IEEE Communications Letters}, vol.~24,
  no.~6, pp. 1284--1288, Mar. 2020.

\bibitem{wu2023split}
W.~Wu, M.~Li, K.~Qu, C.~Zhou, X.~Shen, W.~Zhuang, X.~Li, and W.~Shi, ``Split
  learning over wireless networks: Parallel design and resource management,''
  \emph{IEEE Journal on Selected Areas in Communications}, vol.~41, no.~4, pp.
  1051--1066, Feb. 2023.

\bibitem{saguil2020layer}
D.~Saguil and A.~Azim, ``A layer-partitioning approach for faster execution of
  neural network-based embedded applications in edge networks,'' \emph{IEEE
  Access}, vol.~8, pp. 59\,456--59\,469, Mar. 2020.

\bibitem{kang2022personalized}
J.~Kang, H.~Du, Z.~Li, Z.~Xiong, S.~Ma, D.~Niyato, and Y.~Li, ``Personalized
  saliency in task-oriented semantic communications: Image transmission and
  performance analysis,'' \emph{IEEE Journal on Selected Areas in
  Communications}, vol.~41, no.~1, pp. 186--201, 2022.

\bibitem{yang2021local}
Z.~Yang, R.~Wang, D.~Wu, H.~Wang, H.~Song, and X.~Ma, ``Local trajectory
  privacy protection in {5G} enabled industrial intelligent logistics,''
  \emph{IEEE Transactions on Industrial Informatics}, vol.~18, no.~4, pp.
  2868--2876, Sep. 2021.

\bibitem{zhang2020deephealth}
W.~Zhang, D.~Yang, Y.~Xu, X.~Huang, J.~Zhang, and M.~Gidlund, ``Deephealth: A
  self-attention based method for instant intelligent predictive maintenance in
  industrial {Internet} of things,'' \emph{IEEE Transactions on Industrial
  Informatics}, vol.~17, no.~8, pp. 5461--5473, Oct. 2020.

\bibitem{zhang2021optimizing}
W.~Zhang, D.~Yang, W.~Wu, H.~Peng, N.~Zhang, H.~Zhang, and X.~Shen,
  ``Optimizing federated learning in distributed industrial {IoT}: A
  multi-agent approach,'' \emph{IEEE Journal on Selected Areas in
  Communications}, vol.~39, no.~12, pp. 3688--3703, Oct. 2021.

\bibitem{matsubara2019distilled}
Y.~Matsubara, S.~Baidya, D.~Callegaro, M.~Levorato, and S.~Singh, ``Distilled
  split deep neural networks for edge-assisted real-time systems,'' in
  \emph{Proceedings of the 2019 Workshop on Hot Topics in Video Analytics and
  Intelligent Edges}, Los Cabos, Mexico, Oct. 2019, pp. 21--26.

\bibitem{jiang2021intelligence}
K.~Jiang, C.~Sun, H.~Zhou, X.~Li, M.~Dong, and V.~C. Leung,
  ``Intelligence-empowered mobile edge computing: Framework, issues,
  implementation, and outlook,'' \emph{IEEE Network}, vol.~35, no.~5, pp.
  74--82, Nov. 2021.

\bibitem{sun2021cooperative}
C.~Sun, X.~Wu, X.~Li, Q.~Fan, J.~Wen, and V.~C. Leung, ``Cooperative
  computation offloading for multi-access edge computing in {6G} mobile
  networks via soft actor critic,'' \emph{IEEE Transactions on Network Science
  and Engineering}, pp. 1--1, Apr. 2021.

\bibitem{he2020edge}
X.~He, K.~Wang, H.~Lu, W.~Xu, and S.~Guo, ``Edge {QoE}: Intelligent big data
  caching via deep reinforcement learning,'' \emph{IEEE Network}, vol.~34,
  no.~4, pp. 8--13, Jul. 2020.

\bibitem{guo2019edgeserve}
T.~Guo, R.~J. Walls, and S.~S. Ogden, ``Edgeserve: efficient deep learning
  model caching at the edge,'' in \emph{Proceedings of the 4th ACM/IEEE
  Symposium on Edge Computing}, Arlington, Virginia, Nov. 2019, pp. 313--315.

\bibitem{ogden2021many}
S.~S. Ogden, G.~R. Gilman, R.~J. Walls, and T.~Guo, ``Many models at the edge:
  Scaling deep inference via model-level caching,'' in \emph{Proceedings of the
  IEEE International Conference on Autonomic Computing and Self-Organizing
  Systems}, Washington, DC, Sep. 2021, pp. 51--60.

\bibitem{xu2018deepcache}
M.~Xu, M.~Zhu, Y.~Liu, F.~X. Lin, and X.~Liu, ``Deepcache: Principled cache for
  mobile deep vision,'' in \emph{Proceedings of the 24th Annual International
  Conference on Mobile Computing and Networking}, New Delhi, India, Oct. 2018,
  pp. 129--144.

\bibitem{fuerst2021faascache}
A.~Fuerst and P.~Sharma, ``Faascache: keeping serverless computing alive with
  greedy-dual caching,'' in \emph{Proceedings of the 26th ACM International
  Conference on Architectural Support for Programming Languages and Operating
  Systems}, Virtual Conference, Mar. 2021, pp. 386--400.

\bibitem{zheng2021knowledge}
X.-Y. Zheng, M.-C. Lee, and Y.-W.~P. Hong, ``Knowledge caching for federated
  learning,'' in \emph{2021 IEEE Global Communications Conference
  (GLOBECOM)}.\hskip 1em plus 0.5em minus 0.4em\relax IEEE, 2021, pp. 1--6.

\bibitem{wang2020attention}
X.~Wang, R.~Li, C.~Wang, X.~Li, T.~Taleb, and V.~C. Leung, ``Attention-weighted
  federated deep reinforcement learning for device-to-device assisted
  heterogeneous collaborative edge caching,'' \emph{IEEE Journal on Selected
  Areas in Communications}, vol.~39, no.~1, pp. 154--169, Nov. 2020.

\bibitem{mu2023communication}
Y.~Mu and C.~Shen, ``Communication and storage efficient federated split
  learning,'' \emph{{\rm{arXiv preprint arXiv:2302.05599}}}, Feb. 2023.

\bibitem{yao2022loading}
M.~Yao, L.~Chen, J.~Zhang, J.~Huang, and J.~Wu, ``Loading cost-aware model
  caching and request routing for cooperative edge inference,'' in
  \emph{Proceedings of the IEEE International Conference on Communication},
  Seoul, South Korea, May 2022, pp. 2327--2332.

\bibitem{shi2020communication}
Y.~Shi, K.~Yang, T.~Jiang, J.~Zhang, and K.~B. Letaief,
  ``Communication-efficient edge{AI}: Algorithms and systems,'' \emph{IEEE
  Communications Surveys \& Tutorials}, vol.~22, no.~4, pp. 2167--2191, Jul.
  2020.

\bibitem{xie2022robust}
S.~Xie, Y.~Wu, S.~Ma, M.~Ding, Y.~Shi, and M.~Tang, ``Robust information
  bottleneck for task-oriented communication with digital modulation,''
  \emph{{\rm{arXiv preprint arXiv:2209.10382}}}, Sep. 2022.

\bibitem{ogden2020mdinference}
S.~S. Ogden and T.~Guo, ``Mdinference: Balancing inference accuracy and latency
  for mobile applications,'' in \emph{Proceedings of the IEEE International
  Conference on Cloud Engineering}, NSW, Australia, Apr. 2020, pp. 28--39.

\bibitem{buckler2018eva2}
M.~Buckler, P.~Bedoukian, S.~Jayasuriya, and A.~Sampson, ``Eva$^2$: Exploiting
  temporal redundancy in live computer vision,'' in \emph{2018 ACM/IEEE 45th
  Annual International Symposium on Computer Architecture (ISCA)}, Los Angeles,
  California, Jun. 2018, pp. 533--546.

\bibitem{oakes2018sock}
E.~Oakes, L.~Yang, D.~Zhou, K.~Houck, T.~Harter, A.~Arpaci-Dusseau, and
  R.~Arpaci-Dusseau, ``$\{$SOCK$\}$: Rapid task provisioning with
  serverless-optimized containers,'' in \emph{Proceedings of the $\{$USENIX$\}$
  Annual Technical Conference ($\{$USENIX$\}$$\{$ATC$\}$ 18)}, Boston, MA, Jul.
  2018, pp. 57--70.

\bibitem{chen2020joint}
M.~Chen, Z.~Yang, W.~Saad, C.~Yin, H.~V. Poor, and S.~Cui, ``A joint learning
  and communications framework for federated learning over wireless networks,''
  \emph{IEEE Transactions on Wireless Communications}, vol.~20, no.~1, pp.
  269--283, Oct. 2020.

\bibitem{xu2022privacy}
M.~Xu, D.~Niyato, Z.~Yang, Z.~Xiong, J.~Kang, D.~I. Kim, and X.~Shen,
  ``Privacy-preserving intelligent resource allocation for federated edge
  learning in quantum {Internet},'' \emph{IEEE Journal of Selected Topics in
  Signal Processing}, vol.~17, no.~1, pp. 142--157, Nov. 2023.

\bibitem{chen2021communication}
M.~Chen, N.~Shlezinger, H.~V. Poor, Y.~C. Eldar, and S.~Cui,
  ``Communication-efficient federated learning,'' \emph{Proceedings of the
  National Academy of Sciences}, vol. 118, no.~17, p. e2024789118, Apr. 2021.

\bibitem{wang2021dynamic}
Q.~Wang, Z.~Li, K.~Nai, Y.~Chen, and M.~Wen, ``Dynamic resource allocation for
  jointing vehicle-edge deep neural network inference,'' \emph{Journal of
  Systems Architecture}, vol. 117, p. 102133, Aug. 2021.

\bibitem{yang2022novel}
K.~Yang, P.~Sun, J.~Lin, A.~Boukerche, and L.~Song, ``A novel distributed task
  scheduling framework for supporting vehicular edge intelligence,'' in
  \emph{Proceedings of the IEEE 42nd International Conference on Distributed
  Computing Systems}, Bologna, Italy, Jul. 2022, pp. 972--982.

\bibitem{sun2022meet}
Y.~Sun, B.~Xie, S.~Zhou, and Z.~Niu, ``{MEET}: Mobility-enhanced edge
  intelligence for smart and green {6G} networks,'' \emph{IEEE Communications
  Magazine}, vol.~61, no.~1, pp. 64--70, Oct. 2023.

\bibitem{wang2021resource}
D.~Wang, B.~Song, P.~Lin, F.~R. Yu, X.~Du, and M.~Guizani, ``Resource
  management for edge intelligence ({EI})-assisted {IoV} using quantum-inspired
  reinforcement learning,'' \emph{IEEE {Internet} of Things Journal}, vol.~9,
  no.~14, pp. 12\,588--12\,600, Dec. 2021.

\bibitem{balasubramanian2022venet}
V.~Balasubramanian, S.~Otoum, and M.~Reisslein, ``Venet: hybrid stacked
  autoencoder learning for cooperative edge intelligence in {IoV},'' \emph{IEEE
  Transactions on Intelligent Transportation Systems}, vol.~23, no.~9, pp.
  16\,643--16\,653, May 2022.

\bibitem{dong2021uavs}
C.~Dong, Y.~Shen, Y.~Qu, K.~Wang, J.~Zheng, Q.~Wu, and F.~Wu, ``{UAV}s as an
  intelligent service: Boosting edge intelligence for air-ground integrated
  networks,'' \emph{IEEE Network}, vol.~35, no.~4, pp. 167--175, Aug. 2021.

\bibitem{luo2022keepedge}
H.~Luo, T.~Chen, X.~Li, S.~Li, C.~Zhang, G.~Zhao, and X.~Liu, ``{KeepEdge}: A
  knowledge distillation empowered edge intelligence framework for visual
  assisted positioning in {UAV} delivery,'' \emph{IEEE Transactions on Mobile
  Computing}, pp. 1--1, Mar. 2022.

\bibitem{zhou2019exploiting}
S.~Zhou, Y.~Sun, Z.~Jiang, and Z.~Niu, ``Exploiting moving intelligence:
  Delay-optimized computation offloading in vehicular fog networks,''
  \emph{IEEE Communications Magazine}, vol.~57, no.~5, pp. 49--55, May 2019.

\bibitem{zhu2020millimeter}
L.~Zhu, J.~Zhang, Z.~Xiao, X.~Cao, X.-G. Xia, and R.~Schober, ``Millimeter-wave
  full-duplex {UAV} relay: Joint positioning, beamforming, and power control,''
  \emph{IEEE Journal on Selected Areas in Communications}, vol.~38, no.~9, pp.
  2057--2073, 2020.

\bibitem{du2022performance}
H.~Du, D.~Niyato, Y.-A. Xie, Y.~Cheng, J.~Kang, and D.~I. Kim, ``Performance
  analysis and optimization for jammer-aided multiantenna {UAV} covert
  communication,'' \emph{IEEE Journal on Selected Areas in Communications},
  vol.~40, no.~10, pp. 2962--2979, Oct. 2022.

\bibitem{9959884}
J.~Kang, H.~Du, Z.~Li, Z.~Xiong, S.~Ma, D.~Niyato, and Y.~Li, ``Personalized
  saliency in task-oriented semantic communications: Image transmission and
  performance analysis,'' \emph{IEEE Journal on Selected Areas in
  Communications}, vol.~41, no.~1, pp. 186--201, Nov. 2023.

\bibitem{huynh2022uav}
L.~N. Huynh and E.-N. Huh, ``{UAV}-enhanced edge intelligence: A survey,'' in
  \emph{Proceedings of the 6th International Conference on Computing
  Methodologies and Communication}, Erode, India, Mar. 2022, pp. 42--47.

\bibitem{alsamhi2021drones}
S.~H. Alsamhi, F.~A. Almalki, F.~Afghah, A.~Hawbani, A.~V. Shvetsov, B.~Lee,
  and H.~Song, ``Drones’ edge intelligence over smart environments in {B5G}:
  Blockchain and federated learning synergy,'' \emph{IEEE Transactions on Green
  Communications and Networking}, vol.~6, no.~1, pp. 295--312, Dec. 2021.

\bibitem{wang2022interference}
Z.~Wang, Y.~Zhou, Y.~Shi, and W.~Zhuang, ``Interference management for
  over-the-air federated learning in multi-cell wireless networks,'' \emph{IEEE
  Journal on Selected Areas in Communications}, vol.~40, no.~8, pp. 2361--2377,
  Jun. 2022.

\bibitem{yang2022multi}
T.~Yang, S.~Gao, J.~Li, M.~Qin, X.~Sun, R.~Zhang, M.~Wang, and X.~Li,
  ``Multi-armed bandits learning for task offloading in maritime edge
  intelligence networks,'' \emph{IEEE Transactions on Vehicular Technology},
  vol.~71, no.~4, pp. 4212--4224, Jan. 2022.

\bibitem{wang2021federated}
Z.~Wang, J.~Qiu, Y.~Zhou, Y.~Shi, L.~Fu, W.~Chen, and K.~B. Letaief,
  ``Federated learning via intelligent reflecting surface,'' \emph{IEEE
  Transactions on Wireless Communications}, vol.~21, no.~2, pp. 808--822, Jul.
  2021.

\bibitem{quan2018adaptive}
W.~Quan, N.~Cheng, M.~Qin, H.~Zhang, H.~A. Chan, and X.~Shen, ``Adaptive
  transmission control for software defined vehicular networks,'' \emph{IEEE
  Wireless Communications Letters}, vol.~8, no.~3, pp. 653--656, Nov. 2018.

\bibitem{misra2019soft}
S.~Misra and S.~Bera, ``Soft-{VAN}: Mobility-aware task offloading in
  software-defined vehicular network,'' \emph{IEEE Transactions on Vehicular
  Technology}, vol.~69, no.~2, pp. 2071--2078, Dec. 2019.

\bibitem{sun2020edge}
Y.~Sun, W.~Shi, X.~Huang, S.~Zhou, and Z.~Niu, ``Edge learning with timeliness
  constraints: Challenges and solutions,'' \emph{IEEE Communications Magazine},
  vol.~58, no.~12, pp. 27--33, Dec. 2020.

\bibitem{wang2021green}
J.~Wang, K.~Zhu, and E.~Hossain, ``Green {Internet} of vehicles ({IoV}) in the
  {6G} era: Toward sustainable vehicular communications and networking,''
  \emph{IEEE Transactions on Green Communications and Networking}, vol.~6,
  no.~1, pp. 391--423, Nov. 2021.

\bibitem{huang2021fedparking}
X.~Huang, P.~Li, R.~Yu, Y.~Wu, K.~Xie, and S.~Xie, ``Fedparking: A federated
  learning based parking space estimation with parked vehicle assisted edge
  computing,'' \emph{IEEE Transactions on Vehicular Technology}, vol.~70,
  no.~9, pp. 9355--9368, Jul. 2021.

\bibitem{xu2022secure}
M.~Xu, D.~T. Hoang, J.~Kang, D.~Niyato, Q.~Yan, and D.~I. Kim, ``Secure and
  reliable transfer learning framework for {6G}-enabled {Internet} of
  vehicles,'' \emph{IEEE Wireless Communications}, vol.~29, no.~4, pp.
  132--139, May 2022.

\bibitem{li2020deep}
M.~Li, J.~Gao, L.~Zhao, and X.~Shen, ``Deep reinforcement learning for
  collaborative edge computing in vehicular networks,'' \emph{IEEE Transactions
  on Cognitive Communications and Networking}, vol.~6, no.~4, pp. 1122--1135,
  Jun. 2020.

\bibitem{wu2022delay}
D.~Wu, T.~Liu, Z.~Li, T.~Tang, and R.~Wang, ``Delay-aware edge-terminal
  collaboration in green {Internet} of vehicles: A multi-agent soft
  actor-critic approach,'' \emph{IEEE Transactions on Green Communications and
  Networking}, pp. 1--1, Jun. 2022.

\bibitem{wu2023splitmag}
M.~Wu, G.~Cheng, P.~Li, R.~Yu, Y.~Wu, M.~Pan, and R.~Lu, ``Split learning with
  differential privacy for integrated terrestrial and non-terrestrial
  networks,'' \emph{IEEE Wireless Communications}, pp. 1--1, Apr. 2023.

\bibitem{yao2023split}
J.~Yao, ``Split learning for image classification in {Internet} of drones
  networks,'' in \emph{2023 IEEE 24th International Conference on High
  Performance Switching and Routing (HPSR)}, Albuquerque, NM, Jun. 2023, pp.
  52--55.

\bibitem{zhan2020incentive}
Y.~Zhan and J.~Zhang, ``An incentive mechanism design for efficient edge
  learning by deep reinforcement learning approach,'' in \emph{Proceedings of
  the IEEE INFOCOM}, ON, Canada, Jul. 2020, pp. 2489--2498.

\bibitem{liu2021incentive}
Y.~Liu, L.~Wu, Y.~Zhan, S.~Guo, and Z.~Hong, ``Incentive-driven long-term
  optimization for edge learning by hierarchical reinforcement mechanism,'' in
  \emph{Proceedings of IEEE 41st International Conference on Distributed
  Computing Systems}, DC, USA, Jul. 2021, pp. 35--45.

\bibitem{deng2021fair}
Y.~Deng, F.~Lyu, J.~Ren, Y.-C. Chen, P.~Yang, Y.~Zhou, and Y.~Zhang, ``Fair:
  Quality-aware federated learning with precise user incentive and model
  aggregation,'' in \emph{Proceedings of the IEEE INFOCOM}, BC, Canada, May
  2021, pp. 1--10.

\bibitem{ren2022ai}
X.~Ren, C.~Qiu, X.~Wang, Z.~Han, K.~Xu, H.~Yao, and S.~Guo, ``{AI}-{Bazaar}: A
  cloud-edge computing power trading framework for ubiquitous {AI} services,''
  \emph{IEEE Transactions on Cloud Computing}, pp. 1--1, Aug. 2022.

\bibitem{wang2022infedge}
X.~Wang, Y.~Zhao, C.~Qiu, Z.~Liu, J.~Nie, and V.~C. Leung, ``Infedge: A
  blockchain-based incentive mechanism in hierarchical federated learning for
  end-edge-cloud communications,'' \emph{IEEE Journal on Selected Areas in
  Communications}, vol.~40, no.~12, pp. 3325--3342, Oct. 2022.

\bibitem{zhan2021survey}
Y.~Zhan, J.~Zhang, Z.~Hong, L.~Wu, P.~Li, and S.~Guo, ``A survey of incentive
  mechanism design for federated learning,'' \emph{IEEE Transactions on
  Emerging Topics in Computing}, vol.~10, no.~2, pp. 1035--1044, Mar. 2021.

\bibitem{du2022reconfigurable}
H.~Du, J.~Kang, D.~Niyato, J.~Zhang, and D.~I. Kim, ``Reconfigurable
  intelligent surface-aided joint radar and covert communications:
  Fundamentals, optimization, and challenges,'' \emph{IEEE Vehicular Technology
  Magazine}, vol.~17, no.~3, pp. 54--64, 2022.

\bibitem{chen2022resource}
X.~Chen, Y.~Deng, G.~Zhu, D.~Wang, and Y.~Fang, ``From resource auction to
  service auction: An auction paradigm shift in wireless networks,'' \emph{IEEE
  Wireless Communications}, vol.~29, no.~2, pp. 185--191, Apr. 2022.

\bibitem{wu2022sustainable}
L.~Wu, S.~Guo, Y.~Liu, Z.~Hong, Y.~Zhan, and W.~Xu, ``Sustainable federated
  learning with long-term online vcg auction mechanism,'' in \emph{Proceedings
  of the IEEE 42nd International Conference on Distributed Computing
  Systems}.\hskip 1em plus 0.5em minus 0.4em\relax Bologna, Italy: IEEE, Jul.
  2022, pp. 895--905.

\bibitem{zhan2020learning}
Y.~Zhan, P.~Li, Z.~Qu, D.~Zeng, and S.~Guo, ``A learning-based incentive
  mechanism for federated learning,'' \emph{IEEE {Internet} of Things Journal},
  vol.~7, no.~7, pp. 6360--6368, Jan. 2020.

\bibitem{du2021resource}
J.~Du, W.~Cheng, G.~Lu, H.~Cao, X.~Chu, Z.~Zhang, and J.~Wang, ``Resource
  pricing and allocation in mec enabled blockchain systems: An a3c deep
  reinforcement learning approach,'' \emph{IEEE Transactions on Network Science
  and Engineering}, vol.~9, no.~1, pp. 33--44, Mar. 2021.

\bibitem{ren2019collaborative}
J.~Ren, G.~Yu, Y.~He, and G.~Y. Li, ``Collaborative cloud and edge computing
  for latency minimization,'' \emph{IEEE Transactions on Vehicular Technology},
  vol.~68, no.~5, pp. 5031--5044, Mar. 2019.

\bibitem{tian2022comprehensive}
Z.~Tian, L.~Cui, J.~Liang, and S.~Yu, ``A comprehensive survey on poisoning
  attacks and countermeasures in machine learning,'' \emph{ACM Computing
  Surveys}, vol.~55, no.~8, pp. 1--35, Dec. 2022.

\bibitem{liu2018survey}
Q.~Liu, P.~Li, W.~Zhao, W.~Cai, S.~Yu, and V.~C. Leung, ``A survey on security
  threats and defensive techniques of machine learning: A data-driven view,''
  \emph{IEEE Access}, vol.~6, pp. 12\,103--12\,117, Feb. 2018.

\bibitem{xue2023blockchain}
L.~Xue, J.~Ni, D.~Liu, X.~Lin, and X.~Shen, ``Blockchain-based fair and
  fine-grained data trading with privacy preservation,'' \emph{IEEE
  Transactions on Computers}, pp. 1--1, Mar. 2023.

\bibitem{chen2023challenges}
C.~Chen, Z.~Wu, Y.~Lai, W.~Ou, T.~Liao, and Z.~Zheng, ``Challenges and remedies
  to privacy and security in {AIGC}: Exploring the potential of privacy
  computing, blockchain, and beyond,'' \emph{{\rm{arXiv preprint
  arXiv:2306.00419}}}, Jun. 2023.

\bibitem{kang2023adversarial}
J.~Kang, J.~He, H.~Du, Z.~Xiong, Z.~Yang, X.~Huang, and S.~Xie, ``Adversarial
  attacks and defenses for semantic communication in vehicular metaverses,''
  \emph{{\rm{arXiv preprint arXiv:2306.03528}}}, Jun. 2023.

\bibitem{zhang2023split}
S.~Zhang, W.~Wu, P.~Hu, S.~Li, and N.~Zhang, ``Split federated learning: Speed
  up model training in resource-limited wireless networks,'' \emph{{\rm{arXiv
  preprint arXiv:2305.18889}}}, May 2023.

\bibitem{li2021federated}
J.~Li, Y.~Meng, L.~Ma, S.~Du, H.~Zhu, Q.~Pei, and X.~Shen, ``A federated
  learning based privacy-preserving smart healthcare system,'' \emph{IEEE
  Transactions on Industrial Informatics}, vol.~18, no.~3, pp. 2021--2031, Jul.
  2021.

\bibitem{wang2023privacy}
Z.~Wang, G.~Yang, H.~Dai, and C.~Rong, ``Privacy-preserving split learning for
  large-scaled vision pre-training,'' \emph{IEEE Transactions on Information
  Forensics and Security}, vol.~18, pp. 1539--1553, Feb. 2023.

\bibitem{liu2022wireless}
X.~Liu, Y.~Deng, and T.~Mahmoodi, ``Wireless distributed learning: a new hybrid
  split and federated learning approach,'' \emph{IEEE Transactions on Wireless
  Communications}, vol.~22, no.~4, pp. 2650--2665, Oct. 2022.

\bibitem{kang2022blockchain}
J.~Kang, D.~Ye, J.~Nie, J.~Xiao, X.~Deng, S.~Wang, Z.~Xiong, R.~Yu, and
  D.~Niyato, ``Blockchain-based federated learning for industrial metaverses:
  Incentive scheme with optimal aoi,'' in \emph{2022 IEEE International
  Conference on Blockchain (Blockchain)}, Espoo, Finland, Aug. 2022, pp.
  71--78.

\bibitem{9785702}
J.~Kang, X.~Li, J.~Nie, Y.~Liu, M.~Xu, Z.~Xiong, D.~Niyato, and Q.~Yan,
  ``Communication-efficient and cross-chain empowered federated learning for
  artificial intelligence of things,'' \emph{IEEE Transactions on Network
  Science and Engineering}, vol.~9, no.~5, pp. 2966--2977, May 2022.

\bibitem{cui2021security}
L.~Cui, Y.~Qu, G.~Xie, D.~Zeng, R.~Li, S.~Shen, and S.~Yu, ``Security and
  privacy-enhanced federated learning for anomaly detection in {IoT}
  infrastructures,'' \emph{IEEE Transactions on Industrial Informatics},
  vol.~18, no.~5, pp. 3492--3500, Aug. 2021.

\bibitem{augenstein2019generative}
S.~Augenstein, H.~B. McMahan, D.~Ramage, S.~Ramaswamy, P.~Kairouz, M.~Chen,
  R.~Mathews \emph{et~al.}, ``Generative models for effective {ML} on private,
  decentralized datasets,'' \emph{{\rm{arXiv preprint arXiv:1911.06679}}}, Nov.
  2019.

\bibitem{fan2020federated}
C.~Fan and P.~Liu, ``Federated generative adversarial learning,'' in
  \emph{Proceedings of the Pattern Recognition and Computer Vision}, Nanjing,
  China, Oct. 2020, pp. 3--15.

\bibitem{chung2022federated}
J.~Chung, K.~Lee, and K.~Ramchandran, ``Federated unsupervised clustering with
  generative models,'' in \emph{Proceedings of the AAAI International Workshop
  on Trustable, Verifiable and Auditable Federated Learning}, 2022.

\bibitem{wang2021efficient}
Z.~Wang, Y.~Hu, J.~Xiao, and C.~Wu, ``Efficient ring-topology decentralized
  federated learning with deep generative models for industrial artificial
  intelligent,'' \emph{Electronics}, vol.~11, no.~10, p. 1548, May 2022.

\bibitem{shen2022edgematrix}
S.~Shen, Y.~Ren, Y.~Ju, X.~Wang, W.~Wang, and V.~C. Leung, ``Edgematrix: A
  resource-redefined scheduling framework for sla-guaranteed multi-tier
  edge-cloud computing systems,'' \emph{IEEE Journal on Selected Areas in
  Communications}, vol.~41, no.~3, pp. 820--834, Dec. 2023.

\bibitem{gai2020blockchain}
K.~Gai, J.~Guo, L.~Zhu, and S.~Yu, ``Blockchain meets cloud computing: A
  survey,'' \emph{IEEE Communications Surveys \& Tutorials}, vol.~22, no.~3,
  pp. 2009--2030, Apr. 2020.

\bibitem{lin2022blockchain}
Y.~Lin, Z.~Gao, Y.~Tu, H.~Du, D.~Niyato, J.~Kang, and H.~Yang, ``A
  blockchain-based semantic exchange framework for web 3.0 toward participatory
  economy,'' \emph{{\rm{arXiv preprint arXiv:2211.16662}}}, Nov. 2022.

\bibitem{9711561}
Y.~Lin, Z.~Gao, W.~Shi, Q.~Wang, H.~Li, M.~Wang, Y.~Yang, and L.~Rui, ``A novel
  architecture combining oracle with decentralized learning for {IIoT},''
  \emph{IEEE {Internet} of Things Journal}, vol.~10, no.~5, pp. 3774--3785,
  Mar. 2023.

\bibitem{huang2022blockchain}
C.~Huang, W.~Wang, D.~Liu, R.~Lu, and X.~Shen, ``Blockchain-assisted
  personalized car insurance with privacy preservation and fraud resistance,''
  \emph{IEEE Transactions on Vehicular Technology}, vol.~72, no.~3, pp.
  3777--3792, Mar. 2023.

\bibitem{shen2019privacy}
M.~Shen, X.~Tang, L.~Zhu, X.~Du, and M.~Guizani, ``Privacy-preserving support
  vector machine training over blockchain-based encrypted {IoT} data in smart
  cities,'' \emph{IEEE {Internet} of Things Journal}, vol.~6, no.~5, pp.
  7702--7712, Feb. 2019.

\bibitem{shen2022secure}
M.~Shen, H.~Lu, F.~Wang, H.~Liu, and L.~Zhu, ``Secure and efficient
  blockchain-assisted authentication for edge-integrated
  {Internet}-of-vehicles,'' \emph{IEEE Transactions on Vehicular Technology},
  vol.~71, no.~11, pp. 12\,250--12\,263, Jul. 2022.

\bibitem{shen2020blockchain}
M.~Shen, H.~Liu, L.~Zhu, K.~Xu, H.~Yu, X.~Du, and M.~Guizani,
  ``Blockchain-assisted secure device authentication for cross-domain
  industrial {IoT},'' \emph{IEEE Journal on Selected Areas in Communications},
  vol.~38, no.~5, pp. 942--954, Mar. 2020.

\bibitem{xu2022quantum}
M.~Xu, X.~Ren, D.~Niyato, J.~Kang, C.~Qiu, Z.~Xiong, X.~Wang, and V.~Leung,
  ``When quantum information technologies meet blockchain in web 3.0,''
  \emph{{\rm{arXiv preprint arXiv:2211.15941}}}, Nov. 2022.

\bibitem{9969941}
Y.~Lin, J.~Kang, D.~Niyato, Z.~Gao, and Q.~Wang, ``Efficient consensus and
  elastic resource allocation empowered blockchain for vehicular networks,''
  \emph{IEEE Transactions on Vehicular Technology}, pp. 1--6, Dec. 2022.

\bibitem{dirgantoro2020generative}
K.~P. Dirgantoro, J.~M. Lee, and D.-S. Kim, ``Generative adversarial networks
  based on edge computing with blockchain architecture for security system,''
  in \emph{Proceedings of the International Conference on Artificial
  Intelligence in Information and Communication}, Fukuoka, Japan, Feb. 2020,
  pp. 039--042.

\bibitem{tann2022predicting}
W.~J.-W. Tann, A.~Vuputuri, and E.-C. Chang, ``Predicting non-fungible token
  ({NFT}) collections: A contextual generative approach,'' \emph{{\rm{arXiv
  preprint arXiv:2210.15493}}}, Oct. 2022.

\bibitem{li2020blockchain}
Y.~Li, C.~Chen, N.~Liu, H.~Huang, Z.~Zheng, and Q.~Yan, ``A blockchain-based
  decentralized federated learning framework with committee consensus,''
  \emph{IEEE Network}, vol.~35, no.~1, pp. 234--241, Dec. 2020.

\bibitem{liu2021proof}
Y.~Liu, Y.~Lan, B.~Li, C.~Miao, and Z.~Tian, ``Proof of learning (pole):
  {E}mpowering neural network training with consensus building on
  blockchains,'' \emph{Computer Networks}, vol. 201, p. 108594, Dec. 2021.

\bibitem{zhang2023sustainable}
S.~Zhang, M.~Xu, W.~Y.~B. Lim, and D.~Niyato, ``Sustainable {AIGC} workload
  scheduling of geo-{D}istributed data centers: A multi-agent reinforcement
  learning approach,'' \emph{{\rm{arXiv preprint arXiv:2304.07948}}}, Apr.
  2023.

\bibitem{ma2023reliability}
H.~Ma, R.~Li, X.~Zhang, Z.~Zhou, and X.~Chen, ``Reliability-aware online
  scheduling for {DNN} inference tasks in mobile edge computing,'' \emph{IEEE
  {Internet} of Things Journal}, pp. 1--1, Feb. 2023.

\bibitem{wang2022uplink}
Z.~Wang, J.~Zhang, B.~Ai, C.~Yuen, and M.~Debbah, ``Uplink performance of
  cell-free massive {MIMO} with multi-antenna users over jointly-correlated
  rayleigh fading channels,'' \emph{IEEE Transactions on Wireless
  Communications}, vol.~21, no.~9, pp. 7391--7406, 2022.

\bibitem{wang2023uplink}
Z.~Wang, J.~Zhang, H.~Q. Ngo, B.~Ai, and M.~Debbah, ``Uplink precoding design
  for cell-free massive {MIMO} with iteratively weighted mmse,'' \emph{IEEE
  Transactions on Communications}, vol.~71, no.~3, pp. 1646--1664, 2023.

\bibitem{zhu2021multi}
L.~Zhu, J.~Zhang, Z.~Xiao, X.-G. Xia, and R.~Zhang, ``Multi-uav aided
  millimeter-wave networks: Positioning, clustering, and beamforming,''
  \emph{IEEE Transactions on Wireless Communications}, vol.~21, no.~7, pp.
  4637--4653, 2021.

\bibitem{shi2023task}
Y.~Shi, Y.~Zhou, D.~Wen, Y.~Wu, C.~Jiang, and K.~B. Letaief, ``Task-oriented
  communications for {6G}: Vision, principles, and technologies,''
  \emph{{\rm{arXiv preprint arXiv:2303.10920}}}, Mar. 2023.

\bibitem{cheng2017survey}
Y.~Cheng, D.~Wang, P.~Zhou, and T.~Zhang, ``Model compression and acceleration
  for deep neural networks: The principles, progress, and challenges,''
  \emph{IEEE Signal Processing Magazine}, vol.~35, no.~1, pp. 126--136, Jan.
  2018.

\bibitem{li2022compact}
Z.~Li, W.~Su, M.~Xu, R.~Yu, D.~Niyato, and S.~Xie, ``Compact learning model for
  dynamic off-chain routing in blockchain-based {IoT},'' \emph{IEEE Journal on
  Selected Areas in Communications}, vol.~40, no.~12, pp. 3615--3630, Oct.
  2022.

\bibitem{huang2023ai}
Y.~Huang, M.~Xu, X.~Zhang, D.~Niyato, Z.~Xiong, S.~Wang, and T.~Huang,
  ``{AI}-generated {6G} {Internet} design: A diffusion model-based learning
  approach,'' \emph{{\rm{arXiv preprint arXiv:2303.13869}}}, Mar. 2023.

\bibitem{8424832}
A.~El~Saddik, ``Digital twins: The convergence of multimedia technologies,''
  \emph{IEEE MultiMedia}, vol.~25, no.~2, pp. 87--92, Aug. 2018.

\bibitem{clemm2020toward}
A.~Clemm, M.~T. Vega, H.~K. Ravuri, T.~Wauters, and F.~De~Turck, ``Toward truly
  immersive holographic-type communication: Challenges and solutions,''
  \emph{IEEE Communications Magazine}, vol.~58, no.~1, pp. 93--99, Jan. 2020.

\bibitem{chen2023multi}
J.~Chen, J.~Kang, M.~Xu, Z.~Xiong, D.~Niyato, C.~Chen, A.~Jamalipour, and
  S.~Xie, ``Multi-agent deep reinforcement learning for dynamic avatar
  migration in {AIoT}-enabled vehicular metaverses with trajectory
  prediction,'' \emph{{\rm{arXiv preprint arXiv:2306.14683}}}, Jun. 2023.

\bibitem{chen2023multiple}
J.~Chen, J.~Kang, M.~Xu, Z.~Xiong, D.~Niyato, and Y.~Tong, ``Multiple-agent
  deep reinforcement learning for avatar migration in vehicular metaverses,''
  in \emph{Companion Proceedings of the ACM Web Conference 2023}, Austin, TX,
  Apr. 2023, pp. 1258--1265.

\end{thebibliography}

\end{document}